\pdfoutput=1

\documentclass[reqno,12pt,a4paper]{article}

\usepackage{amsfonts,amssymb,amsthm, 
amsmath,amscd, bbm, bm, mathabx, %bbold,
mathrsfs,delarray,subfigure}

\usepackage[dvipsnames]{xcolor}[svgnames,x11names]

\usepackage{hyperref}
\usepackage{cite}
\usepackage{xypic}
\usepackage{sectsty}

\definecolor{myblue}{rgb}{0.45, 0.45, 1}
\usepackage{fancyhdr,color}
\fancyheadoffset[RE,LO]{0.\textwidth} % to make a top rule of length equal to textwidth
\hypersetup{
  colorlinks,
  linkcolor=blue,
  linktoc=all,
  citecolor=blue,
  urlcolor=blue
}

\usepackage{sectsty}

\colorlet{sectitlecolor}{blue}
\colorlet{sectboxcolor}{white}
\colorlet{secnumcolor}{blue}

\sectionfont{\color{sectitlecolor}}

\makeatletter
\renewcommand\@seccntformat[1]{%
  \colorbox{sectboxcolor}{\textcolor{secnumcolor}{\csname the#1\endcsname}}%
  \quad
}
\makeatother
\usepackage{etoolbox}
\patchcmd{\thebibliography}{\section*{\refname}}{}{}{}

\DeclareSymbolFont{sfletters}{OML}{cmbrm}{m}{it}  %&&&

\DeclareMathSymbol{\sfGamma}{\mathord}{sfletters}{"00}
\DeclareMathSymbol{\sfDelta}{\mathord}{sfletters}{"01}
\DeclareMathSymbol{\sfTheta}{\mathord}{sfletters}{"02}
\DeclareMathSymbol{\sfLambda}{\mathord}{sfletters}{"03}
\DeclareMathSymbol{\sfXi}{\mathord}{sfletters}{"04}
\DeclareMathSymbol{\sfPi}{\mathord}{sfletters}{"05}
\DeclareMathSymbol{\sfSigma}{\mathord}{sfletters}{"06}
\DeclareMathSymbol{\sfUpsilon}{\mathord}{sfletters}{"07}
\DeclareMathSymbol{\sfPhi}{\mathord}{sfletters}{"08}
\DeclareMathSymbol{\sfPsi}{\mathord}{sfletters}{"09}
\DeclareMathSymbol{\sfOmega}{\mathord}{sfletters}{"0A}
\DeclareMathSymbol{\sfalpha}{\mathord}{sfletters}{"0B}
\DeclareMathSymbol{\sfbeta}{\mathord}{sfletters}{"0C}
\DeclareMathSymbol{\sfgamma}{\mathord}{sfletters}{"0D}
\DeclareMathSymbol{\sfdelta}{\mathord}{sfletters}{"0E}
\DeclareMathSymbol{\sfepsilon}{\mathord}{sfletters}{"0F}
\DeclareMathSymbol{\sfzeta}{\mathord}{sfletters}{"10}
\DeclareMathSymbol{\sfeta}{\mathord}{sfletters}{"11}
\DeclareMathSymbol{\sftheta}{\mathord}{sfletters}{"12}
\DeclareMathSymbol{\sfiota}{\mathord}{sfletters}{"13}
\DeclareMathSymbol{\sfkappa}{\mathord}{sfletters}{"14}
\DeclareMathSymbol{\sflambda}{\mathord}{sfletters}{"15}
\DeclareMathSymbol{\sfmu}{\mathord}{sfletters}{"16}
\DeclareMathSymbol{\sfnu}{\mathord}{sfletters}{"17}
\DeclareMathSymbol{\sfxi}{\mathord}{sfletters}{"18}
\DeclareMathSymbol{\sfpi}{\mathord}{sfletters}{"19}
\DeclareMathSymbol{\sfrho}{\mathord}{sfletters}{"1A}
\DeclareMathSymbol{\sfsigma}{\mathord}{sfletters}{"1B}
\DeclareMathSymbol{\sftau}{\mathord}{sfletters}{"1C}
\DeclareMathSymbol{\sfupsilon}{\mathord}{sfletters}{"1D}
\DeclareMathSymbol{\sfphi}{\mathord}{sfletters}{"1E}
\DeclareMathSymbol{\sfchi}{\mathord}{sfletters}{"1F}
\DeclareMathSymbol{\sfpsi}{\mathord}{sfletters}{"20}
\DeclareMathSymbol{\sfomega}{\mathord}{sfletters}{"21}
\DeclareMathSymbol{\sfvarepsilon}{\mathord}{sfletters}{"22}
\DeclareMathSymbol{\sfvartheta}{\mathord}{sfletters}{"23}
\DeclareMathSymbol{\sfvarpi}{\mathord}{sfletters}{"24}
\DeclareMathSymbol{\sfvarrho}{\mathord}{sfletters}{"25}
\DeclareMathSymbol{\sfvarsigma}{\mathord}{sfletters}{"26}
\DeclareMathSymbol{\sfvarphi}{\mathord}{sfletters}{"27}

\DeclareMathSymbol{\spartial}{\mathord}{sfletters}{"40}

\DeclareMathSymbol{\sfA}{\mathord}{sfletters}{"41}
\DeclareMathSymbol{\sfB}{\mathord}{sfletters}{"42}
\DeclareMathSymbol{\sfC}{\mathord}{sfletters}{"43}
\DeclareMathSymbol{\sfD}{\mathord}{sfletters}{"44}
\DeclareMathSymbol{\sfE}{\mathord}{sfletters}{"45}
\DeclareMathSymbol{\sfF}{\mathord}{sfletters}{"46}
\DeclareMathSymbol{\sfG}{\mathord}{sfletters}{"47}
\DeclareMathSymbol{\sfH}{\mathord}{sfletters}{"48}
\DeclareMathSymbol{\sfI}{\mathord}{sfletters}{"49}
\DeclareMathSymbol{\sfJ}{\mathord}{sfletters}{"4A}
\DeclareMathSymbol{\sfK}{\mathord}{sfletters}{"4B}
\DeclareMathSymbol{\sfL}{\mathord}{sfletters}{"4C}
\DeclareMathSymbol{\sfM}{\mathord}{sfletters}{"4D}
\DeclareMathSymbol{\sfN}{\mathord}{sfletters}{"4E}
\DeclareMathSymbol{\sfO}{\mathord}{sfletters}{"4F}
\DeclareMathSymbol{\sfP}{\mathord}{sfletters}{"50}
\DeclareMathSymbol{\sfQ}{\mathord}{sfletters}{"51}
\DeclareMathSymbol{\sfR}{\mathord}{sfletters}{"52}
\DeclareMathSymbol{\sfS}{\mathord}{sfletters}{"53}
\DeclareMathSymbol{\sfT}{\mathord}{sfletters}{"54}
\DeclareMathSymbol{\sfU}{\mathord}{sfletters}{"55}
\DeclareMathSymbol{\sfV}{\mathord}{sfletters}{"56}
\DeclareMathSymbol{\sfW}{\mathord}{sfletters}{"57}
\DeclareMathSymbol{\sfX}{\mathord}{sfletters}{"58}
\DeclareMathSymbol{\sfY}{\mathord}{sfletters}{"59}
\DeclareMathSymbol{\sfZ}{\mathord}{sfletters}{"5A}
\DeclareMathSymbol{\sfa}{\mathord}{sfletters}{"61}
\DeclareMathSymbol{\sfb}{\mathord}{sfletters}{"62}
\DeclareMathSymbol{\sfc}{\mathord}{sfletters}{"63}
\DeclareMathSymbol{\sfd}{\mathord}{sfletters}{"64}
\DeclareMathSymbol{\sfe}{\mathord}{sfletters}{"65}
\DeclareMathSymbol{\sff}{\mathord}{sfletters}{"66}
\DeclareMathSymbol{\sfg}{\mathord}{sfletters}{"67}
\DeclareMathSymbol{\sfh}{\mathord}{sfletters}{"68}
\DeclareMathSymbol{\sfi}{\mathord}{sfletters}{"69}
\DeclareMathSymbol{\sfj}{\mathord}{sfletters}{"6A}
\DeclareMathSymbol{\sfk}{\mathord}{sfletters}{"6B}
\DeclareMathSymbol{\sfl}{\mathord}{sfletters}{"6C}
\DeclareMathSymbol{\sfm}{\mathord}{sfletters}{"6D}
\DeclareMathSymbol{\sfn}{\mathord}{sfletters}{"6E}
\DeclareMathSymbol{\sfo}{\mathord}{sfletters}{"6F}
\DeclareMathSymbol{\sfp}{\mathord}{sfletters}{"70}
\DeclareMathSymbol{\sfq}{\mathord}{sfletters}{"71}
\DeclareMathSymbol{\sfr}{\mathord}{sfletters}{"72}
\DeclareMathSymbol{\sfs}{\mathord}{sfletters}{"73}
\DeclareMathSymbol{\sft}{\mathord}{sfletters}{"74}
\DeclareMathSymbol{\sfu}{\mathord}{sfletters}{"75}
\DeclareMathSymbol{\sfv}{\mathord}{sfletters}{"76}
\DeclareMathSymbol{\sfw}{\mathord}{sfletters}{"77}
\DeclareMathSymbol{\sfx}{\mathord}{sfletters}{"78}
\DeclareMathSymbol{\sfy}{\mathord}{sfletters}{"79}
\DeclareMathSymbol{\sfz}{\mathord}{sfletters}{"7A}

\usepackage[LGR,T1]{fontenc}
\usepackage{amsmath,etoolbox}

\newcommand{\declarebsfgreek}[2]{%
  \protected\csdef{bsf#1}{\mathord{\text{\bsfgreekfont#2}}}%
}
\newcommand{\bsfgreekfont}{\usefont{LGR}{cmss}{bx}{it}}% change the family

\declarebsfgreek{alpha}{a}
\declarebsfgreek{beta}{b}
\declarebsfgreek{gamma}{g}
\declarebsfgreek{delta}{d}
\declarebsfgreek{epsilon}{e}
\declarebsfgreek{zeta}{z}
\declarebsfgreek{eta}{h}
\declarebsfgreek{theta}{j}
\declarebsfgreek{iota}{i}
\declarebsfgreek{kappa}{k}
\declarebsfgreek{lambda}{l}
\declarebsfgreek{mu}{m}
\declarebsfgreek{nu}{n}
\declarebsfgreek{xi}{x}
\declarebsfgreek{omicron}{o}
\declarebsfgreek{pi}{p}
\declarebsfgreek{rho}{r}
\declarebsfgreek{sigma}{s}
\declarebsfgreek{tau}{t}
\declarebsfgreek{upsilon}{u}
\declarebsfgreek{phi}{f}
\declarebsfgreek{chi}{q}
\declarebsfgreek{psi}{y}
\declarebsfgreek{omega}{w}
\declarebsfgreek{varsigma}{c}

\declarebsfgreek{Gamma}{G}
\declarebsfgreek{Delta}{D}
\declarebsfgreek{Upsilon}{U}
\declarebsfgreek{Omega}{W}
\declarebsfgreek{Theta}{J}
\declarebsfgreek{Lambda}{L}
\declarebsfgreek{Xi}{X}
\declarebsfgreek{Pi}{P}
\declarebsfgreek{Sigma}{S}
\declarebsfgreek{Phi}{F}
\declarebsfgreek{Psi}{Y}

\newcommand{\declarebsfitalic}[2]{%
  \protected\csdef{bsf#1}{\mathord{\text{\bsfitalicfont#2}}}%
}
\newcommand{\bsfitalicfont}{\usefont{T1}{cmss}{bx}{it}}% change the family

\declarebsfitalic{a}{a}
\declarebsfitalic{b}{b}
\declarebsfitalic{c}{c}
\declarebsfitalic{d}{d}
\declarebsfitalic{e}{e}
\declarebsfitalic{f}{f}
\declarebsfitalic{g}{g}
\declarebsfitalic{h}{h}
\declarebsfitalic{i}{i}
\declarebsfitalic{j}{j}
\declarebsfitalic{k}{k}
\declarebsfitalic{l}{l}
\declarebsfitalic{m}{m}
\declarebsfitalic{n}{n}
\declarebsfitalic{o}{o}
\declarebsfitalic{p}{p}
\declarebsfitalic{q}{q}
\declarebsfitalic{r}{r}
\declarebsfitalic{s}{s}
\declarebsfitalic{t}{t}
\declarebsfitalic{u}{u}
\declarebsfitalic{v}{v}
\declarebsfitalic{x}{x}
\declarebsfitalic{y}{y}
\declarebsfitalic{z}{z}

\declarebsfitalic{A}{A}
\declarebsfitalic{B}{B}
\declarebsfitalic{C}{C}
\declarebsfitalic{D}{D}
\declarebsfitalic{E}{E}
\declarebsfitalic{F}{F}
\declarebsfitalic{G}{G}
\declarebsfitalic{H}{H}
\declarebsfitalic{I}{I}
\declarebsfitalic{J}{J}
\declarebsfitalic{K}{K}
\declarebsfitalic{L}{L}
\declarebsfitalic{M}{M}
\declarebsfitalic{N}{N}
\declarebsfitalic{O}{O}
\declarebsfitalic{P}{P}
\declarebsfitalic{Q}{Q}
\declarebsfitalic{R}{R}
\declarebsfitalic{S}{S}
\declarebsfitalic{T}{T}
\declarebsfitalic{U}{U}
\declarebsfitalic{V}{V}
\declarebsfitalic{X}{X}
\declarebsfitalic{Y}{Y}
\declarebsfitalic{Z}{Z}

\newcommand{\rmb}{{\mathrm{b}}}
\newcommand{\rmc}{{\mathrm{c}}}

\newcommand{\rme}{{\mathrm{e}}}

\newcommand{\rmw}{{\mathrm{w}}}

\newcommand{\rmA}{{\mathrm{A}}}
\newcommand{\rmB}{{\mathrm{B}}}
\newcommand{\rmC}{{\mathrm{C}}}

\newcommand{\rmE}{{\mathrm{E}}}
\newcommand{\rmF}{{\mathrm{F}}}
\newcommand{\rmG}{{\mathrm{G}}}
\newcommand{\rmH}{{\mathrm{H}}}

\newcommand{\rmK}{{\mathrm{K}}}

\newcommand{\rmP}{{\mathrm{P}}}
\newcommand{\rmQ}{{\mathrm{Q}}}

\newcommand{\rmS}{{\mathrm{S}}}
\newcommand{\rmT}{{\mathrm{T}}}
\newcommand{\rmU}{{\mathrm{U}}}
\newcommand{\rmV}{{\mathrm{V}}}
\newcommand{\rmW}{{\mathrm{W}}}
\newcommand{\rmX}{{\mathrm{X}}}
\newcommand{\rmY}{{\mathrm{Y}}}
\newcommand{\rmZ}{{\mathrm{Z}}}

\newcommand{\msb}{{\mathsans{b}}}

\newcommand{\msg}{{\mathsans{g}}}
\newcommand{\msh}{{\mathsans{h}}}
\newcommand{\msi}{{\mathsans{i}}}

\newcommand{\msk}{{\mathsans{k}}}

\newcommand{\msq}{{\mathsans{q}}}
\newcommand{\msr}{{\mathsans{r}}}

\newcommand{\msv}{{\mathsans{v}}}
\newcommand{\msw}{{\mathsans{w}}}

\newcommand{\msA}{{\mathsans{A}}}

\newcommand{\msC}{{\mathsans{C}}}
\newcommand{\msD}{{\mathsans{D}}}
\newcommand{\msE}{{\mathsans{E}}}
\newcommand{\msF}{{\mathsans{F}}}
\newcommand{\msG}{{\mathsans{G}}}
\newcommand{\msH}{{\mathsans{H}}}
\newcommand{\msI}{{\mathsans{I}}}
\newcommand{\msJ}{{\mathsans{J}}}
\newcommand{\msK}{{\mathsans{K}}}
\newcommand{\msL}{{\mathsans{L}}}
\newcommand{\msM}{{\mathsans{M}}}
\newcommand{\msN}{{\mathsans{N}}}

\newcommand{\msQ}{{\mathsans{Q}}}
\newcommand{\msR}{{\mathsans{R}}}

\newcommand{\msT}{{\mathsans{T}}}
\newcommand{\msU}{{\mathsans{U}}}
\newcommand{\msV}{{\mathsans{V}}}
\newcommand{\msW}{{\mathsans{W}}}

\newcommand{\fkc}{{\mathfrak{c}}}

\newcommand{\fke}{{\mathfrak{e}}}
\newcommand{\fkf}{{\mathfrak{f}}}
\newcommand{\fkg}{{\mathfrak{g}}}
\newcommand{\fkh}{{\mathfrak{h}}}

\newcommand{\fkj}{{\mathfrak{j}}}

\newcommand{\fkl}{{\mathfrak{l}}}
\newcommand{\fkm}{{\mathfrak{m}}}
\newcommand{\fkn}{{\mathfrak{n}}}

\newcommand{\fkq}{{\mathfrak{q}}}

\newcommand{\fkt}{{\mathfrak{t}}}
\newcommand{\fku}{{\mathfrak{u}}}
\newcommand{\fkv}{{\mathfrak{v}}}

\newcommand{\bbC}{{\mathbb{C}}}

\newcommand{\bbR}{{\mathbb{R}}}

\newcommand{\bbZ}{{\mathbb{Z}}}

\newcommand{\scD}{{\matheul{D}}}

\newcommand{\scH}{{\matheul{H}}}
\newcommand{\scI}{{\matheul{I}}}

\newcommand{\clE}{{\mathcal{E}}}
\newcommand{\clF}{{\mathcal{F}}}
\newcommand{\clG}{{\mathcal{G}}}
\newcommand{\clH}{{\mathcal{H}}}
\newcommand{\clI}{{\mathcal{I}}}

\newcommand{\clK}{{\mathcal{K}}}
\newcommand{\clL}{{\mathcal{L}}}

\newcommand{\clO}{{\mathcal{O}}}

\usepackage{accents}

\sectionfont{\fontsize{12}{15}\selectfont}
\subsectionfont{\fontsize{12}{15}\selectfont}

\usepackage{enumitem}

\DeclareMathOperator{\ad}{ad}
\DeclareMathOperator{\Ad}{Ad}
\DeclareMathOperator{\vol}{vol}
\DeclareMathOperator{\im}{im}
 
\DeclareMathOperator{\id}{id}

\DeclareMathOperator{\Hom}{Hom}

\DeclareMathOperator{\Fun}{Fun}
\DeclareMathOperator{\iFun}{\textsc{Fun}} %{\underline{Fun}}
\DeclareMathOperator{\Conn}{Conn}
\DeclareMathOperator{\iConn}{\textsc{Conn}}

\DeclareMathOperator{\Map}{Map}
\DeclareMathOperator{\iMap}{\textsc{Map}}   %{\underline{Map}}
\DeclareMathOperator{\iDFnc}{\textsc{DFnc}}

\DeclareMathOperator{\Emb}{Emb}

\DeclareMathOperator{\Vect}{Vect}
\DeclareMathOperator{\iVect}{\textsc{Vect}}
\DeclareMathOperator{\Gau}{Gau}
\DeclareMathOperator{\iGau}{\textsc{Gau}}

\DeclareMathOperator{\ev}{ev}
\DeclareMathOperator{\tgr}{tgr}

\DeclareMathOperator{\Aut}{Aut}
\DeclareMathOperator{\iAut}{\textsc{Aut}}

\DeclareMathOperator{\Inn}{Inn}
\DeclareMathOperator{\INN}{\textsc{Inn}}
\DeclareMathOperator{\INV}{\textsc{Inv}}
\DeclareMathOperator{\tr}{tr}

\DeclareMathOperator{\ee}{e}

\DeclareMathOperator{\AD}{\textsc{Ad}}     %{AD}

\DeclareMathOperator{\BB}{B}
\DeclareMathOperator{\W}{We\hspace{-1pt}}
\DeclareMathOperator{\iW}{\textsc{We}\hspace{-1pt}}
\DeclareMathOperator{\WO}{OWe\hspace{-1pt}}
\DeclareMathOperator{\iWO}{\textsc{OWe}\hspace{-1pt}}  %{\textsf{OW\hspace{.25pt}{\footnotesize E}}\hspace{-1pt}}
\DeclareMathOperator{\NN}{N\hspace{-1pt}}

\DeclareMathOperator{\WW}{W\hspace{-1pt}}

\DeclareMathOperator{\ZZ}{Z\hspace{-1pt}}
\DeclareMathOperator{\SD}{SD\hspace{-1pt}}
\DeclareMathOperator{\DD}{D\hspace{-1pt}}
\DeclareMathOperator{\FF}{F\hspace{-1pt}}

\DeclareMathOperator{\iOOO}{\textsc{Op}\hspace{-1pt}}     %{\underline{O}\hspace{-1pt}}

\DeclareMathOperator{\Texp}{Texp}
\DeclareMathOperator{\Pexp}{Pexp}
\DeclareMathOperator{\Sexp}{Sexp}

\numberwithin{equation}{subsection} 
\numberwithin{subsection}{section} 

\newcommand{\ceqref}[1]{{\textcolor{blue}{\eqref{#1}}}}
\newcommand{\cref}[1]{{\textcolor{blue}{\ref{#1}}}}
\newcommand{\ccite}[1]{{\textcolor{blue}{\!\cite{#1}}}}

\newcommand{\sss}{{\hbox{\large $\sum$}}}

\newcommand{\ddd}{{\hbox{\large $\bigoplus$}}}

\newcommand{\ul}[1]{{\underline{#1}}}

\newcommand{\sdot}{\hspace{.5pt}\dot{}\hspace{.5pt}}
\newcommand{\hfpt}{\hspace{.75pt}}
\newcommand{\mhfpt}{\hspace{-.75pt}}
\newcommand{\dd}{\text{d}}

\newcommand{\mathsans}[1]{{{\sf #1}}}
\font\euler=eusm10 at 12.8 truept
\font\scripteuler=eusm7
\font\scriptscripteuler=eusm5 
\def\eul{\fam=12}
\newcommand{\matheul}[1]{{{\eul #1}}}

\newtheorem{defi}{{\sf Definition}}[section]
\newtheorem{prop}{{\sf Proposition}}[section]

\newtheorem{lemma}{{\sf Lemma}}[section]

\DeclareMathSymbol{*}{\mathbin}{symbols}{"03} % \ast

\begin{document}

\thispagestyle{empty} % to have an unnumbered page 

%\hrule\vskip.5cm
%\hbox to 14.5 truecm{October 2016 \hfil DIFA 16}
%\hbox to 14.5 truecm{Version 2 \hfil}
%\vskip.5cm\hrule
\vskip1.5cm
\begin{large}
{\flushleft\textcolor{blue}{\sffamily\bfseries Quantum field theoretic representation of Wilson surfaces:}}  
{\flushleft\textcolor{blue}{\sffamily\bfseries I higher coadjoint orbit theory}}  
\end{large}
\vskip1.3cm
\hrule height 1.5pt
\vskip1.3cm
{\flushleft{\sffamily \bfseries Roberto Zucchini}\\
\it Department of Physics and Astronomy,\\
University of Bologna,\\
I.N.F.N., Bologna division,\\
viale Berti Pichat, 6/2\\
Bologna, Italy\\
Email: \textcolor{blue}{\tt \href{mailto:roberto.zucchini@unibo.it}{roberto.zucchini@unibo.it}}, 
\textcolor{blue}{\tt \href{mailto:zucchinir@bo.infn.it}{zucchinir@bo.infn.it}}}

%roberto.zucchini@unibo.it, zucchinir@bo.infn.it}

\vskip.7cm
%\hrule height 1.5pt
\vskip.6cm 
{\flushleft\sc% \sffamily \bfseries 
Abstract:} 
%\par\noindent
This is the first of a series of two papers devoted to the partition function realization
of Wilson surfaces in strict higher gauge theory. A higher version of the 
Kirillov-Kostant-Souriau theory of coadjoint orbits is presented based on the derived geometric
framework, which has shown its usefulness in 4--dimensional higher Chern--Simons theory. 
An original notion of derived coadjoint orbit is put forward. 
A theory of derived unitary line bundles and Poisson structures on regular derived orbits
is constructed.
The proper derived counterpart of the Bohr--Sommerfeld quantization condition is then identified. 
A version of derived prequantization is proposed. The difficulties hindering a full quantization, shared
with other approaches to higher quantization, are pinpointed and a possible way--out is suggested.
The theory we elaborate  provide the geometric underpinning for the field theoretic constructions 
of the companion paper.

%In the present paper, we propose and describe a higher analog of Borel--Weil--Bott
%theory for Lie group crossed module. Applications to higher Chern--Simons theory
%are considered. 
\vspace{2mm}
\par\noindent
MSC: 81T13 81T20 81T45  

\vfil\eject

{\color{blue}\tableofcontents}

\vfil\eject

\vfil\eject

\renewcommand{\sectionmark}[1]{\markright{\thesection\ ~~#1}}

%\markright{\textcolor{blue}{\sffamily 1 ~~Introduction}}

\section{\textcolor{blue}{\sffamily Introduction}}\label{sec:intro}

\vspace{-1.5mm}

Wilson loops were introduced by Wilson in 1974 \ccite{Wilson:1974sk}
as a natural set of gauge invariant variables suitable for the description of the non
perturbative regime of quantum chromodynamics. Since then, they have been widely employed
in lattice gauge theory. 

In the loop formulation of gauge theory \ccite{Gambini:1980wm,Gambini:1986ew}, the quantum Hilbert space consists
of gauge invariant wave functionals on the gauge field configuration space. According to a theorem of 
Giles \ccite{Giles:1981ej}, Wilson loops constitute a basis of the Hilbert space allowing to switch
from the gauge field to the loop representation. 

Wilson loops are 
%were proposed by Rovelli and Smolin in 1987 \ccite{Rovelli:1987df} as
fundamental constitutive elements of a canonical
formulation of quantum gravity as a gauge theory, known as loop quantum gravity \ccite{Rovelli:1987df},
and their incorporation has led to the very %later more
powerful spin network and foam approaches of this latter \ccite{Rovelli:1995ac}.

Wilson loops are relevant also in condensed matter physics at low energy, specifically 
in the study of topologically ordered phases of matter described by topological quantum
field theories. In models of fractional quantum Hall states as well as lattice
models such as Kitaev's toric code \ccite{Kitaev:2003fta}, fractional braiding statistics between quasiparticles
emerges through the correlation function of a pair of Wilson loops forming a Hopf link
\ccite{Levin:2004mi,Walker:2011mda}. 

Wilson loops depend on the topology of the underlying knots
and, as shown in Witten's foundational work \ccite{Witten:1988hf}, 
they can be employed to study knot topology in 3--dimensional Chern--Simons (CS) theory
using basic techniques of quantum field theory.
CS correlators of Wilson loop operators provide a variety of knot and link invariants.

Higher gauge theory is a generalization of ordinary gauge theory where gauge fields
are higher degree forms \ccite{Baez:2010ya,Saemann:2016sis}. 
%(see refs. \ccite{Baez:2010ya,Saemann:2016sis}
%for readable reviews of the subject and extensive referencing).
%Higher gauge theory
It is considered to be a promising candidate for the description of the dynamics
of the higher dimensional extended objects occurring in supergravity and string theory
thought to be the basic constituents of matter and mediators of fundamental interactions
(see \ccite{Jurco:2019woz} for an updated general overview).
Higher gauge theory is relevant also in spin foam theory \ccite{Baez:1999sr}
and condensed matter physics \ccite{Zhu:2018kzd}. 

Wilson surfaces \ccite{Alvarez:1997ma,Chepelev:2001mg,Martins:2007,Schrei:2011, 
Soncini:2014zra,Zucchini:2015wba,Zucchini:2015xba,Zucchini:2019mbz},
2--dimensional counterparts of Wilson loops, emerge naturally in theories with
higher form gauge fields such as those mentioned in the previous paragraph
and are so expected to be relevant in the analysis of various basic aspects 
of them for reasons analogous to those for which Wilson loops are. %, as we briefly exemplify next.

%\ccite{Alekseev:2015hda,Chekeres:2018kmh,Chekeres:2019xit}

In 4 spacetime dimensions, particle--like excitations cannot braid and have only ordinary bosonic/fermionic
statistics. Fractional braiding statistics can still occur through the braiding of
either a point--like and a loop--like or two loop--like excitations. This has been adequately described
through the correlation functions of Wilson loops and surfaces in BF type topological quantum field theories
\ccite{Balachandran:1992qg,Bergeron:1994ym,Szabo:1998ej}.

Wilson surfaces also should be a basic element of any 
field theoretic approach to 4-dimensional 2--knot topology \ccite{CottaRamusino:1994ez,Cattaneo:2002tk}.
Based on Witten's paradigm, 
it should be possible to study surface knot topology in 4--dimensions computing correlators
of Wilson surfaces in an appropriate 4--dimensional version of CS theory
using again techniques of quantum field theory \ccite{Soncini:2014ara,Zucchini:2015ohw,Zucchini:2021bnn}.

%\ccite{Soncini:2014zra,Zucchini:2015wba,Zucchini:2015xba,Alekseev:2015hda,Chekeres:2018kmh,Zucchini:2019mbz}.  

The aim of the present two--part study is constructing a 2--dimensional
topological sigma model whose quantum partition function yields a Wilson surface
in strict higher gauge theory on the same lines as the 1--dimensional topological 
sigma model providing a Wilson loop in ordinary gauge theory.
The path we have in mind is described in the next subsections.

%\vfil\eject

\subsection{\textcolor{blue}{\sffamily Wilson loops as partition functions 
%and their functional integral representation
}}\label{subsec:wilsurf}

The idea of representing a given Wilson loop as the partition function
of a 1--dimensional sigma model has a long history.
In the context of 4--dimensional Yang-Mills theory, this formulation %of $W_R(C)$
can be traced back to the work of Balachandran {\it et al.}
\ccite{Balachandran:1977ub}. The approach was subsequently
developed by Alekseev {\it et al.} in \ccite{Alekseev:1988vx} 
and Diakonov and Petrov in \ccite{Diakonov:1989fc,Diakonov:1996zu}.
More recently, it was applied to the canonical quantization of CS %Chern--Simons
theory by Elitzur {\it et al.} in \ccite{Elitzur:1989nr}.
The functional integral expression of a Wilson loop 
holds in fact in general for any gauge theory in
any dimension. Below, we briefly outline the principles on which this theoretical framework
is based. 
See also \ccite{Witten:1999ams,Beasley:2009mb,Alexandrov:2011ab}
for clear illustrations of the underlying theory and some of its most significant applications.

The definition of ordinary Wilson loops in gauge theory is well--known. 
In a gauge theory with gauge group $\msG$, a Wilson loop 
$W_R(C)$ depends on a representation $R$ of $\msG$ and an oriented loop $C$
in the spacetime manifold $M$ and is given by 
the gauge invariant trace in $R$ of the holonomy of the gauge field $\omega$ along $C$, 
\begin{equation}
W_R(C)=\tr_R\Pexp\left(-\int\nolimits_{\hfpt C}\omega\right)\!.
\label{phys1}
\end{equation}
This description of $W_R(C)$ is intrinsically quantum mechanical \ccite{Witten:1988hf}.
Expression \ceqref{phys1} clearly indicates that $W_R(C)$ may be identified with
the partition function of some auxiliary fictitious quantum system. 
The representation space of $R$ corresponds to the Hilbert space $\clH$ of the system,
the trace over $R$ to the usual trace over $\clH$ and the gauge field $\omega$ specifying 
holonomy to the Hamiltonian operator $H$ governing time-evolution. 
The correspondence established in this way is summarized schematically by 
\begin{equation}
R\,\longleftrightarrow\,\clH,\qquad \omega \,\longleftrightarrow\,iH.
%\Pexp\left(-\int\nolimits_{\hfpt C}\omega\right)\,\longleftrightarrow\,
%\Texp\left(-i \int\nolimits_{\hfpt C}H\right),\quad \tr_R\,\longleftrightarrow\,\tr_\clH. 
\label{phys4}
\end{equation}
From such a standpoint, $W_R(C)$ takes so the form 
\begin{equation}
W_R(C)=\tr_\clH\Texp\left(-i\int\nolimits_{\hfpt C}H\right)\!.
\label{phys3}
\end{equation}

%Anticipating a little bit, at the risk of being somewhat vague, %In fact, e
%is then the system's partition function, % in the Hamiltonian formalism.

The inherent quantum mechanical nature of the Wilson loop $W_R(C)$
rules out any possibility of performing semiclassical path
integral manipulations in gauge theory with Wilson loop insertions 
based solely on the expression \ceqref{phys1}. If we still want to achieve a description
of such a kind, an independent fully semi-classical functional integral expression of $W_R(C)$
is required.

The description of $W_R(C)$ alluded to in the previous paragraph
must necessarily be based upon a 1--dimensional field theory on the loop $C$
compatible with the correspondence \ceqref{phys4} and its implications
upon quantization. The gauge group $\msG$ should further act 
as a symmetry group to account for the gauge invariance of $W_R(C)$.
Therefore, it is plausible that this field theory is a 1--dimensional
sigma model featuring a $\msG$--valued auxiliary bosonic field $g$ coupled to the gauge field 
$\omega$ acting as a background field. \pagebreak The expression of $W_R(C)$ we are aiming to
should then have the schematic form \hphantom{xxxxxxxx}
\begin{equation}
W_R(C)=\int\scD g\exp\left(i\,\sfS_R(g,\omega)\right),
\label{phys2}
\end{equation}
where $\sfS_R(g,\omega)$ is a gauge-invariant action
functional of $g$ and the restriction of $\omega$ to $C$
depending upon the representation $R$ given by an integral
on $C$ of a local Lagrangian density.

%\vfil\eject

\subsection{\textcolor{blue}{\sffamily Wilson loops and coadjoint orbits
}}\label{subsec:wilscoadj}

The quantum system underlying the partition function
realization of a Wilson loop can be described quite explicitly
as we review below. 

We assume that $\msG$ is a compact semisimple Lie group and that
$R$ is an irreducible representation of $\msG$. $R$ is uniquely characterized up to equivalence
by its highest weight $\lambda$ and so we write $R=R_\lambda$.
As is well--known, $\lambda\in\Lambda_{\msw\msG}{}^+$, the lattice
of dominant weights of $\msG$ in the dual space $\fkg^*$ of the Lie algebra $\fkg$ of $\msG$.

In general, with any element $\lambda\in\fkg^*$ there is associated the coadjoint orbit
$\clO_\lambda=\{\Ad^*\gamma(\lambda)|\gamma\in\msG\}$. $\clO_\lambda$ is a
homogeneous space: $\clO_\lambda=\msG/\msG_\lambda$, where $\msG_\lambda$ is the stabilizer
subgroup of $\lambda$. $\msG$ is in this way structured as a principal $\msG_\lambda$--bundle over $\clO_\lambda$. 
%Forms on $\clO_\lambda$ are so representable as horizontal and invariant forms on $\msG$. 
Forms on $\clO_\lambda$ are thus representable as forms on $\msG$ which are %basic, that is
horizontal and invariant with respect to the multiplicative right $\msG_\lambda$--action.

The left multiplicative action of $\msG$ on itself induces owing
to its commutativity with the right $\msG_\lambda$--action a $\msG$--action
on the coadjoint orbit $\clO_\lambda$. This action constitutes a primal property
of $\clO_\lambda$.

In Kirillov--Kostant--Souriau (KKS) theory \ccite{Kirillov:2004lom}, 
the coadjoint orbit $\clO_\lambda$ is promoted to a symplectic manifold by %endowing
equipping it with the symplectic 2--form
\begin{equation}
\nu_\lambda=\frac{1}{2}\langle\lambda,[\gamma^{-1}d\gamma,\gamma^{-1}d\gamma]\rangle,
\label{phys5}
\end{equation}
where $\langle\cdot,\cdot\rangle$ is the duality pairing of $\fkg$ and $\fkg^*$.
In this way, $\clO_\lambda$ is endowed with a Poisson bracket structure $\{\cdot,\cdot\}$.
%and so may be regarded as a phase space.
$\nu_\lambda$ is invariant under the $\msG$--action. This latter is actually Hamiltonian.
Its moment map $q_\lambda:\fkg\rightarrow\rmC^\infty(\clO_\lambda)$ is given by
\begin{equation}
q_\lambda(x)=\langle\Ad^*\gamma(\lambda),x\rangle
\label{phys6}
\end{equation}
with $x\in\fkg$ and satisfies the Poisson bracket
\begin{equation}
\{q_\lambda(x),q_\lambda(y)\}=q_\lambda([x,y])
\label{phys7}
\end{equation}
for $x,y\in\fkg$. As $q$ is a Lie algebra morphism, the Hamiltonian governing %the dynamics of 
the classical system underlying the Wilson loop $W_{R_\lambda}(C)$ is assumably %identified with
\begin{equation}
H_\lambda=q_\lambda(\varsigma^*\omega),%\langle\lambda,\Ad\hamma^{-1}(\varsigma^*\omega)\rangle
\label{phys8}
\end{equation}
where $\varsigma:C\rightarrow M$ is the embedding of $C$ in the spacetime manifold $M$.
The quantization of the coadjoint orbit $\clO_\lambda$ can now be carried in two distinct though related ways.
Below, we restrict for simplicity to the case where $\lambda$ is regular, that is
the stabilizer subgroup $\msG_\lambda$ of $\lambda$ is a maximal torus of $\msG$.
%(See \ccite{Beasley:2009mb} for a fully general treatment.)

The techniques %methods
of geometric quantization \ccite{Kostant:1970qur,Souriau:1970sds,Woodhouse:1992de}
can be applied to the coadjoint orbit $\clO_\lambda$ if the Bohr--Sommerfeld quantization condition
is satisfied requiring that the cohomology class $[\nu_\lambda/2\pi]\in H^2(\clO_\lambda,\bbR)$
lies in the image of $H^2(\clO_\lambda,\bbZ)$. 
This happens precisely when $\lambda\in\Lambda_{\rmw\msG}$, the weight lattice of $\msG$, just
the situation we are interested in. %It can be shown that 
By the integrality of $\nu_\lambda/2\pi$, there is a unitary line bundle
$\clL_\lambda$ on $\clO_\lambda$ with curvature form $i\nu_\lambda$. $\clL_\lambda$ is endowed with 
a canonical Hermitian metric and the $\msG$--invariant unitary connection
\begin{equation}
D_\lambda=d-i\langle\lambda,\gamma^{-1}d\gamma\rangle.
\label{phys12}
\end{equation}
It turns out that there is a $\msG$--invariant complex structure $J_\lambda$ on $\clO_\lambda$ 
compatible with $\nu_\lambda$. $J_\lambda$ and $\nu_\lambda$ constitute in this way an invariant Kaehler
structure on $\clO_\lambda$ providing a complex polarization. Since the curvature $i\nu_\lambda$
of $\clL_\lambda$ equals the Kaehler form, $\clL_\lambda$ is also a holomorphic line bundle.
All the elements of geometric quantization are in place in this way. 
The quantum Hilbert space is then identified with the space of holomorphic
sections of $\clL_\lambda$ 
\begin{equation}
\clH_\lambda=H^0_{\overline \partial}(\clO_\lambda,\clL_\lambda).
\label{phys9}
\end{equation}
The Hilbert inner product is the one naturally induced by the Hermitian structure of
$\clL_\lambda$ and the symplectic  form $\nu_\lambda$. The quantization map is constructed based on the connection $D_\lambda$.
A unitary $\msG$--action on $\clH_\lambda$ is associated with the $\msG$--action of $\clO_\lambda$. 
The quantization of the moment map $q_\lambda$ is the corresponding infinitesimal generator.
Geometric quantization provides in this manner the quantum set-up required for expressing the 
Wilson loop $W_{R_\lambda}(C)$ according to \ceqref{phys3}. 

The  Borel--Weil--Bott theorem \ccite{Bott:1957abc,Kirillov:1976etr} connects the quantization 
of $\clO_\lambda$ described above 
to the representation $R_\lambda$. It states that $H^0_{\overline \partial}(\clO_\lambda,\clL_\lambda)\neq 0$ precisely when
$\lambda\in\Lambda_{\rmw\msG}{}^+$ and that in that case under the $\msG$--action
$H^0_{\overline \partial}(\clO_\lambda,\clL_\lambda)$ is the representation
space of the representation $R_\lambda$.

%\vfil\eject

%belongs to the image of the second integer cohomology $H^2(\clO_\lambda,\bbZ)$
%of $\clO_\lambda$ in the second real cohomology $H^2(\clO_\lambda,\bbR)$.

The methods of functional integral quantization \ccite{Zinn-Justin:2010piq} can also be applied to
$\clO_\lambda$. To this end, we need to begin with an action $\sfS_\lambda$.
This has the standard form $\sfS_\lambda=\int_C(\varpi_\lambda-H_\lambda)$, where $\varpi_\lambda$ is a symplectic
potential of $\nu_\lambda$ satisfying $\nu_\lambda=d\varpi_\lambda$, 
\begin{equation}
\varpi_\lambda=-\langle\lambda,\gamma^{-1}d\gamma\rangle,
\label{phys10}
\end{equation}
and $H$ is the Hamiltonian \ceqref{phys8}. The action thus reads
\begin{equation}
\sfS_\lambda(g,\omega)=\int\nolimits_C\langle\lambda,\Ad g^{-1}(\varsigma^*\omega)+g^{-1}dg\rangle
\label{phys11}
\end{equation}
after a conventional overall sign redefinition. Here, $g$ is a $\msG$ valued field on the closed oriented curve $C$. 
Hence, the action $\sfS_\lambda$ is a functional on the mapping space $\Map(C,\msG)$ that generically describes
a 1--dimensional sigma model with target space $\msG$. Since we are interested in the coadjoint orbit
$\clO_\lambda=\msG/\msG_\lambda$ instead than $\msG$, $\sfS_\lambda$, or better its exponentiated form
$\exp(i\sfS_\lambda)$ relevant in functional integral quantization, should rather be a functional on the mapping
space $\Map(C,\msG/\msG_\lambda)$. This requires that $\sfS_\lambda(g',\omega)=\sfS_\lambda(g,\omega)$ mod $2\pi\bbZ$ 
for $g'=g\upsilon$ with $\upsilon\in\Map(C,\msG_\lambda)$ resulting again in the condition that
$\lambda\in\Lambda_{\rmw\msG}$. The quantization of $\clO_\lambda$
is now given formally by the partition function 
\begin{equation}
Z_\lambda(C)=\int_{\Map(C,\rmG/\rmG_\lambda)}\scD g\exp\left(i\,\sfS_\lambda(g,\omega)\right).
\label{phys13}
\end{equation}
The Wilson loop $W_{R_\lambda}(C)$ equals $Z_\lambda(C)$ up to an overall normalization in
accordance with \ceqref{phys2}. This identification can be tested in a number of ways
by verifying that $Z_\lambda(C)$ enjoys the properties which $W_{R_\lambda}(C)$ does.
In fact, as a functional of the gauge field $\omega$, $Z_\lambda(C)$ is gauge invariant. Furthermore,
when $\omega$ is flat, \pagebreak $Z_\lambda(C)$ is also invariant under smooth variations of the embedding
$\varsigma$ as a consequence of certain Schwinger--Dyson relations.

The 1--dimensional sigma model summarily described in the previous paragraph has properties analogous to those
of 3--dimensional CS theory. It is in fact a Schwarz type topological quantum field theory. In this paper, we shall call it
topological coadjoint orbit (TCO) model for reference.

%\vfil\eject

\subsection{\textcolor{blue}{\sffamily Wilson surfaces as partition functions}}\label{subsec:wilssurf}

The natural question arises about the degree to which the above analysis can be extended and adapted
to Wilson surfaces. In this paper, we shall consider Wilson surfaces in the simplest version
of higher gauge theory, the strict one. As this is the only form of higher gauge theory that
we shall deal with, we shall omit the specification 'strict' in the rest of our discussion. 

Higher gauge symmetry hinges on Lie group crossed modules.
For the purpose of this brief outline, it is sufficient to recall that a Lie group crossed module
$\msM$ consists of two Lie groups $\msG$, $\msE$ together with two structure maps
relating them and enjoying certain properties \ccite{Baez5,Baez:2003fs}. 
A higher gauge theory with structure crossed module $\msM$ features a 1--form gauge field $\omega$
and a 2--form gauge field $\varOmega$ valued respectively in the Lie algebras $\fkg$, $\fke$ of
$\msG$, $\msE$ \ccite{Baez:2004in,Baez:2005qu}. 

A Wilson surface is naturally given by a straightforward
extension of the familiar Wilson formula \ceqref{phys1},
\begin{equation}
W_R(N)=\tr_R\Sexp\left(-\int\nolimits_{\hfpt N}\varOmega\right)\!.
\label{phys15}
\end{equation}
Above, $N$ is an oriented closed surface. $\tr_R$ denotes an invariant trace.
$\Sexp$ signifies sweep ordered exponentiation, a 2--dimensional counterpart of
path ordered exponentiation
\footnote{$\vphantom{\dot{\dot{\dot{a}}}}$ The notion of sweep ordering is purely heuristic and is used
here only to stress its analogy to path ordering. Strictly speaking, in higher gauge theory it is not possible
to define 1-- and 2--dimensional holonomies independently. $\Sexp\left(-\int\nolimits_{\hfpt N}\varOmega\right)$
depends not only on the 2--form gauge field $\varOmega$ but 
secretly also on the 1--form gauge field $\omega$, even though this is not shown by our notation,
and is well defined only if $\omega$, $\varOmega$ satisfy the so--called vanishing fake curvature condition. 
See \ccite{Martins:2007,Schrei:2011} and in particular sect. 2.1 of \ccite{Soncini:2014zra}
for a rigorous treatment of this point.}. % The notion of invariant trace $\tr_R$ is also similarly heuristic.}. 
$\Sexp\left(-\int\nolimits_{\hfpt N}\varOmega\right)$ depends on two choices:
the gauge of $\varOmega$ and the marking of $N$ by a homotopically
trivial base loop (analogous to the pointing of a loop by a base point). The invariance of 
$\tr_R$ ensures that the value of $W_R(N)$ is unaffected by
the variation of $\Sexp\left(-\int\nolimits_{\hfpt N}\varOmega\right)$ resulting 
from a change of such data. 
The above elements are only qualitatively sketched. A more precise definition together with 
an in--depth analysis of their properties 
can be found in refs. \ccite{Zucchini:2015wba,Zucchini:2015xba,Zucchini:2019mbz}. 

%intuitively
%brought about by a change of
%All the elements of the
%expression \ceqref{phys15} are admittedly only intuitively sketched here.

Based on the characterization of Wilson loops illustrated in subsect. \cref{subsec:wilsurf}, 
it is natural to ask whether a Wilson surface can be expressed as the partition function
of an auxiliary quantum system in a manner analogous to that a Wilson loop does. Assuming anyway that
this is indeed possible, the issue is then posed about the precise description of the partition function
as the Hilbert space trace of the evolution operator of the system and the path integral of an
associated sigma model on the lines of subsect. \cref{subsec:wilsurf}. 
The 2--dimensional nature of the underlying surface entails in any case
that we have to investigate this matter in the realm of 2--dimensional quantum
field theory.

%\vfil\eject

\subsection{\textcolor{blue}{\sffamily Our approach to Wilson surfaces}}\label{subsec:wsoutline}

The problem of obtaining a functional integral realizations of a Wilson surface, raised
at the end of the previous subsection, has already been tackled in the literature from
different perspectives \ccite{Alekseev:2015hda,Chekeres:2018kmh,Chekeres:2019xit}. 
In this subsection, we shall outline our approach to the subject firmly framed in 
higher gauge theory.

%$\msM$ consists of two
%Lie groups $\msE$, $\msG$ together with a Lie group action
%$\mu:\msG\times\msE\rightarrow\msE$
%of $\msG$ on $\msE$ by automorphisms and an equivariant target 
%map $\tau:\msE\rightarrow\msG$ obeying a certain relation \ccite{Baez5,Baez:2003fs}.
%A simple prototypical example is provided by a Lie group $\msG$, a normal sub\-group $\msE$ of $\msG$,
%the conjugation action $\mu$ of $\msG$ on $\msE$ and the inclusion $\tau$ of $\msE$ into $\msG$. 

As already anticipated, higher gauge symmetry rests on Lie group crossed modules. 
Our handling of crossed modules is based on the derived set--up
originally worked out in refs. \ccite{Zucchini:2019rpp,Zucchini:2019pbv}. 
It is in essence a superfield formalism providing an efficient way
of encoding most of the structural features of crossed modules
and proceeds by associating with any crossed module a derived Lie group, a graded Lie group
with a structure determined by that of the crossed module.
At the infinitesimal level, higher gauge symmetry is described by Lie algebra crossed modules.
With any such module there is similarly attached a derived Lie algebra. The whole
derived set--up is compatible with Lie differentiation. % both in its crossed modular 

%group.
%The derived Lie group $\DD\msM$ of a crossed module $\msM$ consists
%of the formal expressions $\mathrm{P}(\alpha)=\ee^{\alpha P}p$, where $\alpha\in\mathbb{R}[-1]$, 
%and $p\in\msG$, $P\in\fke[1]$, with a group structure defined by the crossed module structure
%of $\msM$. Its Lie algebra $\DD\fkm$, the derived Lie algebra,  
%consists correspondingly of the formal expressions $\rmX(\alpha)=x+\alpha X$, 
%with $\bar\alpha\in\mathbb{R}[-1]$, 
%where $x\in\fkg$, $X\in\fke[1]$, with the associated Lie bracket.

The relevant fields of a higher gauge theory are just crossed module valued
inhomogeneous form fields. They can be dealt with in the derived framework
in a very elegant and compact manner as derived fields, that is derived Lie group and algebra valued
fields. The resulting derived field formalism allows one 
to cast any higher gauge theory as a derived gauge theory,
basically an ordinary gauge theory with the derived group as gauge group.
The higher gauge fields and gauge transformations, once expressed in derived form,
%as derived gauge fields and gauge transformations,
can then be manipulated very much as their ordinary counterparts.
In this manner, by highlighting the close correspondence of the higher to the ordinary 
setting, the derived formalism enables one to import many ideas and
techniques of the latter to the former.

The derived field formalism has been successfully applied in ref. \ccite{Zucchini:2021bnn}
to the formulation of 4--dimensional CS theory, a higher gauge theoretic enhancement
of familiar 3--dimensional CS theory. The tight formal relationship
of the 4-dimensional theory to the 3--dimensional one brought out by the derived design
has shown itself to be very useful in the analysis of the properties of the model.
It is reasonable to expect that the derived set--up could be the most appropriate formal framework
also for the investigation of Wilson surfaces construed as higher gauge theoretic extension of Wilson loops. 

%It is reasonable to expect that the derived set--up could be the most appropriate formal framework
%also for the formulation of the 2--dimensional topological
%quantum field theory underlying the partition function realization
%of Wilson surfaces and that much may be learnt relying on the formal analogies to the well--understood
%corresponding 1--dimensional theory %underpinning the  partition function expression of
%for Wilson loops. % reviewed in subsect. \cref{subsec:wilscoadj}.

Our approach to the realization of Wilson surfaces as partition functions consists therefore
in extending the ordinary geometric or functional integral quantization schemes of the coadjoint
orbits reviewed in subsect. \cref{subsec:wilscoadj} to a higher crossed module theoretic
setting by relying on the derived formal set--up.
This however is not simply a matter of a straightforward
derived rewriting of these well-established approaches to coadjoint orbit quantization.
There are in fact very basic elements of such schemes which do not have any fitting
higher counterparts for reasons which we are going to survey momentarily. 
These issues are inevitably going to come to the surface in some form
also in the derived approach.

%in all its articulations %well established theory

To the best of our knowledge, there are no obvious counterparts of the notions of coadjoint action
and orbit for Lie group crossed modules. Further, there is no fully developed representation theory
and no analog of the highest weight theorem for crossed modules. While the derived formulation should 
provide in principle the definition of these higher objects and describe their properties,
the way this is achieved in practice is far from clear.  

Lie crossed modules belong to the realm of higher Lie theory. So, the geometric quantization
of a derived coadjoint orbit, no matter the way it is conceived, presumably must be formulated
in the framework of multisymplectic geometry \ccite{Rogers:2011zc}. Higher geometric quantization
of multisymplectic manifolds is however a subject not fully understood yet
(see however refs. \ccite{Hawkins:2006jbu,Bunk:2016rta,Fiorenza:2013kqa,Fiorenza:2013lpq}
for a variety of approaches to this issue). Since a 2--dimensional field theory must be the end
result of quantization as already noticed, geometric quantization may alternatively be based on a
symplectic loop space \ccite{Gawedzki:1987ak,Brylinski:1993ab}, proceeding via transgression
from a finite dimensional 2--plectic space on the lines of ref. \ccite{Saemann:2012ab}.
The infinite dimensional geometry involved in this approach is however problematic to deal with. 

The uncertainties affecting a workable theory of derived coadjoint orbits render problematic
also the construction of the derived TCO sigma model on which the functional integral quantization
of one such orbit should be based, if as expected crucial elements of the orbit's geometry are
required by the model's formulation.

Higher gauge theory possesses in addition to a gauge symmetry also a gauge for gauge symmetry.
The latter should emerge in the derived TCO model as a novel gauge symmetry with non counterpart
in the ordinary model and requiring a special handling. 

In fact, all the themes discussed above  eventually emerge and are dealt with
in the derived formulation, but in a novel and unified manner.

%\vfil\eject

\subsection{\textcolor{blue}{\sffamily Plan of the endeavour}}\label{subsec:wsproject}

The present endeavour is naturally divided in two parts, henceforth referred to as I and II,
of which the present paper is the first.

In I, a higher version of the KKS theory of coadjoint orbits is presented
based on the derived geometric framework. An original notion of derived coadjoint orbit is proposed. 
A theory of derived unitary line bundles and Poisson structures on regular derived orbits
is constructed.
The proper derived counterpart of the Bohr--Sommerfeld quantization condition is then identified. 
A version of derived prequantization is put forward. The difficulties hindering a full quantization
are discussed and a possible way--out is suggested.
The theory elaborated and the results 
obtained, mainly of a geometric nature, provide a basic underpinning for the field theoretic constructions 
of II.

In II, the derived TCO sigma model is presented and studied. Its manifold symmetries
are described. Its quantization is analyzed in the functional integral framework. Strong
evidence is provided that the model does indeed underlie the partition function realization of a Wilson
surface. The origination of the vanishing fake curvature condition is explained 
and homotopy invariance for flat derived gauge field is shown.
The model's Hamiltonian formulation is elaborated and through this the its close relationship to the
derived KKS theory developed in I is highlighted.

\vfil\eject

\renewcommand{\sectionmark}[1]{\markright{\thesection\ ~~#1}}

%\markright{\textcolor{blue}{\sffamily 2 ~~Part I: derived KKS theory}}

\section{\textcolor{blue}{\sffamily Part I: derived KKS theory}}\label{sec:wspartone}

The present paper, which constitutes part I of our endeavour, is devoted to derived KKS theory.
In this section, we provide an introductory overview of
this subject and an outlook on future developments. 

%In this section, we provide an introductory overview of part I of our endeavour, devoted to derived
%KKS theory, and provide an outlook on future developments. 

%Our work is written to a substantial extent in 

Our presentation of derived KKS theory employs the language of graded differential
geometry. The reader is referred to the appendixes for useful background and e.g.
ref. \ccite{Cattaneo:2010re} for a thorough exposition of this subject.
The graded geometric set--up is naturally suited for the description of the
higher geometric structures dealt with in this paper. It subsumes the ordinary differential geometric one, but
at the same times it enriches and broadens it allowing for a series of non standard constructions which otherwise
would not be possible. %be impossible. %be hardly attainable or even possible. 

The organizing principle of our construction of derived KKS theory is operational calculus, a formal
extension of classic Cartan calculus that has found a wide range of applications in
differential geometry and topology \ccite{Greub:1973ccc}. The operational framework
furnishes indeed an efficient and elegant means of describing the basic geometry of the 
principal bundles occurring in KKS theory.

%\vfil\eject

\subsection{\textcolor{blue}{\sffamily Plan of the part I}}\label{subsec:planone}

Paper I is organized in a number of sections and appendixes as follows.

In sect. \cref{sec:derform}, we survey the basic notions of the derived theory of Lie group crossed modules
and review the main results of the derived field formalism used throughout the present work.  

In sect. \cref{sec:clkks}, we provide an overview of standard KKS theory and geometric
quantization of coadjoint orbits. Our presentation of the subject is unconventional and partial: it relies on
an operational description and touches the subject of quantization only marginally. It is however
designed in a way that directly points to the derived extension constructed in the following section. 

In sect. \cref{sec:hkks}, we construct a derived KKS theory drawing inspiration from
ordinary KKS theory as exposed in sect. \cref{sec:clkks}, exploiting the advantages
provided by the operational set--up and employing the full power
of the derived approach. In particular, we introduce an appropriate definition
of derived coadjoint orbit. We further elaborate a suitable notion of
derived prequantum line bundle and connection thereof and use a connection's curvature
to construct a derived presymplectic structure satisfying the appropriate
Bohr–Sommerfeld quantization condition. We also lay the foundations for derived orbit prequantization. 
The section ends with the determination of 
the derived counterpart of the classic KKS symplectic structure. 

%The structure of the associated curvature suggests what the form of a derived symplectic structure
%satisfying the Bohr–Sommerfeld quantization condition should be.

Finally, in the appendixes of sect. \cref{app:app}, we collect basic notions of graded geometry
and operation theory used throughout the paper.

\subsection{\textcolor{blue}{\sffamily Overview of derived KKS theory}}\label{subsec:kksproject}

Since the remarks of subsect. \cref{subsec:wsoutline} are merely qualitative, we provide
in this subsection a somewhat more formal introduction to derived KKS theory to more precisely delineate
the subject and facilitate the reading of the paper.
A more rigorous and complete analysis of the material surveyed below is available
in the main body of this paper. 
%(see subsect. \cref{subsec:planone} above for an illustration of its organization).

As we indicated earlier, a Lie group crossed module $\msM$ features a pair of Lie groups
$\msE$ and $\msG$, the module's source and target groups. 
To these there are added an equivariant morphism $\tau:\msE\rightarrow\msG$ and an action
$\mu:\msG\times\msE\rightarrow\msE$ of $\msG$ on $\msE$ by automorphisms, 
the module's target and action structure maps, with certain natural properties.

%, called the module's target
%and action structure map respectively, 
%The reader is referred to the appendixes of ref. \ccite for more details

The derived Lie group $\DD\msM$ of $\msM$ is a graded Lie group built out of $\msG$ and the degree shifted
variant $\fke[1]$ of the Lie algebra $\fke$ of $\msE$, viz the semidirect product
\begin{equation}
\DD\msM=\fke[1]\rtimes_{\mu\sdot}\msG, 
\label{kksproject1}
\end{equation}
where here and below the dot $\sdot$ denotes Lie differentiation with respect to the relevant
variable. 

The derived formalism on which our formulation of higher KKS theory is based
is graded geometric in nature. A derived map $\Phi$ is a map from the degree shifted
tangent bundle $T[1]X$ of the relevant manifold $X$ into either the derived Lie group $\DD\msM$
or its Lie algebra $\DD\fkm$ or one of its degree shifted modifications. 
It always has two components $\phi$ and $\varPhi$.
$\phi$ is valued in either the Lie group $\msG$ or its Lie algebra
$\fkg$ or a degree shifted variant of it; $\varPhi$ in either the Lie algebra $\fke$ of $\msE$
or a degree shifted variant thereof. Moreover, one has $\deg\varPhi=\deg\phi+1$. 
A degree 1 nilpotent derived differential $\dd$ extending the ordinary de Rham
differential $d$ is also available.

The derived formalism can be employed to describe the derived group $\DD\msM$ itself.
Analogously to any ordinary Lie group,
$\DD\msM$ is characterized by a derived group variable $\Gamma$ and the associated derived
Maurer--Cartan form $\Sigma=\Gamma^{-1}\dd\Gamma$. 

%Integration is expressed by means of the Berezinian $\varrho_X$
%of $X$. Upon equipping $\msM$ with an invariant pairing, it is possible
%to construct a degree 1 graded symmetric pairing $(\cdot,\cdot)$ of derived fields.
%The basic assumption of our approach is simple:
%a derived coadjoint orbit should be a derived homogeneous space, that is a graded manifold
%of the form $\DD\msM/\DD\msM'$, where $\msM$ is the relevant symmetry crossed module and $\msM'$
%is a suitable submodule of $\msM$. In fact, we are able to formulate a derived
%KKS theory in this respect featuring appropriate derived counterparts of the ordinary KKS
%symplectic structure and prequantum line bundle and involving 
%derived analogs of various familiar Lie theoretic notions such as those of %one--parameter subgroup,
%centralizer, maximal torus and character among others.

With any element $\varLambda\in\fke$, there is associated a crossed submodule $\ZZ\msM_\varLambda$ of 
$\msM$, the centralizer crossed module of $\varLambda$, which is the largest crossed submodule of $\msM$
whose associated adjoint and $\mu$ actions both leave $\varLambda$ invariant.

%upon acting by the adjoint $\msE$-- and the $\mu$ $\msG$--actions. 

%in the appropriate sense. 

In derived theory, we attach derived groups $\DD\msM$ and $\DD\ZZ\msM_\varLambda$ to $\msM$
and $\ZZ\msM_\varLambda$ respectively. Since  $\ZZ\msM_\varLambda$ is a crossed submodule of $\msM$,
$\DD\ZZ\msM_\varLambda$ is a subgroup of $\DD\msM$. The derived coadjoint orbit of $\varLambda$
is the derived homogeneous space %of derived groups 
\begin{equation}
\clO_\varLambda=\DD\msM/\DD\ZZ\msM_\varLambda.
\label{kksproject2}
\end{equation}
$\clO_\varLambda$ is a non negatively graded manifold of degree 1, as both the derived groups $\DD\msM$
and $\DD\ZZ\msM_\varLambda$ themselves are. The notion of derived coadjoint orbit given here
manifestly replicates in a derived key that of ordinary coadjoint orbit as homogeneous space
recalled in subsect. \cref{subsec:wilscoadj}. 

A Lie group crossed module $\msM$ is said to be compact if the target Lie group $\msG$ is compact.
For a compact crossed module, a notion of maximal toral crossed submodule can be defined 
analogous to that of maximal torus of a compact group. Indeed, when $\msJ$ is a maximal toral crossed submodule
of a compact crossed module $\msM$, the target group $\msT$ of $\msJ$ is a maximal torus of the target
group $\msG$ of $\msM$. The regular elements $\varLambda\in\fke$ are those for which the centralizer
$\ZZ\msM_\varLambda$ is a maximal toral crossed submodule $\msJ$ of $\msM$ and so the associated derived coadjoint
orbit $\clO_\varLambda$ is the regular homogeneous space $\DD\msM/\DD\msJ$, just as in the ordinary
theory of subsect. \cref{subsec:wilscoadj}. 

The study of derived coadjoint orbits is most effectively done by studying general derived homogeneous spaces
of the form $\DD\msM/\DD\msM'$, where $\msM'$ is a cross\-ed submodule of a Lie group crossed module $\msM$.
The operational approach advocated in refs. \ccite{Zucchini:2019rpp,Zucchini:2019pbv} 
turns out to be quite natural and useful to this end, as $\DD\msM$ can \linebreak 
be regarded as a principal $\DD\msM'$--bundle
over $\DD\msM/\DD\msM'$, whose basic geometry is aptly described by a
$\DD\msM'$--operation. In this way, derived maps on $\DD\msM/\DD\msM'$ of a given kind can be regarded 
as basic maps on $\DD\msM$ of the same kind. 
The operational set--up allows further to efficiently construct such basic maps by 
employing the derived variable $\Gamma$ and Maurer Cartan
element $\Sigma$ of $\DD\msM$. 

When $\msM$ is a compact Lie group crossed module and $\msJ$ is a maximal toral crossed submodule of $\msM$,
the regular derived homogeneous space $\DD\msM/\DD\msJ$ can be considered in particular. A derived theory of 
unitary line bundles and connections thereof can then be worked out. Indeed, as for an ordinary compact Lie group
with a maximal torus, characters of $\msJ$ can be defined and with any character $\beta$ of $\msJ$ one can associate
a derived unitary line bundle $\clL_\beta$ on $\DD\msM/\DD\msJ$ whose typic\-al fiber is a proper derived version
$\DD\bbC$ of the complex line $\bbC$. A derived unitary connection $\rmA$ of $\clL_\beta$ is a derived
$\msU(1)$ gauge field with certain properties in the underlying
$\DD\msJ$--operation. $\rmA$ has a 1--form component $a$ and a 2--form component $A$.
Its curvature $\rmB=\dd\rmA$ has therefore a 2--form component $b$ and 3--form component $B$. 

The curvature $\rmB$ of a connection $\rmA$ of a derived line bundle $\clL_\beta$ is $\dd$--closed
by virtue of the derived Bianchi identity. In this way, $-i\rmB$ constitutes a derived
presymplectic form, which can be used to construct a derived Poisson structure
$\{\cdot,\cdot\}_\rmA$ on a suitable space of derived Hamiltonian functions $\iDFnc_{\rmA}(\DD\msM)$.
The functions of $\iDFnc_{\rmA}(\DD\msM)$ are basic and therefore represent
functions on $\DD\msM/\DD\msJ$. A derived presymplectic structure of this kind
belongs to the realm of multisymplectic geometry, since the 3--form component $-iB$ is
closed in the usual sense. They further obey a derived Bohr--Sommerfeld quantization
condition, as they arise from connections of $\clL_\beta$. 

For a suitably non singular derived presymplectic structure $-i\rmB$ of the above type, it is possible
to define a natural derived prequantization map. There exists a distinguished subspace $\iDFnc_{\rmA\msh}(\DD\msM)$
of the derived Hamiltonian function space $\iDFnc_{\rmA}(\DD\msM)$ closed under derived Poisson
bracketing constituted by the prequantizable functions. The prequantization map assigns to any function
$\rmF\in\iDFnc_{\rmA\msh}(\DD\msM)$
a first order differential operator $\widehat{\rmF}$ acting on a space
$\DD\Omega^0{}_\msh(\clL_\beta)$ of 0--form derived sections of $\clL_\beta$. 
The map is such that 
\begin{equation}
[\widehat{\rmF},\widehat{\rmG}]=i\widehat{\{\rmF,\rmG\}}_\rmA\vphantom{\bigg]}
\label{kksproject3}
\end{equation}
for any two functions $\rmF,\rmG\in\iDFnc_{\rmA\msh}(\DD\msM)$. 
However, there is no prequantum Hilbert space structure with respect to which the operators
yielded by derived prequantization are formally Hermitian.

%\begin{equation}
%\widehat{\rmF}\rmS=ij_{P_\rmF}\dd_\rmA\rmS+\rmF\rmS, 
%\label{kksproject}
%\end{equation}

Let $\msM$ again be a compact Lie group crossed module and equipped with a non singular
bilinear pairing $\langle\cdot,\cdot\rangle:\fkg\times\fke\rightarrow \bbR$ invariant under
the adjoint and $\mu$ $\msG$--actions. If $\varLambda\in\fke$ is a regular element, 
the associated derived coadjoint orbit is $\clO_\varLambda=\DD\msM/\DD\msJ$
for some maximal toral crossed submodule $\msJ$ of $\msM$. If $\varLambda$ satisfies furthermore 
a certain quantization condition, $\xi_\varLambda:=\langle\cdot,\varLambda\rangle$ is an element
of the dual integral lattice of the maximal toral subalgebra $\fkt$ of $\fkg$.
$\xi_\varLambda$ yields in turn a character $\beta_\varLambda$ of $\msJ$ and so
a derived line bundle $\clL_\varLambda:=\clL_{\beta_\varLambda}$. $\clL_\varLambda$ possesses a canonical connection
$\rmA_\varLambda$. The curvature $\rmB_\varLambda$ of $\rmA_\varLambda$ furnishes a derived symplectic structure
with 2- and 3--form component 
\begin{align}
&-ib_\varLambda=\frac{1}{2}\langle[\sigma,\sigma],\varLambda\rangle,
\vphantom{\Big]}
\label{kksproject4}
\\
%\end{align}
%\begin{align}
&-iB_\varLambda=\langle\dot\tau(\varLambda),\sdot\mu\sdot(\sigma,\varSigma)\rangle,
\vphantom{\Big]}
\label{kksproject5}
\end{align}
where $\sigma$ and $\varSigma$ are the degree 1-- and 2--form components of the Maurer--Cartan form $\Sigma$. 
This is the derived KKS presymplectic structure, the sought for extension of the classic KKS symplectic
structure \ceqref{phys5}.
%It is non singular and so a symplectic structure 
%only when the underling crossed module 
%$\msM$ is quasi injective, i.e. such that $\ker\tau$ a discrete subgroup of $\msE$
%and so that $\ker\dot\tau=0$.

%\vfil\eject \label{}

\subsection{\textcolor{blue}{\sffamily Outlook}}\label{subsec:outlookone}

Though we have gone a long way toward reaching the goal of constructing a higher KKS theory,
as essential geometrical underpinning of the realization of Wilson surfaces as partition functions,
a few basic issues remain unsolved.

Derived geometric prequantization ostensibly does not admit a prequantum Hilbert space structure
with respect to which the operators prequantizing derived Hamiltonian
functions are formally Hermitian. This is due to the fact
that the manifold on which the would be wave functions should be defined is a non negatively graded
one of positive degree. On manifolds of this kind, only Dirac delta distributional integral forms can
be integrated \ccite{bernstein:1977ifs}. The derived formulation does not yield anything of this kind. 
The question arises about whether this is an essential impossibility or else new elements
can be added to the theory allowing for the construction of a natural prequantum Hilbert space structure. 

A definition of a derived analog of polarization, assuming that such a thing exists at all, is still
to be achieved. The absence of a prequantum Hilbert space structure precludes in any case 
going beyond derived geometric prequantization into quantization proper along the familiar 
lines of ordinary quantization. 

%As we have noticed above, derived prequantization requires the derived KKS presymplectic structure
%\ceqref{kksproject4}, \ceqref{kksproject5} to be truly
%symplectic. This requirement is met only when the underling crossed module
%$\msM$ is quasi injective. This restricts much the range of crossed modules which
%can be employed. It is not clear whether such a restriction can be
%dropped through a more extensive reformulation of the theory, which is left to future work.

We believe that the limitations pointed out above of which derived prequantization suffers 
indicate that the appropriate geometric quantization of derived KKS theory cannot have 
some kind of quantum mechanical model, albeit exotic, as its end result but a two
dimensional quantum field theory. This is in line with standard expectations to the extent to which 
derived KKS theory can be regarded as some kind of categorification of the ordinary theory. 
The derived TCO model studied in paper II is an attempt to concretize these intuitions. 

The derived KKS theory formulated in this paper provides the geometric backdrop against which the
derived TCO model is built and by virtue of which it has the form it does. In turn, the TCO model furnishes
the physical motivation for the elaboration of the derived KKS set--up carried out in the present paper I.

%In any case, derived prequantization suffers the limitation we have pointed out.
%We believe that the reason for this is ultimately that

\vfil\eject

\vfil

\vfill\eject

\renewcommand{\sectionmark}[1]{\markright{\thesection\ ~~#1}}

%\markright{\textcolor{blue}{\sffamily 3 ~~Derived geometric framework}}

\section{\textcolor{blue}{\sffamily Derived geometric framework}}\label{sec:derform}

In this section, we review the derived geometric framework originally
elaborated in refs. \ccite{Zucchini:2019rpp,Zucchini:2019pbv}.
By design, the derived set--up allows reformulating any theory whose symmetry is
specified by a Lie group crossed module as one with a symmetry codified by
a graded group, the associated derived group, e.g. 
4--dimensional higher CS theory \ccite{Zucchini:2015wba}.
The derived framework also enables one to structure the higher KKS theory
worked out in sect. \cref{sec:hkks} on the model of the standard theory as presented in sect.
\cref{sec:clkks} and lies at the heart of the construction of the higher TCO model
in close analogy to the ordinary one in II. 
By its many virtues, it will  employed throughout our endeavour.

The derived set--up belongs to the realm of graded differential geometry
in a deeper way than the set--up of employed in standard KKS and TCO theory.
In fact, some of the structures featured in it cannot be expressed %rewritten
in the language of ordinary differential geometry in any straightforward if 
cumbersome way. 
 
%e.g. the operational set--up already encountered in sect. \cref{sec:clkks}.

% and must be appreciated in this light

%The graded geometric perspective 
%provides an especially natural language for the description of the
%higher geometric structures dealt with in this paper. 

%we formulate higher gauge theory %field theory, and in particular higher gauge theory,
%in the derived geometric framework worked out in refs. \ccite{Zucchini:2019rpp,Zucchini:2019pbv}.
%By design, the derived set--up structures higher gauge theory as an ordinary gauge theory 
%with an exotic

%\vfil\eject

\subsection{\textcolor{blue}{\sffamily Lie group and algebra crossed modules and invariant pairings}}\label{subsec:liecrmod}

Crossed modules encode the symmetry of higher gauge theory both at the finite and the infinitesimal level.
A geometric formulation of higher KKS and TCO theory must necessarily set forth from them. In this subsection, we
review the theory of Lie group and algebra crossed modules and module morphisms.
A more comprehensive treatment complete with detailed definitions and relevant relations
is provided in refs. \ccite{Baez5,Baez:2003fs} and the appendixes of ref. \ccite{Zucchini:2021bnn}. 
%The precise definitions and properties of crossed modules are collected in app. \cref{app:def}. 
%The reader familiar with this
%subject matter can safely skip to the next subsection.

The structure of finite Lie crossed module %, which appeared for the first time in homotopy theory,
abstracts and extends the set--up consisting of a Lie group $\msG$ and 
a normal subgroup $\msE$ of $\msG$ acted upon by $\msG$ by conjugation.
A Lie group crossed module $\msM$ features indeed  two Lie groups $\msE$, $\msG$ 
together with an action $\mu:\msG\times\msE\rightarrow\msE$
of $\msG$ on $\msE$ by automorphisms and an equivariant morphism %Lie group 
map $\tau:\msE\rightarrow\msG$
intertwining $\mu$ and the adjoint action of $\msE$.
%satisfying the finite Peiffer identity %(cf. eqs. \ceqref{liecrmod1}, \ceqref{liecrmod2}).
$\msE$, $\msG$ and 
$\tau$, $\mu$ are called the source and target groups and the target and action structure maps of
$\msM$, respectively. We shall often write $\msM=(\msE,\msG,\tau,\mu)$
to specify the crossed module through its data. 

%Taking the direct product of the relevant constituent data in the Lie group category, 
%it is possible to construct the direct product $\msM_1\times\msM_2$ 
%of two Lie group crossed modules $\msM_1$, $\msM_2$ and the direct product $\beta_1\times\beta_2$
%of two Lie group crossed module morphisms $\beta_1$, $\beta_2$ in straightforward fashion.
%Complicated crossed modules and module morphisms can sometimes be analyzed by factorizing them
%into direct products of simpler modules and module morphisms. 

There exist many examples of Lie group crossed modules. % and crossed module morphisms.
In particular, Lie groups and automorphisms, representations and central extensions of Lie groups can be
described as instances of Lie group crossed modules.
We mention here two basic examples of crossed modules for illustrative purposes.
Other examples will be presented later. 
They are defined for any Lie group $\msG$. 
The first is the inner automorphism crossed module of $\msG$,
$\INN\msG=(\msG,\msG,\id_{\msG},\Ad_{\,\msG})$.
%where $\Ad_{\,\msG}$ is the conjugation action of $\msG$. 
The second is the (finite) coadjoint action crossed module of $\msG$,
$\AD^*\msG=(\fkg^*,\msG,1_{\msG},\Ad_{\msG}{}^*)$, where 
$\fkg$ is the Lie algebra of $\msG$ and its dual space 
$\fkg^*$ is viewed as an Abelian group and $1_{\msG}:\fkg^*\rightarrow\msG$
is the trivial  morphism. 
%and $\Ad_{\msG}{}^*$ is the coadjoint action of $\msG$.
%A crossed module morphism
%$\rho:\INN\msG'\rightarrow\INN\msG$ reduces to a
%group morphism $\chi:\msG'\rightarrow\msG$. A 
%crossed module morphism
%$\alpha:\AD^*\msG'\rightarrow\AD^*\msG$ is specified by 
%a group morphism $\lambda:\msG'\rightarrow\msG$ and an intertwiner
%$\varLambda:\fkg'^*\rightarrow\fkg^*$ of $\Ad_{\msG'}{}^*$ to
%$\Ad_{\msG}{}^*\circ\lambda$.

%Lie group morphisms can be employed to
%construct morphisms of such crossed modules.

%There are two basic model crossed modules to which a
%variety of crossed modules can be related to.

%\vspace{2.5mm}
%
%\noindent
%{\it Lie algebra crossed modules}

Crossed module morphisms will occur only occasionally in our analysis, but they nevertheless
play an important role at some point in it. 
A morphism of finite crossed modules is a map of crossed modules respecting the module structure
expressing so a relationship of likeness of the modules involved. 
More explicitly, a morphism $\beta:\msM'\rightarrow\msM$ of Lie group crossed modules
consists of two group morphisms $\varPhi:\msE'\rightarrow\msE$
and $\phi:\msG'\rightarrow\msG$ intertwining in the obvious way the structure 
maps $\tau'$, $\mu'$, $\tau$, $\mu$. %In particular, $\phi\circ\tau'=\tau\circ\varPhi$. 
We shall normally write $\beta:\msM'\rightarrow\msM=(\varPhi,\phi)$
to indicate the constituent morphisms of the crossed module morphism.

A crossed module morphism
$\rho:\INN\msG'\rightarrow\INN\msG$ reduces to a
group morphism $\chi:\msG'\rightarrow\msG$. A 
crossed module morphism
$\alpha:\AD^*\msG'\rightarrow\AD^*\msG$ is specified similarly by 
a group morphism $\lambda:\msG'\rightarrow\msG$ and an intertwiner
$\varLambda:\fkg'^*\rightarrow\fkg^*$ of $\Ad_{\msG'}{}^*$ to
$\Ad_{\msG}{}^*\circ\lambda$.

The structure of infinitesimal Lie crossed module 
axiomatizes likewise the set--up consisting of a Lie algebra $\fkg$ and 
an ideal $\fke$ of $\fkg$ equipped with the adjoint action of $\fkg$.
It is therefore the differential version of that of finite Lie crossed module. A Lie algebra
crossed module $\fkm$ consists so of two Lie algebras $\fke$, $\fkg$ 
together with an action $m:\fkg\times\fke\rightarrow\fke$
of $\fkg$ on $\fke$ by derivations and an equivariant morphism 
$t:\fke\rightarrow\fkg$ intertwining $m$ and the adjoint action of $\fke$.
%satisfying the infinitesimal Peiffer identity (cf. eqs. \ceqref{liecrmod5}, \ceqref{liecrmod6}).
$\fke$, $\fkg$ and
$t$, $m$ are called the source and target algebras and the target and action structure maps of $\fkm$,
respectively. We shall often write $\fkm=(\fke,\fkg,t,m)$
to specify the crossed module through its data.

%Similarly to the Lie group case, taking the direct sum of the relevant
%constituent data in the Lie algebra category, it is possible to define 
%the direct sum $\fkm_1\oplus\fkm_2$ 
%of two Lie algebra crossed modules $\fkm_1$, $\fkm_2$ and direct sum $p_1\oplus p_2$
%of two Lie algebra crossed module morphisms $p_1$, $p_2$ in obvious fashion.
%These notions answer at the differential level to those of direct products
%of finite crossed modules and module morphisms. They allow to analyze
%crossed modules and module morphisms by decomposing them
%as direct sums of more elementary modules and module morphisms as we shall see in particular in
%subsect. \cref{subsec:crmodinv} below. 

Many examples of Lie algebra crossed modules and crossed module morphisms
are also available. They match precisely with the examples of
Lie group crossed modules and crossed module morphisms recalled above.
Ordinary Lie algebras and derivations, representations and central extensions of Lie algebras can
be described as instances of Lie algebra crossed modules. 
In particular, there are two crossed modules defined for any Lie algebra
$\fkg$ corresponding to the inner
automorphism and coadjoint action crossed modules introduced above. 
The first is the inner derivation crossed module of $\fkg$,
$\INN\fkg=(\fkg,\fkg,\id_{\fkg},\ad_{\,\fkg})$.
%where $\ad_{\,\fkg}$ is the adjoint action of $\fkg$. 
The second is the (infinitesimal) coadjoint action crossed module of $\fkg$,
$\AD^*\fkg=(\fkg^*,\fkg,0_{\fkg},\ad_{\fkg}{}^*)$,
where $\fkg^*$ is regarded as an Abelian algebra and
$0_{\fkg}:\fkg^*\rightarrow\fkg$ is the vanishing Lie algebra morphism.
%and $\ad_{\fkg}{}^*$ is the coadjoint action of $\fkg$.
%A crossed module morphism
%$r:\INN\fkg'\rightarrow\INN\fkg$ reduces to an
%algebra morphism $x:\fkg'\rightarrow\fkg$. A 
%crossed module morphism
%$a:\AD^*\fkg'\rightarrow\AD^*\fkg$ is specified by 
%an algebra morphism $l:\fkg'\rightarrow\fkg$ and an intertwiner
%$L:\fkg'^*\rightarrow\fkg^*$ of $\ad_{\fkg'}{}^*$ to
%$\ad_{\fkg}{}^*\circ l$.

%and Lie algebra morphisms can be assembled
%variously to construct morphisms of such crossed modules.

%Generic Lie algebra crossed modules
%are often related through isomorphisms to one of these model crossed modules.
%More specific examples will be considered later.

A morphism of infinitesimal Lie crossed modules is a map of crossed modules preserving the %infinitesimal
module structure. It describes a way such crossed modules are concordant. 
They constitute therefore the differential counterpart of the  morphisms of finite Lie
group crossed modules introduced above. 
More explicitly, a morphism $p:\fkm'\rightarrow\fkm$ of Lie algebra crossed modules
consists of two algebra morphisms $h:\fkg'\rightarrow\fkg$ and 
$H:\fke'\rightarrow\fke$ intertwining in the appropriate sense the structure 
maps $t'$, $m'$, $t$, $m$. 
We shall use often the notation $p:\fkm'\rightarrow\fkm=(H,h)$
to indicate constituent morphisms of the crossed module morphism.

A crossed module morphism
$r:\INN\fkg'\rightarrow\INN\fkg$ reduces to an
algebra morphism $x:\fkg'\rightarrow\fkg$. A 
crossed module morphism
$a:\AD^*\fkg'\rightarrow\AD^*\fkg$ is specified likewise by 
an algebra morphism $l:\fkg'\rightarrow\fkg$ and an intertwiner
$L:\fkg'^*\rightarrow\fkg^*$ of $\ad_{\fkg'}{}^*$ to
$\ad_{\fkg}{}^*\circ l$. 

Lie differentiation plays the same important role in Lie crossed module theory
as it does in Lie group theory.
With any Lie group crossed module $\msM=(\msE,\msG,\tau,\mu)$
there is associated the Lie algebra crossed module
$\fkm=(\fke,\fkg,\dot\tau,\sdot\mu{}\sdot\hfpt)$, where 
$\fke$, $\fkg$ are the Lie algebras of Lie groups $\msE$, $\msG$ respectively
and the dot notation $\dot{}$ denotes Lie differentiation along the relevant Lie group,  
much as a Lie algebra is associated with a Lie group
\footnote{$\vphantom{\dot{\dot{\dot{a}}}}$ Note here that $\mu$ has three Lie differentials,
 $\sdot\mu$, $\mu\sdot$, $\sdot\mu\sdot$, according to whether the $\msG$, the $\msE$ and both
the $\msG$ and $\msE$ arguments are subject to differentiation, respectively \ccite{Zucchini:2021bnn}.}.
Similarly, with any Lie group crossed module morphism
$\beta:\msM'\rightarrow\msM=(\varPhi,\phi)$ there is associated the Lie algebra crossed module morphism
$\dot\beta:\fkm'\rightarrow\fkm=(\dot\varPhi,\dot\phi)$, 
just as a Lie algebra morphism is associated with a Lie group morphism.

As examples, we mention that the Lie algebra crossed modules of the Lie group crossed modules
$\INN\msG$ and $\AD^*\msG$ we introduced above for any Lie group $\msG$ 
are precisely $\INN\fkg$ and $\AD^*\fkg$, respectively, as expected. 

Crossed modules with invariant pairing enter in many higher gauge theoretic
constructions. % \ccite{Zucchini:2021bnn}. %4--dimensional CS and 2--dimensional TCO actions.  
Indeed, invariant pairings play in higher gauge theory a role
similar to that of invariant traces in ordinary gauge theory
and are basic structural elements of e.g. kinetic terms enjoying the appropriate symmetries.
For similar reasons, they appear prominently also in derived KKS and TCO theory. %, as we shall see in due course. 

%In fact, on general grounds, in order to construct the kinetic term of the Lagrangian of a field theory,
%a non singular bilinear pairing is required. When further the field theory is characterized by
%certain symmetries, the same symmetries must be enjoyed by the pairing, which so is in addition
%invariant.

Following ref. \ccite{Zucchini:2021bnn}, we define an invariant pairing 
on a Lie algebra crossed module $\fkm=(\fke,\fkg,t,m)$ as a non singular bilinear form
$\langle\cdot,\cdot\rangle:\fkg\times\fke\rightarrow\mathbb{R}$
enjoying the invariance property
\begin{equation}
\langle\ad z(x),X\rangle+\langle x,m(z,X)\rangle=0
\label{crmodinv2/rep}
\end{equation}
for $z,x\in\fkg$, $X\in\fke$ and obeying the symmetry relation
\begin{equation}
\langle t(X),Y\rangle=\langle t(Y),X\rangle.
\label{crmodinv1/rep}
\end{equation}
for $X,Y\in\fke$. Other definitions of invariant pairing involving only either
$\fkg$ or $\fke$ turn out to be both unnatural and non viable
\footnote{$\vphantom{\dot{\dot{\dot{a}}}}$ Viewing a Lie algebra crossed module $\fkm$ as
a strict 2--term $L_\infty$ algebra $\fkl$ as described in ref. \ccite{Baez:2003fs},
an invariant pairing on $\fkm$ corresponds to a cyclic structure on $\fkl$.}. 

%Since the infinitesimal higher gauge symmetry described by $\fkm$ is ultimately reduced 
%to the adjoint action of $\fkg$ on itself and the module action $m$ of $\fkg$
%on $\fke$, %. The kinetic term will so have the required invariance properties if
%the pairing $\langle\cdot,\cdot\rangle$ obeys in addition

%
%It remains to clarify how the pairing $\langle\cdot,\cdot\rangle$ behaves with respect to the
%module target map $t$. This reduces to find an appropriate requirement for the difference 
%$\langle t(X),Y\rangle-\langle t(Y),X\rangle$ with for $X,Y\in\fke$. The minimal choice
%avoiding introducing further structures consists in demanding that 

A Lie algebra crossed module with invariant pairing is a
Lie algebra crossed module $\fkm=(\fke,\fkg,t,m)$
furnished with an invariant pairing 
$\langle\cdot,\cdot\rangle$ %:\fkg\times\fke\rightarrow\mathbb{R}$ 
of the kind defined in the previous paragraph.
%enjoying properties \ceqref{crmodinv2/rep}, \ceqref{crmodinv1/rep} above.

A Lie group crossed module with invariant pairing is a Lie group crossed module
$\msM=(\msE,\msG,\tau,\mu)$ whose associated  Lie algebra crossed module 
$\fkm=(\fke,\fkg,\dot\tau,\sdot\mu{}\sdot\hfpt)$ is one with invariant pairing.
The invariance property however is required to hold not only at
the infinitesimal level as in eq. \ceqref{crmodinv2/rep} but also at the finite one, viz
%so that %It is further required 
\begin{equation}
\langle\Ad a(x),\mu\sdot(a,X)\rangle=\langle x,X\rangle \vphantom{\bigg]}
\label{crmodinv14/rep}
\end{equation}
for $a\in\mathsans{G}$, $x\in\mathfrak{g}$, $X\in\mathfrak{e}$.

A Lie algebra crossed module $\fkm$ with invariant pairing is balanced,
that is such $\dim\fkg=\dim\fke$, by the non singularity of the pairing.
This is not too restrictive. %as it may appear at first sight.
In fact, any Lie algebra crossed module $\fkm$ can always be trivially extended to a balanced crossed module
$\fkm^c$ \ccite{Soncini:2014ara}.  %=(\fke',\fkg')$, where
Similarly a Lie group crossed module $\msM$ with invariant pairing is balanced,
as $\dim\msG=\dim\msE$. Further, any Lie group crossed module $\msM$ 
can always be trivially  extended to a balanced crossed module
$\msM^c$.

%for which depending on cases one has either 
%$\fke'=\fke\oplus\mathfrak{p}$ and $\fkg'=\fkg$
%or $\fke'=\fke$ and $\fkg'=\fkg\oplus\mathfrak{q}$ 
%for suitable Abelian Lie algebras $\mathfrak{p}$, $\mathfrak{q}$.

%A morphism $p:\fkm'\rightarrow\fkm=(H,h)$ of Lie algebra crossed modules with
%invariant pairings $\langle\cdot,\cdot\rangle'$, $\langle\cdot,\cdot\rangle$ is naturally
%defined as a crossed module morphism that preserves the pairings (cf. eq. \ceqref{crmodinv3}).
%Such a morphism describes a stronger form of sameness or likeness of the crossed modules
%concerning not only their algebraic structures but
%involving also to their invariant pairings. 

%We shall now explore the implications
%of having an invariant pairing structure attached to the crossed module.

%\vfil\eject

%\vspace{.5mm}

\subsection{\textcolor{blue}{\sffamily Derived Lie groups and algebras}}\label{subsec:dergralg}

%\vspace{.25mm}

The notion of derived Lie group of a Lie group crossed module and the corresponding infinitesimal notion of
derived Lie algebra of a Lie algebra crossed module were originally introduced in refs.
\ccite{Zucchini:2019rpp,Zucchini:2019pbv}.

%As anticipated in the introduction, they play a basic role in the formulation
%of 4--dimensional CS theory of ref. \ccite{Zucchini:2021bnn}.
%They do so also in the derived KKS theory and TCO model we present in sects.
%\cref{sec:hkks} and \cref{sec:hafsmod}. 

%, they play a role analogous to that of the gauge group of ordinary gauge theory

%Before proceeding to the illustration of this topic a few introductory comments are useful.
The formal set--up of derived Lie groups and algebras is an elegant and convenient way
of handling certain structural elements of the Lie group and algebra crossed modules
%entering in the formulation of
appearing in higher gauge theory. %As adumbrated earlier in the introduction, 
It is a compact superfield formalism not unlike the analogous 
formalisms broadly used in supersymmetric field theories and particularly suited for
a higher gauge theoretic setting.
%capable of representing higher gauge theory 
%as an ordinary gauge theory with an exotic graded gauge group or algebra.
%For the very same reason, it is particularly suited for the models studied in the preset work. 
%in a way that suits particularly well this latter.

The derived Lie group of a Lie group crossed module does not fully
encode this latter, but it only describes an approximation of it in the sense of synthetic geometry.
In fact, the target map of the crossed module is not involved in the definition of the derived group, 
nor could it be because, roughly speaking, the approximation is such to push
the range of the target map away out of reach. Only the action map of the crossed module  
and its properties not implicating the target map are presupposed. All the algebraic structure hinging
on the target map is in this way forgotten by the derived construction. 
Similar considerations apply to the derived Lie algebra of a Lie algebra crossed module.
The reader is referred to ref. \ccite{Zucchini:2019rpp} for a more precise discussion. 
%of this point. 

%The deep reasons why the derived set--up allows for such a natural formulation of 4--dimensional CS theory
%are still not completely clear. %Future work will perhaps shed light on this point.

Consider a Lie group crossed module $\msM=(\msE,\msG,\tau,\mu)$.
%At the algebraic level, t
The derived Lie group $\DD\msM$ of $\msM$ is the semidirect product group 
\begin{equation}
\DD\msM=\fke[1]\rtimes_{\mu\sdot}\msG, 
\label{liecm13}
\end{equation}
where $\fke[1]$ is regarded as a $\msG$--module through the $\msG$--action
$\mu\sdot$. $\DD\msM$ is therefore a graded Lie group concentrated in degrees $0,~1$.

%The derived Lie group $\DD\msM$ has a superfield description. Each element 
%$\rmP\in\DD\msM$ is represented through the formal expression \hphantom{xxxxxxxxx}

Each element $\rmP\in\DD\msM$ has the formal representation 
\begin{equation}
\mathrm{P}(\alpha)=\ee^{\alpha P}p \vphantom{\ul{\ul{\ul{\ul{g}}}}}
\label{liecmx1}
\end{equation}
with $\alpha\in\bbR[-1]$
\footnote{$\vphantom{\dot{\dot{\dot{a}}}}$ 
In this paper, $\alpha\in\bbR[p]$ has degree $p$,
because its geometrical degree is tacitly considered. %, viz $p$. 
In refs. \ccite{Zucchini:2019rpp,Zucchini:2019pbv}, $\alpha\in\bbR[p]$ has degree $-p$,
because its algebraic degree is considered instead. In this paper, we 
work with the geometric degree rather than the algebraic one, as is more natural
in a geometrical framework as the one illustrated here.
See app. \cref{subsec:gradgeo} for more details.},
where $p\in\msG$, $P\in\fke[1]$. $p$, $P$ are called the components of $\rmP$.
In this description, the group operations of $\DD\msM$ read as %take a simple form,  
\begin{align}
&\mathrm{PQ}(\alpha)=\ee^{\alpha(P+\mu\sdot(p,Q))}pq,
\vphantom{\Big]}
\label{liecmx2}
\\
&\mathrm{P}^{-1}(\alpha)=\ee^{-\alpha\mu\sdot(p^{-1},P)}p^{-1},
\vphantom{\Big]}
%\nonumber %
\label{liecmx3}
\end{align}
where $\mathrm{P},\mathrm{Q}\in\DD\msM$ are any two group elements with
$\mathrm{P}(\alpha)=\ee^{\alpha P}p$, $\mathrm{Q}(\alpha)=\ee^{\alpha Q}q$. 

%where $\fke$ is regarded as an Abelian Lie group 
%and $\fke[1]\rtimes_{\mu\sdot}\msG$
%denotes the semidirect product of the Lie groups $\fke[1]$ and $\msG$ 
%with respect to the $\msG$--action $\mu\sdot$.
%The operator $\DD$ has nice functorial properties. 

A morphism $\beta:\msM'\rightarrow\msM$
of Lie group crossed modules induces by means of its constituent group morphisms
$\varPhi:\msE'\rightarrow\msE$, $\phi:\msG'\rightarrow\msG$
a Lie group morphism $\DD\beta:\DD\msM'\rightarrow\DD\msM$ 
\footnote{$\vphantom{\dot{\dot{\dot{a}}}}$ 
This property might be expressed in the language of category theory
by characterizing $\DD$ as a functor of the appropriate Lie crossed module and group categories,
but we shall not do so here.}. 

%This property underlies finite crossed modules being organized  as a category. 

%Further, if $\msM_1$, $\msM_2$ are Lie group crossed modules, then
%$\DD(\msM_1\times\msM_2)=\DD\msM_1\times\DD\msM_2$. 
%\footnote{$\vphantom{\dot{\dot{\dot{a}}}}$ 
%The map $\DD$ associating with each Lie group crossed module its derived Lie group and with each 
%Lie group module morphism its derived Lie group morphism is 
%a functor of the monoidal category of Lie group crossed modules and module morphisms
%into that of graded Lie groups and group morphisms. However, we shall not rely on this categorical characterization
%in the following.}. 
%from the Lie group crossed module category
%$\bfs{\mathrm{Lgcm}}$
%into the graded Lie group category. 

The notion of derived Lie group has an evident infinitesimal counterpart.
Consider a Lie algebra crossed module $\fkm=(\fke,\fkg,t,m)$.
The derived Lie algebra of $\fkm$ is the semidirect product algebra 
\begin{equation}
\DD\fkm\simeq\fke[1]\rtimes_m\fkg,
\label{liecm14}
\end{equation}
where $\fke[1]$ is regarded as a $\fkg$--module through the $\fkg$--action
$m$. $\DD\fkm$ is so a graded Lie algebra concentrated in degree $0,~1$.

%The derived Lie algebra $\DD\fkm$ has also a superfield description.

Analogously, each element $\rmU\in\DD\fkm$ has the formal representation 
\begin{equation}
\rmU(\alpha)=u+\alpha U \vphantom{\bigg]}
\label{liecmx5}
\end{equation}
with $\alpha\in\bbR[-1]$, \pagebreak where $u\in\fkg$, $U\in\fke[1]$ are the components of $\rmU$.
The Lie bracket read in this set--up as 
\begin{equation}
[\rmU,\rmV](\alpha)=[u,v]+\alpha(m(u,V)-m(v,U))
\label{liecmx6}
\end{equation}
with $\rmU,\rmV\in\DD\fkm$ any two algebra elements with %such that 
$\rmU(\alpha)=u+\alpha U$, $\rmV(\alpha)=v+\alpha V$. 

%where %$\fke[1]$ is $\fke$ shifted to degree $+1$ and 
%$\fke$ is regarded as an Abelian Lie algebra and  $\fke[1]\rtimes_m\fkg$
%denotes the semidirect product of the Lie algebras
%$\fke[1]$ and $\fkg$ with respect to the $\fkg$--action $m$.
%Also this construction is functorial:
%As in the finite case above, the operator $\DD$ has functorial properties. 

A morphism $p:\fkm'\rightarrow\fkm$ 
of Lie algebra crossed modules induces through its underlying algebra morphisms
$H:\fke'\rightarrow\fke$, $h:\fkg'\rightarrow\fkg$, 
a Lie algebra morphism $\DD p:\DD\fkm'\rightarrow\DD\fkm$, analogously to the finite case.

%Further, if $\fkm_1$, $\fkm_2$ are Lie algebra crossed modules, 
%$\DD(\fkm_1\oplus\fkm_2)=\DD\fkm_1\oplus\DD\fkm_2$. 
%\footnote{$\vphantom{\dot{\dot{\dot{a}}}}$ 
%Again, the map $\DD$ associating with each Lie algebra crossed module its derived Lie algebra 
%and with each Lie algebra crossed module morphism its derived Lie algebra morphism 
%constitutes a functor from the monoidal category of Lie algebra crossed modules and
%module morphisms.}.
%$\bfs{\mathrm{Lacm}}$ into  that of graded Lie algebras and algebra morphisms.}. 
%$\bfs{\mathrm{Lacm}}$ $\bfs{\mathrm{gLa}}$.}.

%In the above constructions $\fke$ behaves as if it were an Abelian Lie algebra, albeit in general
%it is not. This can be understood in two interrelated ways. First, if $\fke$ were not regarded as 
%Abelian, the expression \ceqref{liecmx2} of the group product of $\DD\msM$ 
%and \ceqref{liecmx6} of the bracket of $\DD\fkm$ would have to incorporate extra terms belonging
%to the spaces $\fke[n]$ with $n\geq 2$ incompatible with their being valued in $\DD\msM$ and
%$\DD\fkm$, respectively. Second, the elements of $\fke[1]$ always multiply the degree 1
%parameter $\alpha$. Thus, the Lie bracket of any pair of elements of $\fke[1]$ necessarily
%multiplies $\alpha^2$ and so is effectively zeroed out. 

The derived construction  introduced above is fully compatible with Lie differentiation.
This property is in fact essential for its viability. 
If $\msM=(\msE,\msG,\tau,\mu)$ is a Lie group crossed module and 
$\fkm=(\fke,\fkg,\dot\tau,{}\dot{}\mu{}\dot{})$ 
is its associated Lie algebra crossed module, % (cf. subsect. \cref{subsec:liecrmod}), 
then $\DD\fkm$ is the Lie algebra of $\DD\msM$. 
Further, if $\beta:\msM'\rightarrow\msM$ is a Lie group crossed module morphism
and $\dot\beta:\fkm'\rightarrow\fkm$ is the corresponding Lie algebra
crossed module morphism, then $\dot\DD\beta=\DD\dot\beta$.

\subsection{\textcolor{blue}{\sffamily Derived field formalism}}\label{subsec:superfield}

In this subsection, we shall survey the main spaces of Lie group and algebra crossed module valued fields
using a derived field framework. 
%This allows for a very compact geometrically transparent
%formulation of 4--dimensional CS theory studied in later sections. 

We assume that the fields propagate on a general non negatively graded manifold $X$. Later, we shall add
the restriction that $X$ is an ordinary orientable and compact manifold, possibly with boundary.
To include also differential forms without renouncing to a convenient graded geometric description,
the fields will be maps from the shifted tangent bundle $T[1]X$ of $X$ into some graded target manifold
$T$. Below, we denote by $\iMap(T[1]X,T)$ the space of non negative internal degree internal maps from $T[1]X$ into $T$.
The more restricted space  $\Map(T[1]X,T)$ of ordinary maps from $T[1]X$ to $T$
can be also considered and treated much in the same way. See apps. \cref{subsec:gradgeo}, \cref{subsec:difform} for
more details. % about this matter. 

%Later in subsect. \cref{subsec:internal}, we shall consider . 
%When $Y$ is a graded vector space, group, Lie algebra etc.,
%so is $\bfs{{\iMap}}(X,Y)$ with the pointwise operations induced by those of $Y$.

The fields we shall consider will be valued either in the derived Lie group $\DD\msM$ of
a Lie group crossed module $\msM=(\msE,\msG,\tau,\mu)$
or in the derived Lie algebra $\DD\fkm$ of the associated Lie algebra crossed module
$\fkm=(\fke,\fkg,\dot\tau,\sdot\mu{}\sdot\hfpt)$ (cf. subsect. \cref{subsec:dergralg}).
A more comprehensive treatment of this kind of fields is given in ref. \ccite{Zucchini:2019rpp}.

We consider first $\DD\msM$--valued fields. Fields of this kind are elements of the mapping space
$\iMap(T[1]X,\DD\msM)$. If $\mathrm{U}\in\iMap(T[1]X,\DD\msM)$, then 
\begin{equation}
\rmU(\alpha)=\ee^{\alpha U}u
\label{superfield1}
\end{equation}
with $\alpha\in\bbR[-1]$,
where $u\in\iMap(T[1]X,\msG)$, $U\in\iMap(T[1]X,\fke[1])$.
$u$, $U$ are the components of $\mathrm{U}$. 
$\iMap(T[1]X,\DD\msM)$ has a Lie 
group structure induced by that of $\DD\msM$: if $\mathrm{U}\in\iMap(T[1]X,\DD\msM)$,
$\mathrm{V}\in\iMap(T[1]X,\DD\msM)$, then
%$\mathrm{UV},\mathrm{U}^{-1}\in\iMap(T[1]X,\DD\msM)$ are given by, 
\begin{equation}
\mathrm{UV}(\alpha)=\ee^{\alpha(U+\mu\sdot(u,V))}uv, \qquad \mathrm{U}^{-1}(\alpha)
=\ee^{-\alpha\mu\sdot(u^{-1},U))}u^{-1}. 
\label{superfield2}
\end{equation}

Next, we consider first $\DD\fkm$--valued fields. Fields of this kind are elements of the mapping space
$\iMap(T[1]X,\DD\fkm)$. If $\Phi\in\iMap(T[1]X,\DD\fkm)$, then 
\begin{equation}
\Phi(\alpha)=\phi+\alpha\varPhi  
\label{superfield3}
\end{equation}
with $\alpha\in\bbR[-1]$, where $\phi\in\iMap(T[1]X,\fkg)$, $\varPhi\in\iMap(T[1]X,\fke[1])$. 
Again, $\phi$, $\varPhi$ are the components of $\Phi$. $\iMap(T[1]X,\DD\fkm)$ has a Lie 
algebra structure induced by that of $\DD\fkm$: if 
$\Phi\in\iMap(T[1]X,\DD\fkm)$, $\Psi\in\iMap(T[1]X,\DD\fkm)$, then 
%$[\Phi,\Psi]\in\iMap(T[1]X,\DD\fkm)$ is given by 
\begin{equation}
[\Phi,\Psi](\alpha)=[\phi,\psi]
+\alpha\!\left(\hfpt\sdot\mu\sdot(\phi,\varPsi)-\sdot\mu\sdot(\psi,\varPhi)\right). 
\label{superfield4}
\end{equation}
$\iMap(T[1]X,\DD\fkm)$ is the virtual Lie algebra of $\iMap(T[1]X,\DD\msM)$. (For an explanation
of this terminology, see ref. \ccite{Zucchini:2019rpp}). 

As it turns out, the $\DD\fkm$--valued fields introduced above are not enough for our proposes.
One also needs to incorporate fields that are valued in the degree shifted linear spaces
$\DD\fkm[p]$ with $p$ some integer. Together, they constitute the mapping space $\iMap(T[1]X,\SD\fkm)$
\footnote{$\vphantom{\dot{\dot{\dot{h}}}}$\label{foot:degext} Throughout this paper, we denote by $\rmS\hfpt E$
the degree extended form of a possibly graded vector space $E$. Explicitly, one has $\rmS\hfpt E=\bigoplus_{p=-\infty}^\infty E[p]$.
In refs. \ccite{Zucchini:2019rpp,Zucchini:2019pbv}, %a related object, viz $\bigoplus_{p=-\infty}^\infty E[p]$,
the very same object is denoted instead as $\rmZ\hfpt E$,
a notation that we shall employ in later sections with a different meaning.}. 
If $\Phi\in\iMap(T[1]X,\DD\fkm[p])$, then 
\begin{equation}
\Phi(\alpha)=\phi+(-1)^p\alpha\varPhi  
\label{superfield5}
\end{equation}
with components $\phi\in\iMap(T[1]X,\fkg[p])$, $\varPhi\in\iMap(T[1]X,\fke[p+1])$. 
There is a bilinear bracket that associates with a pair of fields 
$\Phi\in\iMap(T[1]X,\DD\fkm[p])$, $\Psi\in\iMap(T[1]X,\DD\fkm[q])$
a field $[\Phi,\Psi]\in\iMap(T[1]X,\DD\fkm[p+q])$ given by 
\begin{equation}
[\Phi,\Psi](\alpha)=[\phi,\psi]
+(-1)^{p+q}\alpha\!\left(\hfpt\sdot\mu\sdot(\phi,\varPsi)
-(-1)^{pq}\hfpt\sdot\mu\sdot(\psi,\varPhi)\right).
\label{superfield6}
\end{equation}
$\iMap(T[1]X,\SD\fkm)$ becomes in this way a graded Lie algebra. This contains
the Lie algebra $\iMap(T[1]X,\DD\fkm)$ as its degree $0$ subalgebra. 

An adjoint action of $\iMap(T[1]X,\DD\msM)$ on the Lie algebra $\iMap(T[1]X,\DD\fkm)$
and more generally on the graded Lie algebra $\iMap(T[1]X,\SD\fkm)$ is defined. For 
$\mathrm{U}\in\iMap(T[1]X,\DD\msM)$, $\Phi\in\iMap(T[1]X,\DD\fkm[p])$, one has 
\begin{align}
&\Ad\mathrm{U}(\Phi)(\alpha)
=\Ad u(\phi)+(-1)^p\alpha(\mu\sdot(u,\varPhi)-\sdot\mu\sdot(\Ad u(\phi),U)),
\vphantom{\Big]}
\label{superfield7}
\\
&\Ad\mathrm{U}^{-1}(\Phi)(\alpha)=\Ad u^{-1}(\phi)+(-1)^p\alpha\mu\sdot(u^{-1},\varPhi+\sdot\mu\sdot(\phi,U)).
\vphantom{\Big]}
\label{superfield8}
\end{align}
The adjoint action preserves Lie brackets as in ordinary Lie theory. Indeed, for 
$\mathrm{U}\in\iMap(T[1]X,\DD\msM)$, $\Phi\in\iMap(T[1]X,\DD\fkm[p])$,
$\Psi\in\iMap(T[1]X,\DD\fkm[q])$, 
\begin{equation}
[\Ad\mathrm{U}(\Phi),\Ad\mathrm{U}(\Psi)]=\Ad\mathrm{U}([\Phi,\Psi]).
\label{superfield9}
\end{equation}
%\vfil\eject

As is well--known, in the graded geometric formulation we adopt, the nilpotent de Rham differential
$d$ is a degree $1$ homological vector field on $T[1]X$, $d^2=0$. $d$ induces a natural degree $1$ 
derived differential $\dd$ on the graded vector space $\iMap(T[1]X,\SD\fkm)$.
Concisely, $\dd=d+\dot\tau d/d\alpha$. In more ore explicit terms, for 
$\Phi\in\iMap(T[1]X,\DD\fkm[p])$, the field $\dd\Phi\in\iMap(T[1]X,\DD\fkm[p+1])$
reads as 
\begin{equation}
\dd\Phi(\alpha)=d\phi+(-1)^p\dot\tau(\varPhi)+(-1)^{p+1}\alpha d\varPhi.
\label{superfield10}
\end{equation}
It can be straightforwardly verified that %shown that 
\begin{equation}
\dd [\Phi,\Psi]=[\dd \Phi,\Psi]+(-1)^p[\Phi,\dd \Psi]
\label{superfield11}
\end{equation}
for $\Phi\in\iMap(T[1]X,\DD\fkm[p])$, $\Psi\in\iMap(T[1]X,\DD\fkm[q])$
and that
\begin{equation}
\dd^2=0.
\label{superfield12}
\end{equation}
In this way, $\iMap(T[1]X,\SD\fkm)$ becomes a differential graded Lie algebra. 

On several occasions, the pull--backs $\mathrm{dUU}^{-1},\mathrm{U}^{-1}\mathrm{dU}\in\iMap(T[1]X,\DD\fkm[1])$
of the Maurer--Cartan forms of $\DD\msM$ by a
$\DD\msM$ field $\mathrm{U}\in\iMap(T[1]X,\DD\msM)$ will enter our considerations.  
For these, there exist explicit expressions, 
\begin{align}
&\mathrm{dUU}^{-1}(\alpha)=duu^{-1}+\dot\tau(U)
\vphantom{\Big]}
\nonumber
\\
&\hspace{3.5cm}
-\alpha\!\left(dU+\tfrac{1}{2}[U,U]-\sdot\mu\sdot(duu^{-1}+\dot\tau(U),U)\right),
\vphantom{\Big]}
\label{superfield13}
\\
&\mathrm{U}^{-1}\mathrm{dU}(\alpha)=\Ad u^{-1}\!\left(duu^{-1}+\dot\tau(U)\right)
-\alpha\mu\sdot\!\left(u^{-1},dU+\tfrac{1}{2}[U,U]\right). 
\vphantom{\Big]}
\label{superfield14}
\end{align}
By the relation $\dd=d+\dot\tau d/d\alpha$, 
\ceqref{superfield13}, \ceqref{superfield14} follow from \ceqref{superfield1} and %straightforwardly from the well--known
the variational identities 
$\delta\ee^{\alpha X}\ee^{-\alpha X}=\frac{\exp(\alpha\ad X)-1}{\alpha\ad X}\delta(\alpha X)$,
$\ee^{-\alpha X}\delta\ee^{\alpha X}=\frac{1-\exp(-\alpha\ad X)}{\alpha\ad X}\delta(\alpha X)$ 
with $\delta=\dot\tau d/d\alpha$, owing to the nilpotence of $\alpha$. 

Next, let  the Lie group crossed module $\msM$ be equipped with
an invariant pairing $\langle\cdot,\cdot\rangle$
(cf. subsect. \cref{subsec:liecrmod}). A pairing on the graded Lie algebra
$\iMap(T[1]X,\SD\fkm)$ is induced in this way: for 
$\Phi\in\iMap(T[1]X,\DD\fkm[p])$, $\Psi\in\iMap(T[1]X,\DD\fkm[q])$
\begin{equation}
(\Phi,\Psi)=\langle\phi,\varPsi\rangle+(-1)^{pq}\langle\psi,\varPhi\rangle. 
\label{superfield15}
\end{equation}
Note that $(\Phi,\Psi)\in\iMap(T[1]X,\bbR[p+q+1])$. The field pairing $(\cdot,\cdot)$
therefore has degree $1$. $(\cdot,\cdot)$ is bilinear. More generally, when scalars with non trivial
grading are involved, the left and right brackets $($ and $)$ behave as if they had respectively
degree $0$ and $1$. For instance, $(c\Phi,\Psi)=c(\Phi,\Psi)$ whilst $(\Phi,\Psi c)=(-1)^k(\Phi,\Psi)c$
if the scalar $c$ has degree $k$. $(\cdot,\cdot)$ is further graded symmetric, 
\begin{equation}
(\Phi,\Psi)=(-1)^{pq}(\Psi,\Phi).
\label{superfield16}
\end{equation}
$(\cdot,\cdot)$ is also non singular. 

The field pairing $(\cdot,\cdot)$ has several other properties which make it a very natural ingredient in the field
theoretic constructions of later sections. 
%Another relevant property of the field pairing $(\cdot,\cdot)$
First, $(\cdot,\cdot)$ is $\DD\msM$--invariant.
If $\Phi\in\iMap(T[1]X,\DD\fkm[p])$, $\Psi\in\iMap(T[1]X,\DD\fkm[q])$, we have 
\begin{equation}
\left(\Ad\mathrm{U}(\Phi),\Ad\mathrm{U}(\Psi)\right)=(\Phi,\Psi)
\label{superfield19}
\end{equation}
for $\mathrm{U}\in\iMap(T[1]X,\DD\msM)$. By Lie differentiation, $(\cdot,\cdot)$
enjoys also $\DD\fkm$ invariance. This latter, however, admits a graded extension, because of which 
\begin{equation}
([\Xi,\Phi],\Psi)+(-1)^{pr}(\Phi,[\Xi,\Psi])=0
\label{superfield17}
\end{equation}
for $\Xi\in\iMap(T[1]M,\DD\fkm[r])$. 

Second, $(\cdot,\cdot)$ is compatible %Finally, the field pairing $(\cdot,\cdot)$ is compatible
with the derived differential $\dd$, i. e. %meaning that
the de Rham vector field $d$ differentiates $(\cdot,\cdot)$ through $\dd$, 
\begin{equation}
d(\Phi,\Psi)=(\dd \Phi,\Psi)+(-1)^p(\Phi,\dd \Psi). 
\label{superfield18}
\end{equation}

Let $\msM$, $\msM'$ be Lie group crossed modules with associated Lie algebra crossed modules
$\fkm$, $\fkm'$. Suppose that $\msM'$ is a submodule of $\msM$ and that,
consequently, $\fkm'$ is a submodule of $\fkm$. % (cf. subsect. \cref{subsec:crosumo}). 
As $\DD\msM'$ is a Lie subgroup of $\DD\msM$, $\iMap(T[1]X,\DD\msM')$ 
is a Lie subgroup of $\iMap(T[1]X,\DD\msM)$. Similarly, as $\DD\fkm'$ is a Lie subalgebra
of $\DD\fkm$, $\iMap(T[1]X,\DD\fkm')$ is a Lie subalgebra of $\iMap(T[1]X,\DD\fkm)$.
%As to its graded extension,
What is more, $\iMap(T[1]X,\SD\fkm')$ is a differential graded Lie subalgebra of
$\iMap(T[1]X,\SD\fkm)$, since it is invariant under the action of $\dd$ as is evident from 
\ceqref{superfield10}.
%If $\msM$ is also equipped
%with invariant pairing $\langle\cdot,\cdot\rangle$ with respect to which $\msM'$
%is isotropic (cf. subsect. \cref{subsec:crosumo}),
%then $\fkm'$ is isotropic and thus
%the Lie algebra $\iMap(T[1]X,\SD\fkm')$ is isotropic, that is
%$(\Phi,\Psi)=0$ for $\Phi\in\iMap(T[1]X,\DD\fkm'[p])$, $\Psi\in\iMap(T[1]X,\DD\fkm'[q])$. 

In concrete field theoretic analyses, one deals with functionals of the relevant derived fields 
on some compact manifold $X$. These are given as integrals on $T[1]X$ of certain functions of
$\Fun(T[1]X)$ constructed using the derived fields. 
Integration is carried out using the Berezinian $\varrho_X$ of $X$.

\subsection{\textcolor{blue}{\sffamily Ordinary geometric framework as a special case 
}}\label{subsec:derord}

The geometric framework employed in ordinary gauge theory as well as in the formulation
of ordinary KKS theory and the TCO model is in fact a special case of the derived geometric framework.
We devote this final subsection to the illustration of this point.

Let $\msG$ be a Lie group. There exists a unique Lie group crossed module with target group $\msG$
and trivial source group $\msE=1$, since the target morphism $\tau$ and the action map $\mu$ in this case
can be only the trivial ones. With a harmless abuse of notation, we shall denote this crossed module also
by $\msG$, since it codifies the Lie group structure of $\msG$ in a manner equivalent to the usual one. 
Similarly, for a Lie algebra $\fkg$, there exists a unique Lie algebra crossed module with target algebra $\fkg$
and trivial source algebra $\fke=0$, since the target morphism $t$ and the action map $m$ again 
can be only the trivial ones. We shall denote this crossed module also by $\fkg$, since it provides
an equivalent codification of the Lie algebra structure of $\fkg$. This crossed module reinterpretation 
of ordinary Lie theory is compatible with Lie differentiation: $\fkg$ is the Lie algebra of the Lie
group $\msG$ if and only if $\fkg$ is the Lie algebra crossed module of the Lie group crossed module
$\msG$.

We note however that while a Lie algebra $\fkg$ can support an invariant pairing $\langle\cdot,\cdot\rangle$ as such, it
does not support any invariant pairing $\langle\cdot,\cdot\rangle$ of the kind defined in subsect.
\cref{subsec:liecrmod} as a Lie algebra crossed module because of the vanishing of the source Lie algebra
of this latter. 

The derived functor $\DD$ relates the crossed module and the ordinary formulation
of Lie theory. From \ceqref{liecm13} and \ceqref{liecmx1}--\ceqref{liecmx3} with $\fke=0$, it appears that 
$\DD\msG\simeq\msG$ for any Lie group $\msG$: the derived Lie group $\DD\msG$ of $\msG$
as a Lie group crossed module is just $\msG$ as a Lie group. 
From \ceqref{liecm14} and \ceqref{liecmx5}, \ceqref{liecmx6} with $\fke=0$, it similarly appears that 
$\DD\fkg\simeq\fkg$ for any Lie algebra $\fkg$: the derived Lie algebra $\DD\fkg$ of $\fkg$
as a Lie algebra crossed module is just $\fkg$ as a Lie algebra. 

For a Lie group $\msG$ with Lie algebra $\fkg$, the mapping spaces $\iMap(T[1]X,\DD\msG)$
and $\iMap(T[1]X,\msG)$ are in this way identified and so are also the mapping spaces $\iMap(T[1]X,\DD\fkg)$
and $\iMap(T[1]X,\fkg)$ and their degree shifted versions. Further,
the derived adjoint action defined in \ceqref{superfield7}, \ceqref{superfield8}
reduces to the ordinary one and likewise the derived differential
$\dd$ defined in \ceqref{superfield10} reproduced the ordinary de Rham differential.

\vfil\eject

\renewcommand{\sectionmark}[1]{\markright{\thesection\ ~~ #1}}

%\markright{\textcolor{blue}{\sffamily 4 ~~Standard KKS theory, a review}}

\section{\textcolor{blue}{\sffamily Standard KKS theory, a review}}\label{sec:clkks}

\vspace{-.35mm}
The Kirillov--Kostant--Souriau (KKS) construction provides the coadjoint orbit of an
element of the dual of a Lie algebra with a natural symplectic structure, which under certain integrality
conditions can be quantized. Since our aim is to
generalize the KKS construction to a crossed module theoretic setting
in a derived perspective, it is appropriate
to begin our path by reviewing the standard KKS theory. 
Our survey of this latter is biased on one hand and incomplete
on the other. It is biased, because it is purposefully patterned in a way that plainly alludes to the
derived extension presented in sect. \cref{sec:hkks}, based as it is on an operational description,
a formal refinement of the classic Cartan calculus, which 
is not strictly necessary in the ordinary theory but it is essential in the higher derived version.
It is also incomplete, because we have deliberately avoided to delve in depth into quantization proper, aware
that the problem we aim to eventually solve is to some extent related, 
although not fully equivalent, to one of quantization of a 2--plectic manifold, a delicate issue that is  
not completely settled yet in the literature
and whose solution lies beyond the scope of the present work
%our scope
\ccite{Saemann:2012ab,Bunk:2016rta,Fiorenza:2013kqa,Fiorenza:2013lpq}.

The main notions of the operational framework mentioned in the previous paragraph are briefly reviewed in
app. \cref{subsec:clkks}. An exhaustive exposition of the subject is furnished in ref. \ccite{Greub:1973ccc}. 
In outline, if $P$ is a manifold and $\fkf$ is a Lie algebra, an $\fkf$--operation on $P$
is a collection of graded derivations of $\Fun(T[1]P)$ formed by a degree $-1$ derivation $j_x$
and a degree $0$ derivation $l_x$ for each $x\in\fkf$ and a degree $1$ derivation $d$
obeying the six Cartan relations \ceqref{hkksop1}--\ceqref{hkksop4}. If $P$ is a manifold carrying a right
action of a Lie group $\msF$ with Lie algebra $\fkf$, in particular a principal $\msF$--bundle, 
then an $\fkf$--operation on $P$ is defined such that $j_x$, $l_x$ are the contraction
and Lie derivative along the vertical vector field  $S_x\in\Vect(P)$ of the action
corresponding to $x\in\fkf$ and $d$ is the Rham differential of $P$. 
The operational framework furnishes a powerful method for the description of the geometry of the base
manifolds of principal bundles, the so called basic geometry,
in particular of homogeneous manifolds such as the coadjoint orbits
studied in KKS theory, in an essentially algebraic fashion. \vspace{-2mm} \eject

%This is the case in particular when $P$ is a principal $\msF$--bundle, but the
%operational method applies also to manifolds with non free group actions.
%For $\msF=\Diff(P)$, we have $\fkf=\Vect(P)$ and we recover the familiar Cartan calculus.

\vfil\eject

\subsection{\textcolor{blue}{\sffamily Coadjoint orbits
}}\label{subsec:coadorb} 

Let $\msG$ be a Lie group and $\fkg$ be its Lie algebra. For
any element $\lambda\in\fkg^*$ of the dual vector space of $\fkg$,
the coadjoint orbit $\clO_\lambda$ of $\lambda$ is the submanifold of $\fkg^*$
spanned by the coadjoint action of $\msG$ on $\lambda$. Explicitly, 
$\clO_\lambda=\{\Ad^*\gamma(\lambda)|\gamma\in\msg\}$,
where $\Ad^*$ denotes the coadjoint representation of $\msG$, the dual of the
adjoint representation $\Ad$ with respect to the canonical
duality pairing of $\fkg$ and $\fkg^*$. %$\clO_\lambda$ is  a submanifold of $\fkg^*$. 

The orbit $\clO_\lambda$ is a homogeneous $\msG$--manifold. Indeed, 
\begin{equation}
\clO_\lambda=\msG/\ZZ\msG_\lambda,
\label{coadorb1}
\end{equation}
where $\ZZ\msG_\lambda$ is the invariance subgroup of $\lambda$ in $\msG$.
Suppose that $\msG$ is compact and connected. Then, $\ZZ\msG_\lambda$ always contains a maximal torus $\msT$
of $\msG$. If $\ZZ\msG_\lambda=\msT$, $\lambda$ is said to be regular. In that case, $\clO_\lambda=\msG/\msT$. 
%In the following treatment,
Below, we shall rely heavily on the homogeneous space description of
$\clO_\lambda$ concentrating on the regular case.

%\vfil\eject

\subsection{\textcolor{blue}{\sffamily Operational description of homogeneous spaces
}}\label{subsec:ophomsp}

%\vspace{1mm}

The fact that coadjoint orbits are instances of homogeneous spaces opens the possibility
of an operational formulation of KKS theory. We shall however consider the problem we intend to study
from a wider perspective as follows. 
If $\msG$ is a Lie group and $\msG'$ is a subgroup of $\msG$, then $\msG$ can be regarded as a principal
$\msG'$--bundle over the homogeneous space $\msG/\msG'$. As such, $\msG$ is amenable to the operational
description, from which much information about $\msG/\msG'$ can be extracted.

%As already anticipated, with KKS theory in mind, we employ the operational approach to describe a standard principal
%$\msG'$--bundle $\msG\rightarrow\msG/\msG'$, where $\msG'$ is a subgroup of a Lie group $\msG$.
For the operational analysis, 
we need to begin with appropriate coordinates of the shifted tangent bundle $T[1]\msG$. By the isomorphism
$T[1]\msG\simeq\msG\times\fkg[1]$, we can use as coordinates of $T[1]\msG$ variables
$\gamma\in\msG$ and $\sigma\in\fkg[1]$. They obey
\begin{equation}
\gamma^{-1}d\gamma=\sigma,  \quad d\sigma=-\frac{1}{2}[\sigma,\sigma],
\label{ophomsp1}
\end{equation}
where $d$ is the de Rham differential regarded as a homological vector field
on $T[1]\msG$. $\sigma$ is therefore identified with the Maurer--Cartan form of $\msG$.

The right $\msG'$--action of $\msG$ induces an action on $T[1]\msG$, which reads as 
\begin{equation}
\gamma^{\msR q}=\gamma q,\quad \sigma^{\msR q}=\Ad q^{-1}(\sigma)
\label{ophomsp2}
\end{equation}
with $q\in\msG'$ in terms of the coordinates $\gamma$, $\sigma$. 
From here, we can readily read off the action of the derivations of the associated right 
$\fkg'$--operation of $\msG$ on $\gamma$, $\sigma$, % viz
\begin{equation}
\gamma^{-1}j_{\msR x}\gamma=0,\quad \gamma^{-1}l_{\msR x}\gamma=x,
\quad j_{\msR x}\sigma=x,\quad l_{\msR x}\sigma=-[x,\sigma], 
\label{ophomsp3}
\end{equation}
where $x\in\fkg'{}$.

As a group, $\msG$ is characterized also by a left $\msG$--action, which we shall write in right form
for convenience. In terms of the coordinates $\gamma$, $\sigma$, it reads as 
\begin{equation}
\gamma^{\msL e}=e^{-1}\gamma,\quad \sigma^{\msL e}=\sigma
\label{ophomsp4}
\end{equation}
with $e\in\msG$. Its main feature is the invariance of the Maurer--Cartan form. % $\sigma$.
%The left action allows us to view $\msG$ as a principal $\msG$--bundle on the singleton $*$.
%We shall not use this perspective though, but
Below, we shall employ extensively the associated left
$\fkg$--operation. The derivations of this latter act as 
\begin{equation}
j_{\msL h}\gamma\gamma^{-1}=0,\quad l_{\msL h}\gamma\gamma^{-1}=-h,
\quad j_{\msL h}\sigma=-\Ad\gamma^{-1}(h),\quad l_{\msL h}\sigma=0
\label{ophomsp5}
\end{equation}
with $h\in\fkg$.

%Above, we use the same notation for the right $\msG'$-- and left $\msG$--action on $\msG$
%and the derivations of the associated $\fkg'$-- and $\fkg$--operations. It should be clear
%from the context which is which, wherever they both appear.

The right $\msG'$-- and left $\msG$--actions commute, as is evident from their coordinate
expressions. Consequently, the derivations of the right $\fkg'$-- and left $\fkg$--operations also
commute in the graded sense.

Though automorphism symmetry is rarely mentioned in standard presentations of KKS theory,
it is nevertheless a feature of the KKS set--up inherent in its bundle theoretic nature.
We consider also this aspect in the above wider perspective. 
The automorphisms of the principal $\msG'$--bundle $\msG$ are the fiber preserving
invertible maps of $\msG$ into itself compatible the right $\msG'$--action.
Concretely, they are %realized as
maps $\psi\in\Map(\msG,\msG')$ with certain basicness properties under the right $\msG'$--action
and so naturally constitute a distinguished subgroup $\Aut_{\msG'}(\msG)$ of the infinite
dimensional Lie group $\Map(\msG,\msG')$. $\Aut_{\msG'}(\msG)$ is selected by the condition 
\begin{equation}
\psi^{\msR q}=q^{-1}\psi q
\label{ophomsp6}
\end{equation}
with $q\in\msG'$, where $\psi^{\msR q}$ denotes the pull--back of $\psi$ by the 
right $\msG'$--action, i.e. $\psi^{\msR q}(\gamma)=\psi(\gamma^{\msR q^{-1}})$. % for $\gamma\in\msG$. 
In the right $\fkg'$--operation, so, $\psi$ obeys
\begin{equation}
j_{\msR x}\psi\psi^{-1}=0,\quad j_{\msR x}\psi\psi^{-1}=-x+\Ad\psi(x)
\label{ophomsp7}
\end{equation}
for $x\in\fkg'$, where we view $\psi$ as a map of $\Map(T[1]\msG,\msG')$
relying on the identity $\Map(\msG,\msG')=\Map(T[1]\msG,\msG')$. 

The automorphism group $\Aut_{\msG'}(\msG)$ of the $\msG'$--bundle $\msG$ acts on $\msG$, 
the action being a left one. In terms of the coordinates $\gamma$, $\sigma$, 
the action of an automorphism $\psi\in\Aut_{\msG'}(\msG)$ on $T[1]\msG$ reads as 
\begin{equation}
{}^\psi\gamma=\gamma\psi^{-1},\quad {}^\psi\sigma=\Ad\psi(\sigma)-d\psi\psi^{-1}.
\label{ophomsp8}
\end{equation}
The second relation follows form the first one and the first relation \ceqref{ophomsp1}.
Notice that $({}^\psi\gamma)^{\msR q}={}^{\psi^{\msR q}}\gamma^{\msR q}$ as required by compatibility. 

By \ceqref{ophomsp4} and \ceqref{ophomsp8}, the left $\msG$-- and automorphism actions 
of the bundle $\msG$ commute and consequently are automatically compatible. %This entails that
In fact, automorphisms transform trivially under the action being 
\begin{equation}
\psi^{\msL e}=\psi
\label{ophomsp9}
\end{equation}
for $e\in\msG$. In the left $\fkg$--operation, therefore, for $h\in\fkg$ we have
\begin{equation}
j_{\msL h}\psi\psi^{-1}=0,\quad j_{\msL h}\psi\psi^{-1}=0.
\label{ophomsp10}
\end{equation}

%\vfil\eject

\subsection{\textcolor{blue}{\sffamily Unitary line bundles on a regular homogeneous space
}}\label{subsec:stkkspreq}

The construction of the appropriate prequantum line bundle is an essential step of the geometric
prequantization of KKS theory. In this subsection, we examine this problem from a broader point
of view again. We consider a homogeneous space of the form $\msG/\msT$, where
$\msG$ is a compact Lie group and $\msT$ is a maximal torus of $\msG$, modelling 
a regular coadjoint orbit and for this reason called regular. Relying on the operational
set--up of subsect. \cref{subsec:ophomsp},
we then describe the unitary line bundles over $\msG/\msT$, their sections and their unitary connections. 

%The construction of the appropriate prequantum line bundle is an essential step of the geometric
%prequantization of KKS theory. We consider again this problem from a broader point of view relying on
%the operational set--up of subsect. \cref{subsec:ophomsp}.
%Here however we restrict ourselves to the relevant case where $\msG$ is a compact Lie group and
%$\msG'$ is a maximal torus $\msT$ of $\msG$.  The operational approach then describes
%the homogeneous space $\msG/\msT$ and through a standard construction provides a family
%of line bundles over $\msG/\msT$. 

Recall that a character of $\msT$ is a Lie group morphism of $\msT$ into $\msU(1)$, so an element 
$\xi\in\Hom(\msT,\msU(1))$. 
With the character $\xi$, we can associate a unitary line bundle $\clL_\xi\rightarrow\msG/\msT$,
viz $\clL_\xi=\msG\times_\xi\bbC$. By definition, $\msG\times_\xi\bbC=\msG\times\bbC\,/\!\sim_\xi$, where 
$\sim_\xi$ denotes the equivalence relation on $\msG\times \bbC$ yielded by the identification
\begin{equation}
(\gamma,z)\sim_\xi(\gamma q^{-1},\xi(q)z)
\label{stkkspreq1}
\end{equation}
with $\gamma\in\msG$, $z\in\bbC$ and $q\in\msT$. %\pagebreak 

A $p$--form section of $\clL_\xi$ is a map $s\in\Map(T[1]\msG,\bbC[p])$  
%\begin{equation}
%s(\gamma^{\msR q^{-1}},\sigma^{\msR q^{-1}})(\gamma,\sigma)
%s^{\msR q}=\xi(q)s
%\label{stkkspreq2}
%\end{equation}
%for any $q\in\msT$, where $s^{\msR q}$ denotes the pull--back of $s$ by the 
%right $\msT$--action, i.e. $s^{\msR q}(\gamma,\sigma)=s(\gamma^{\msR q^{-1}},\sigma^{\msR q^{-1}})$.
%for $\gamma\in\msG$, $\sigma\in\fkg[1]$.
obeying  
\begin{equation}
j_{\msR x}s=0,\quad l_{\msR x}s=\dot\xi(x)s
\label{stkkspreq3}
\end{equation}
with $x\in\fkt$ in the right $\fkt$--operation, where $\dot\xi\in\Hom(\fkt,\fku(1))$ denotes the Lie differential of $\xi$.
$p$--form section of $\clL_\xi$ form a space $\Omega^p(\clL_\xi)$.

The operational framework allows for a total space description of 
$\msU(1)$ gauge theory, in particular of connections. 
A unitary connection of the line bundle $\clL_\xi$ is a map $a\in\Map(T[1]\msG,\fku(1)[1])$ obeying 
\begin{equation}
j_{\msR x}a=-\dot\xi(x),\quad l_{\msR x}a=0
\label{stkkspreq4}
\end{equation}
for $x\in\fkt$. The connection $a$ defines a covariant derivative $d_a$ of $p$--form sections of $\clL_\xi$.
The curvature of $a$ is the map $b\in\Map(T[1]\msG,\fku(1)[2])$ satisfying
\begin{equation}
b=da,\quad db=0
\label{stkkspreq5}
\end{equation}
the second relation being the Bianchi identity. $b$ obeys 
\begin{equation}
j_{\msR x}b=0,\quad l_{\msR x}b=0. 
\label{stkkspreq6}
\end{equation}
By \ceqref{stkkspreq5}, \ceqref{stkkspreq6}, $-ib$ is a closed 2--form on $\msG/\msT$. 
The unitary connections of $\clL_\xi$ form an affine space $\Conn(\clL_\xi)$.

The de Rham cohomology class $c(\clL_\xi)=[-ib/2\pi]\in H^2(\msG/\msT,\bbR)$ of the normalized curvature $-ib/2\pi$
of a connection $a$ of $\clL_\xi$ does not depend on $a$ and so characterizes the line bundle $\clL_\xi$ topologically.
%As is well--known,
$c(\clL_\xi)$ lies in a cohomology lattice, the image of the integral
cohomology $H^2(\msG/\msT,\bbZ)$ in $H^2(\msG/\msT,\bbR)$.  

%Connections of the line bundle $\clL_\xi$ are best defined directly in .

The gauge transformations of the line bundle $\clL_\xi$ can be straightforwardly
characterized in the operational framework too. A gauge transformations is a map
$u\in\Map(\msG,\msU(1))$ obeying the basicness conditions
\begin{equation}
j_{\msR x}uu^{-1}=0, \quad l_{\msR x}uu^{-1}=0
\label{stkkspreq7}
\end{equation}
for $x\in\fkt$, where we view $u$ as a map of $\Map(T[1]\msG,\msU(1))$ by virtue of the identity
$\Map(T[1]\msG,\msU(1))=\Map(\msG,\msU(1))$. Gauge transformations form a subgroup $\Gau(\clL_\xi)$
of the infinite dimensional Lie group $\Map(\msG,\msU(1))$. 

A gauge transformation $u\in\Gau(\clL_\xi)$ act on a $p$--form section $s$
of $\clL_\xi$ as %according to 
\begin{equation}
{}^us=us. 
\label{stkkspreq8}
\end{equation}
It acts also on a connection $a$ of $\clL_\xi$ in the familiar manner
\begin{equation}
{}^ua=a-duu^{-1}. 
\label{stkkspreq9}
\end{equation}
The curvature $b$ of $a$ is then gauge invariant
\begin{equation}
{}^ub=b. 
\label{stkkspreq10}
\end{equation}

%\vfill\eject

\subsection{\textcolor{blue}{\sffamily Poisson structures on a regular homogeneous space
%KKS symplectic structure
}}\label{subsec:stkkssympl}

In KKS theory, the prequantization of a coadjoint orbit requires that its symplectic structure matches
the curvature of a suitable prequantum line bundle. In this subsection, working from a broader perspective
as done before, we study the presymplectic structures on a regular homogeneous space $\msG/\msT$
%modelling a regular coadjoint orbit
of the kind considered in subsect. \cref{subsec:stkkspreq}
which enjoy this essential compatibility property
as well as the conjoined Poisson bracket structures and Hamiltonian function algebras.

Below, %With all this in mind, 
we shall rely to a large extent on the Cartan calculus of $\msG$, the $\Vect(\msG)$ --operation on $\msG$ 
featuring the contractions $j_V$ and Lie derivatives $l_V$ along the vector fields $V\in\Vect(\msG)$ and
the de Rham differential $d$. 
The derivations $j_{\msR x}$, $l_{\msR x}$ with $x\in\fkt$ of the right $\fkt$--operation 
are just the derivations $j_{S_{\msR x}}$, $l_{S_{\msR x}}$ associated with 
the vertical vector field $S_{\msR x}$ of the right $\msT$--action corresponding to $x$ and similarly
for the derivations $j_{\msL h}$, $l_{\msL h}$ with $h\in\fkg$ of the left $\fkg$--operation
with regard to the vertical vector fields $S_{\msL h}$ of the left $\msG$--action
corresponding to $h$.

%This features a set of graded derivations of $\iFun(T[1]\msG)$ comprising
%the degree $-1$ contractions
%$j_V$ and Lie derivatives $l_V$ along the vector fields $V\in\Vect(\msG)$ and
%the de Rham differential $d$ and obeying the Cartan calculus relations \ceqref{hkksop1}--\ceqref{hkksop4}.

The closed 2--form $-ib$ associated with the curvature of a unitary connection
$a$ of the unitary line bundle $\clL_\xi$ on $\msG/\msT$ of a character $\xi$ of $\msT$
constitutes a presymplectic 2--form on $\msG/\msT$
and therefore define a Poisson bracket structure on an associated Hamiltonian function algebra.
In the operational description, the Hamiltonian functions are the degree 0 elements 
$f\in\Fun(T[1]\msG)$ which are invariant under the right $\fkt$--action, so that 
\begin{equation}
l_{\msR x}f=0 \vphantom{\ul{\ul{\ul{g}}}}
\label{stkkssympl1}
\end{equation}
for $x\in\fkt$, and have the property that there is a vector field $P_f\in\Vect(\msG)$ with 
\begin{equation}
df-ij_{P_f}b=0. 
\label{stkkssympl2}
\end{equation}
The Hamiltonian vector field $P_f$ of $f$ is defined modulo vector fields $V\in\Vect(\msG)$ such that
$j_Vb=0$ and obeys $j_{l_{S_{\msR x}}P_f}b=0$. The Hamiltonian functions form a degree $0$ 
subalgebra $\Fun_a(T[1]\msG)$ of $\Fun(T[1]\msG)$ depending on $a$. The Poisson bracket of two functions
$f,g\in\Fun_a(T[1]\msG)$ is given by 
\begin{equation}
\{f,g\}_a=-ij_{P_g}j_{P_f}b.
\label{stkkssympl3}
\end{equation}
By the right invariance condition property \ceqref{stkkssympl1}, the Hamiltonian function
algebra $\Fun_a(T[1]\msG)$ can be identified with a subalgebra $\Fun_a(\msG/\msT)$ of
$\Fun(\msG/\msT)$ and $\{\cdot\hfpt,\cdot\}_a$ with a Poisson bracket structure on $\Fun_a(\msG/\msT)$. 

%all right invariant functions are automatically Hamiltonian.  

%For the time being we do not assume the non singularity of $-ib$ as a 2--form on $\msG/\msT$.
%Therefore, $\Fun_a(T[1]\msG)$ may be considerably smaller than 
%subalgebra of $\Fun(T[1]\msG)$ spanned by the degree $0$ elements $f\in\Fun(T[1]\msG)$
%obeying \ceqref{stkkssympl1}. 

With standard KKS theory in mind, we concentrate now on the case where the connection
$a$ of $\clL_\xi$ is invariant  under the left $\msG$--action so that 
\begin{equation}
l_{\msL h}a=0
\label{stkkssympl4}
\end{equation}
for $h\in\fkg$. The associated curvature $b$ is then also invariant
\begin{equation}
l_{\msL h}b=0.
\label{stkkssympl5}
\end{equation}
We can then expect that the Poisson bracket structure $\{\cdot\hfpt,\cdot\}_a$ will exhibit
left symmetry properties. In fact, the left $\msG$--action is Hamiltonian. The Hamiltonian function
$q_a(h)\in\Fun_a(T[1]\msG)$ corresponding the Lie algebra element $h\in\fkg$ is %the Hamiltonian function
\begin{equation}
q_a(h)=-ij_{\msL h}a.
\label{stkkssympl6}
\end{equation}
The Hamiltonian property of $q_a(h)$ follows from the left invariance of $a$ 
and the commutativity of the right $\msT$-- and left $\msG$--actions, which imply that 
$q_a(h)$ satisfies \ceqref{stkkssympl1} and \ceqref{stkkssympl2} with
$P_{q_a(h)}=S_{\msL h}$. 
%$l_{\msR x}q_a(h)=0$ and $dq_a(h)-ij_{\msL h}b=0$ for $x\in\fkt$ and so
%is Hamiltonian with Hamiltonian vector field $S_{\msL h}$. 
%\begin{equation}
%dq_a(h)-ij_{\msL h}b=0.
%\label{stkkssympl7}
%\end{equation}
Crucially, the map $q_a:\fkg\rightarrow\Fun_a(T[1]\msG)$ is equivariant and constitutes
a representation of $\fkg$, as for $h,k\in\fkg$,
\begin{equation}
l_{\msL h}q_a(k)=q_a([h,k])=\{q_a(h),q_a(k)\}_a,
\label{stkkssympl8}
\end{equation}
a relation that characterizes $q_a$ as the moment map of the left $\msG$--symmetry. 

The Poisson bracket structure $\{\cdot\hfpt,\cdot\}_a$ has simple gauge invariance properties.
Since the connection $a$ is restricted to be left invariant by \ceqref{stkkssympl4},
the gauge transformations allowed must be correspondingly left invariant, viz
\begin{equation}
l_{\msL h}uu^{-1}=0
\label{stkkssympl9}
\end{equation}
for $h\in\fkg$. It is immediate that the Poisson bracket $\{\cdot\hfpt,\cdot\}_a$ is gauge invariant,
since it is defined via \ceqref{stkkssympl2}, \ceqref{stkkssympl3} in terms of the curvature $b$ of $a$
which is gauge invariant. It is also readily checked from \ceqref{stkkssympl6}, \ceqref{stkkssympl9}
that the Hamiltonian functions $q_a(h)$, $h\in\fkg$, are gauge invariant.

We conclude this subsection by observing that the Poisson bracket structure
$\{\cdot\hfpt,\cdot\}_a$ in general is not induced 
by a genuine symplectic structure on $\msG/\msT$ unless the right $\msT$--action vertical vector fields
$S_{\msR x}$ with $x\in\fkg$ are the only vector fields $V\in\Vect(\msG)$ such that $j_Vb=0$.
In KKS theory, this is the situation customarily considered because in such a case %as under such assumption
$\Fun_a(\msG/\msT)=\Fun(\msG/\msT)$ and standard geometric prequantization is possible.

%\vfil\eject

\subsection{\textcolor{blue}{\sffamily Prequantization
}}\label{subsec:prequstkks} %pqpq

The datum on which prequantization is based is the Poisson structure of the regular homogeneous
space $\msG/\msT$ associated with a unitary connection $a$ of the line bundle $\clL_\xi$
of a character $\xi$ of $\msT$. Prequantization requires that the underlying presymplectic structure
$-ib$ is a symplectic one. We therefore assume that the curvature $b$
of $a$ is non singular, so that $j_Vb=0$ with $V\in\Vect(\msG)$ only if $V=S_{\msR x}$ for some $x\in\fkg$. 

Prequantization takes its start by setting 
\begin{equation}
\widehat{f}s=ij_{P_f}\dd_as+fs
\label{prequstkks1}
\end{equation}
for any Hamiltonian function $f\in\Fun_a(\msG/\msT)$ and
0--form section $s\in\Omega^0(\clL_\xi)$. 
It is immediately checked using \ceqref{stkkspreq3} and \ceqref{stkkssympl1}
that $\widehat{f}s\in\Omega^0(\clL_\xi)$. $\widehat{f}$ is manifestly an endomorphism of the
vector space $\Omega^0(\clL_\xi)$. The symplectic structure $-ib$ being proportional to the 
the curvature $b$ of the connection $a$ ensures that  
\begin{equation}
[\widehat{f},\widehat{g}]=i\widehat{\{f,g\}}_a
\label{prequstkks2}
\end{equation}
for $f,g\in\Fun_a(\msG/\msT)$. The basic requirement of prequantization is thus met. 

By virtue of \ceqref{stkkssympl6},
for $h\in\fkg$ the operator $\widehat{q}_a(h)$ associated with the left $\msG$--symmetry
moment map $q_a(h)$ takes the form 
\begin{equation}
\widehat{q}_a(h)=il_{\msL h},
\label{prequstkks3}
\end{equation}
in agreement with the interpretation
of $\widehat{q}_a$ as infinitesimal generator of the symmetry. From \ceqref{stkkssympl8} and
\ceqref{prequstkks2}, one has 
\begin{equation}
[\widehat{q}_a(h),\widehat{q}_a(k)]=i\widehat{q}_a([h,k])
\label{prequstkks4}
\end{equation}
for $h,k\in\fkg$, as expected.  

The space $\Omega^0(\clL_\xi)$ of 0--form sections of $\clL_\xi$ has a Hilbert structure:
for any two sections $s,t\in\Omega^0(\clL_\xi)$
\begin{equation}
\langle s,t\rangle=\frac{1}{n!}\int_{T[1](\msG/\msT)}\varrho_{\msG/\msT}(-ib)^n s^*t,
\label{prequstkks5}
\end{equation}
where $n=\dim(\msG/\msT)/2$. The operator $\widehat{f}$ associated with each Hamiltonian
function $f\in\Fun_a(\msG/\msT)$ is formally Hermitian with respect to this Hilbert structure
by virtue of the unitarity of the connection $a$.

%\vfil\eject

\subsection{\textcolor{blue}{\sffamily Coadjoint orbit quantization and Borel--Weil theorem
}}\label{subsec:stkksorb}

The presymplectic structures %of $\msG/\msT$
studied in subsect. \cref{subsec:stkkssympl} obey a Bohr--Som\-merfeld type quantization condition since
they come from the curvature of connections of line bundles. % on $\msG/\msT$.
This important fact plays an important role
in KKS theory as it circumscribes the range of quantizable coadjoint orbits as symplectic manifolds. 
This introduces us to the central part of our review of KKS theory:
the quantization of coadjoint orbits and the related Borel--Weil theorem.

%We note that the presymplectic structure $-ib$ of $\msG/\msT$
%discussed in subsect. \cref{subsec:stkkssympl} obeys a Bohr--Sommerfeld type quantization condition since
%it comes from the curvature of a connection of $\clL_\xi$. This important fact plays an important role
%in KKS theory as it restricts the range of quantizable coadjoint orbits. 
%This introduces us to the main topic of our review of KKS theory.
%%: the quantization of coadjoint orbits and the related Borel--Weil theorem.

%of the standard KKS theory
%Although somewhat unnatural from the purely mathematical perspective,

%Let $\lambda\in\fkg^*$. The coadjoint orbit $\clO_\lambda$ of $\lambda$
%is then the set of all elements $x\in\fkg$ of the form $x=\Ad\gamma(\lambda)$ with $\gamma\in\msG$.
%(Actually, this is the adjoint orbit, but this makes no difference for the reasons explained
%in the previous paragraph.) 
%Geometrically, $\clO_\lambda$ is the homogeneous space $\msG/\ZZ\msG_\lambda$, where $\ZZ\msG_\lambda$
%is the invariance subgroup of $\lambda$ in $\msG$. $\clO_\lambda$ is the base of the canonical
%principal $\ZZ\msG_\lambda$--bundle $\msG\rightarrow\msG/\ZZ\msG_\lambda$.

The basic data of the KKS construction are a compact semisimple Lie group $\msG$ and a regular element
$\lambda\in\fkg^*$ of the dual of the Lie algebra $\fkg$ of $\msG$. So, the coadjoint orbit $\clO_\lambda=\msG/\msT$
for some maximal torus $\msT$ of $\msG$. We can so %in this way
rely on the results we obtained in the previous subsections. 

The Lie algebra $\fkt$ of $\msT$ is a maximal toral subalgebra of $\fkg$.
By restricting $\lambda$ to $\fkt$, we obtain an element $\lambda\in\fkt^*$ denoted
in the same way for simplicity. If $\lambda/2\pi\in\Lambda_\msG{}^*$, the dual of the
integral lattice $\Lambda_\msG$ of $\fkt$, then there exists a character $\xi_\lambda\in\Hom(\msT,\msU(1))$
given by  \hphantom{xxxxxxxxxxxx} 
\begin{equation}
\xi_\lambda(\ee^t)=\ee^{i\langle\lambda,t\rangle}
\label{stkksorb1}
\end{equation}
for $t\in\fkt$. With $\lambda$, we can thus associate a unitary line bundle
$\clL_\lambda:=\clL_{\xi_\lambda}$ using the construction of subsect.
\cref{subsec:stkkspreq}.

$\clL_\lambda$ is equipped with a canonical unitary connection %$a_\lambda$, 
\begin{equation}
a_\lambda=-i\langle\lambda,\sigma\rangle. 
\label{stkksorb2}
\end{equation}
By \ceqref{stkkspreq5} and \ceqref{ophomsp1}, the curvature of $a_\lambda$ is 
\begin{equation}
b_\lambda=\frac{i}{2}\langle\lambda,[\sigma,\sigma]\rangle.
\label{stkksorb3}
\end{equation}
Both $a_\lambda$ and $b_\lambda$ are left invariant, since $\sigma$ is on account of  \ceqref{ophomsp4}.
The presymplectic structure $-ib_\lambda$ is furthermore non singular on $\msG/\msT$ and thus symplectic.
This is the KKS symplectic structure of $\clO_\lambda$, the KKS $\lambda$--structure.  
The Hamiltonian functions of the left $\msG$-symmetry also have a simple expression
\begin{equation}
q_\lambda(h)=\langle\Ad\gamma^*(\lambda),h\rangle.
\label{stkksorb4}
\end{equation}

The whole above structure has natural invariance properties under the automorphism group action of the principal
$\msT$--bundle $\msG$. With any automorphism $\psi\in\Aut_\msT(\msG)$, there is associated a map 
$u_\lambda\in\iMap(T[1]\msG,\msU(1))$ by 
\begin{equation}
u_\lambda=\exp\left(-i\int_{1_\msG}^{\,\cdot}\langle\lambda,d\psi\psi^{-1}\rangle\right).
\vphantom{\Big]}
\label{scrlbreg5}
\end{equation}
As $d\psi\psi^{-1}\in\iMap(T[1]\msG,\fkt[1])$ represents an element of $\Omega^1(\msG,\fkt)$ 
with periods in the integral lattice $\Lambda_\msG$ and $\lambda\in\Lambda_\msG{}^*$,
$u_\lambda$ is singlevalued as required.
%The choice of $1_\msG$ as base point of integration is conventional but natural.
By \ceqref{ophomsp7}, $u_\lambda$ obeys relations \ceqref{stkkspreq7}
%with respect to the character components $\xi_\lambda$, $\varXi_\lambda$
and so is a gauge transformation, $u_\lambda\in\Gau(\clL_\lambda)$. 
A group morphism $\psi\rightarrow u_\lambda$ from the automorphism group $\Aut_\msT(\msG)$ of $\msG$ to
the gauge transformation group $\Gau(\clL_\lambda)$ of $\clL_\lambda$
is established by \ceqref{scrlbreg5} in this way. 

For an automorphism $\psi\in\Aut_\msT(\msG)$, let ${}^\psi a_\lambda$ be the connection 
 given by \ceqref{stkksorb2} with $\sigma$ replaced by its transform ${}^\psi\sigma$
(cf. eqs. \ceqref{ophomsp8}). Then, ${}^\psi a_\lambda={}^{u_\lambda}a_\lambda$
%\begin{equation}
%{}^\psi a_\lambda={}^ua_\lambda,
%\vphantom{\Big]}
%\label{scrlbreg7}
%\end{equation}
where the expression in the right hand side is given by
\ceqref{stkkspreq9}. \pagebreak Consequently, one has ${}^\psi b_\lambda={}^{u_\lambda}b_\lambda=b_\lambda$
%\begin{equation}
%{}^\psi b_\lambda={}^ub_\lambda,
%\vphantom{\Big]}
%\label{scrlbreg9}
%\end{equation}
by \ceqref{stkkspreq10}. Similarly, one has ${}^\psi q_\lambda=q_\lambda$. %\vfil\eject

%Because of the way $\clL_\lambda$ is defined, it is automatically equipped with a Hermitian metric
%with respect to which the connection $a_\lambda$ is unitary. The symplectic form $-ib_\lambda$
%furnishes in addition an integration measure on $\msG/\msT$. Both the metric and the measure are
%left $\msG$--invariant.
%Since The presymplectic 2--form $-ib_\lambda$ is non singular and therefore symplectic.

By the above findings, it appears that $\clL_\lambda$ is a prequantum line bundle on $\msG/\msT$.
A prequantum Hilbert space structure is then defined as described in subsect. \cref{subsec:prequstkks}. 
%The square integrable sections of $\clL_\lambda$ form the prequantum Hilbert space.
To get the quantum
Hilbert space, one needs a polarization. This is obtained by endowing $\msG/\msT$ with a 
complex structure $J$ realized as an integrable complex splitting of the
complexified tangent bundle $T_\bbC(\msG/\msT)$.  
Because of the left $\msG$--invariance of the whole geometric set--up, it is enough to provide the splitting
at the identity coset $\msT$ of $\msG/\msT$. Since $T_{\bbC\msT}(\msG/\msT)\simeq(\fkg\ominus\fkt)\otimes\bbC$,
the complexification of the orthogonal complement of $\fkt$ 
with respect to the Cartan form of $\fkg$, %pairing $\langle\cdot\hfpt,\hfpt\cdot\rangle$,
we may choose $(\fkg\hfpt\ominus\hfpt\fkt)^{1,0}=\bigoplus_{\alpha\in\Delta_+}\mhfpt\fke_\alpha$,
$(\fkg\hfpt\ominus\hfpt\fkt)^{0,1}=\bigoplus_{\alpha\in\Delta_+}\mhfpt\fke_{-\alpha}$, where $\Delta_+$
is a set of posit\-ive roots of $\fkg$ and the $\fke_\alpha$ are the root spaces
of $\fkg\otimes\bbC$. 
Further, $J$ is compatible with the symplectic structure $-ib_\lambda$ in the sense that
$-ib_\lambda(\hfpt\cdot,J\cdot\hfpt)$
constitutes a left $\msG$--invariant Kaehler metric on $\msG/\msT$ whose Kaehler form is precisely $-ib_\lambda$. 
In this way, once endowed with the complex structure $J$, $\msG/\msT$ is a Kaehler manifold.
$\clL_\lambda$ then turns out to be a holomorphic line bundle on $\msG/\msT$.
With the above polarization of $\msG/\msT$ available, it is possible to define the quantum Hilbert space
$\scH_\lambda$ as the space of square integrable holomorphic sections of $\clL_\lambda$,
\begin{equation}
\scH_\lambda=H^0_{\overline\partial}(\msG/\msT,\clL_\lambda), 
\label{}
\end{equation}
provided that this latter is non vanishing. 
According to the Borel--Weil--Bott theorem \ccite{Bott:1957abc,Kirillov:1976etr},
$H^0_{\overline\partial}(\msG/\msT,\clL_\lambda)$ is non zero precisely when $\lambda\in\Lambda_{\msw\msG}{}^+$,
the lattice of dominant weights of $\msG$. The highest weight theorem establishes a one--to--one
correspondence between $\Lambda_{\msw\msG}{}^+$ and the set of equivalence classes of irreducible
representations of $\msG$. The theorem provides in this way the grounding for the relation between coadjoint orbit 
geometric quantization and partition function description of Wilson loops to be discussed in II.

%is the Hilbert space of the geometrically quantized coadjoint orbit of $\lambda$.
%where $\ominus$ denotes orthogonal complement with respect to the Cartan form of $\fkg$,

%When $\msG$ is simply connected, as it is normally assumed in standard KKS theory, the dual integral lattice 
%$\Lambda_\msG{}^*$ equals the weight lattice $\Lambda_{\msw\fkg}$ of $\fkg$, since the 1st homotopy group
%$\pi_1(\msG)=\Lambda_{\msw\fkg}/\Lambda_\msG{}^*=0$. In this case, then, 
%$\lambda\in\Lambda_{\msw\fkg}$. % and $\lambda$ is highest weight of a representation of $\msG$. 

%The basic data of the KKS construction are a compact semisimple Lie group $\msG$ 
%together with an invariant positive definite symmetric non singular bilinear form
%$\langle\cdot\hfpt,\hfpt\cdot\rangle$ on the Lie algebra $\fkg$ of $\msG$. The form
%allows equating the dual vector space $\fkg^*$ of $\fkg$ with $\fkg$ itself. 
%The identification, which we shall tacitly use below, 
%will make the relationship of the derived KKS theory illustrated in sect. \cref{sec:hkks}
%to the standard one more readily evident.%

\vfil\eject

\renewcommand{\sectionmark}[1]{\markright{\thesection\ ~~#1}}
%\markright{\textcolor{blue}{\sffamily 5 ~~Derived KKS theory}}

\section{\textcolor{blue}{\sffamily Derived KKS theory}}\label{sec:hkks}

\vspace{-.63mm}

In this section, working systematically within the derived framework
of sect. \cref{sec:derform},
we shall elaborate a derived KKS theory on the model of the standard KKS theory 
expounded in sect. \cref{sec:clkks}. In ref. \ccite{Zucchini:2021bnn}, we showed that
higher 4--dimensional CS theory admits a derived formulation that very closely
parallels that of its familiar ordinary 3--dimensional counterpart. The idea is to follow the same path
and formulate higher KKS theory by reproducing the constructions of its ordinary
counterpart in a derived perspective. This approach will lead us far but it will not solve all the problems.
The decisive step of geometric quantization, the construction of a prequantum Hilbert space and
a polarization, remains elusive even though the immediate reasons for this seem clear. 

%is apparently out of reach. 
%is yet to be solved. 
%The issue of identifying the appropriate higher analogue of polarization, a necessary step
%in geometric quantization, is yet to be settled. 

In outline, %proceeding along the lines of sect. \cref{sec:clkks},
we propose a definition of a derived coadjoint orbit as an instance of derived homogeneous space. 
In general, a derived homogeneous space is a graded manifold of the form $\DD\msM/\DD\msM'$, where $\msM$
is a Lie group crossed module and $\msM'$ is a crossed submodule of $\msM$ and $\DD\msM$ and
$\DD\msM'$ are their derived Lie groups, which is very aptly described
in an appropriate operational framework (cf. app. \cref{subsec:clkks}).
We concentrate on the important regular case, where $\msM'$ 
is a maximal toral crossed submodule of $\msM$. We then show that with any character $\beta$ of $\msJ$,
there is associated a derived unitary line bundle
$\clL_\beta$ on $\DD\msM/\DD\msJ$ and define the notion of derived
connection of $\clL_\beta$ and curvature thereof. The curvature of a connection
provides a derived presymplectic form on $\DD\msM/\DD\msJ$ obeying an integrality condition. 
Since a regular derived coadjoint orbit is just a manifold such as $\DD\msM/\DD\msJ$ equipped with a
derived prequantum line bundle such as $\clL_\beta$, this analysis indicates which
features the derived KKS symplectic form of a regular derived coadjoint orbit with a built in 
Bohr--Sommerfeld quantization condition should have. This is not sufficient for a complete
derived KKS geometric quantization, but it paves the way to other more conventional
quantization schemes such as that of derived TCO theory in sect. 5 of II.
%\cref{sec:hafsmod}. 

The content of this section is mostly %to a considerable extent
technical. The nature of the topics
covered makes this essentially unavoidable. We collect anyway the relevant component identities at
the end of each subsection, to make a first reading easier.

\vfil\eject

\subsection{\textcolor{blue}{\sffamily Derived coadjoint orbits and regular elements
%Crossed module centralizer of a crossed submodule
}}\label{subsec:cmcentr}

Our construction of derived KKS theory must necessarily begin with the definition of
the notions of derived coadjoint orbit and regularity. As anticipated, we shall do so
by taking the corresponding notions of ordinary KKS theory as a model (cf. subsect. \cref{subsec:coadorb})
and using firmly the derived set--up of sect. \cref{sec:derform} as our framework.  %expounded in.

In ordinary KKS theory, a coadjoint orbit always refers to some element $\lambda\in\fkg^*$
of the dual space of the Lie algebra $\fkg$ of the given Lie group $\msG$.
In derived KKS theory, the relevant Lie group crossed module
$\msM=(\msE,\msG,\tau,\mu)$ is as a rule equipped with an invariant 
pairing $\langle\cdot,\cdot\rangle$ (cf. subsect. \cref{subsec:liecrmod}).
In this way, $\fke$, as a vector space, can be identified
with $\fkg^*$. Further, by \ceqref{crmodinv14/rep}, the $\msG$--action of $\fke$,
$\mu\sdot$, is dual with respect to the pairing to the adjoint $\msG$--action of $\fkg$, $\Ad$.
It is so sensible that a derived coadjoint orbit should refer
to some element $\varLambda\in\fke$. It could do so 
even in the absence of an invariant pairing, $\mu\sdot$ playing the role of the coadjoint action. 

With the above considerations in mind, we introduce a few pertinent concepts of crossed module 
theory. 
Let $\msM=(\msE,\msG,\tau,\mu)$ be a Lie group crossed module
and let $\msM'=(\msE',\msG')$ %,\tau',\mu')$
be a crossed submodule of $\msM$. We can associate with $\msM'$ 
two subgroups of $\msG$ and one of $\msE$.
The centralizer $\ZZ\msG'$ of $\msG'$ is the subgroup
of $\msG$ constituted by those elements $a\in\msG$ such that $aba^{-1}=b$ for 
$b\in\msG'$. Similarly, the $\mu$--centralizer of $\mu\ZZ\msE'$ of $\msE'$ is the subgroup
of $\msG$ of the elements $a\in\msG$ such that $\mu(a,B)=B$ for 
$B\in\msE'$. Finally, the $\mu$--trivializer $\mu\FF\msG'$ of $\msG'$
is the Lie subgroup of $\msE$ formed by those elements $A\in\msE$ obeying $\mu(b,A)=A$
for $b\in\msG'$. It can be easily verified that the structure
$\ZZ\msM'=(\mu\FF\msG',\ZZ\msG'\cap\mu\ZZ\msE')$
%,\tau|_{\mu\FF\msG'},\mu|_{\ZZ\msG'\cap\mu\ZZ\msE'\times\mu\FF\msG'})$
is a crossed submodule of $\msM$, that we call the centralizer crossed module of $\msM'$.

%The above notion has an obvious infinitesimal counterpart. 
%Let $\fkm=(\fke,\fkg,t,m)$ be a Lie algebra crossed module
%and let $\fkm'=(\fke',\fkg')$  %,t',m')$
%be a crossed submodule
%of $\fkm$. We can associate with $\fkm'$ 
%two Lie subalgebras of $\fkg$ and one of $\fke$.
%The centralizer $\ZZ\fkg'$ of $\fkg'$ is the Lie subalgebra
%of $\fkg$ constituted by those elements $x\in\fkg$ such that $[x,y]=0$ for 
%$y\in\fkg'$. Similarly, the $m$--centralizer of $m\ZZ\fke'$ of $\fke'$ is the Lie subalgebra
%of $\fkg$ of the elements $x\in\fkg$ such that $m(x,Y)=0$ for 
%$Y\in\fke'$. Finally, the $m$--trivializer $m\FF\fkg'$ of $\fkg'$
%is the Lie subalgebra of $\fke$ formed by those elements $X\in\fke$ obeying $m(y,X)=0$
%for $y\in\fkg'$. It can be easily verified that the structure
%$\ZZ\fkm'=(m\FF\fkg',\ZZ\fkg'\cap m\ZZ\fke')$
%%,t|_{m\FF\fkg'},m|_{\ZZ\fkg'\cap m\ZZ\fke'\times m\FF\fkg'})$
%is a crossed submodule of $\fkm$, which we call the centralizer crossed module of $\fkm'$. 

%If $\msM$ is a Lie group crossed module and $\msM'$ is a crossed submodule
%of $\msM$, then the Lie algebra crossed module of the centralizer $\ZZ\msM'$
%of $\msM'$ is $\ZZ\fkm'$ as expected.

The case we have in mind features %, we shall work mainly in a set--up consisting of Lie group 
a crossed module $\msM$ and
a crossed submodule $\msM'$ of $\msM$ that is a submodule of its centralizer $\ZZ\msM'$.
This property implies restrictively that both $\msE'$ and $\msG'$ are Abelian and that
the $\msG'$--action on $\msE'$ is trivial.
%very restrictively t
%\footnote{$\vphantom{\dot{\dot{\dot{a}}}}$ 
%It implies also that $\ZZ\msM'$ normalizes $\msM'$ so that the central Weyl crossed module
%$\ZZ\WW\msM'$ $=\ZZ\msM'/\msM'$ is defined (see ref. \ccite{Zucchini:2021bnn} for background).\vfil\eject}. 

We now have at our disposal all the elements required for the definition of the derived coadjoint orbit   
of an element $\varLambda\in\fke$. We denote by $\msE_\varLambda$ and $\msG_\varLambda$ the subgroups 
of $\msE$ and $\msG$ constituted by the elements of the form $\ee^{x\varLambda}$ and $\ee^{y\dot\tau(\varLambda)}$
with $x,y\in\bbR$, respectively. Then, the structure
$\msM_\varLambda=(\msE_\varLambda,\msG_\varLambda)$ %,\tau|_{\msE_\varLambda},\mu|_{\msG_\varLambda\times\msE_\varLambda})$
is a crossed submodule of $\msM$, the 1--parameter submodule generated by $\varLambda$. 
The centralizer crossed module of $\msM_\varLambda$,
which we shall call the centralizer crossed submodule of $\varLambda$ in $\msM$ for clarity, 
is %can be explicitly characterized: %one has
$\ZZ\msM_\varLambda=(\ZZ\msE_\varLambda,\mu\ZZ\msE_\varLambda)$, 
%,\tau|_{\ZZ\msE_\varLambda},\mu|_{\mu\ZZ\msG_\varLambda\times\ZZ\msE_\varLambda})$, 
where $\ZZ\msE_\varLambda$ and $\mu\ZZ\msE_\varLambda$ are here the subgroups
of $\msE$ and $\msG$ of the elements $A\in\msE$ and $a\in\msG$ such that $\Ad A(\varLambda)=\varLambda$
and $\mu\sdot(a,\varLambda)=\varLambda$, respectively. As $\msM_\varLambda$ is a crossed submodule
of $\ZZ\msM_\varLambda$, %$\ZZ\msM_\varLambda$ has the Abelian features cited in the previous paragraph.
this is a case of the kind considered in the previous paragraph.

The derived KKS coadjoint orbit of $\varLambda$ is the derived homogeneous space
\begin{equation}
\clO_\varLambda=\DD\msM/\DD\ZZ\msM_\varLambda.
\label{cmcentr1}
\end{equation}
Since the centralizer $\ZZ\msM_\varLambda$ of $\msM_\varLambda$ can be considered 
as the stability crossed submodule of $\varLambda$,
$\clO_\varLambda$ can legitimately be regarded as a
derived model of the coadjoint orbit of $\varLambda$ in $\msM$ in analogy to a coadjoint orbit
of ordinary KKS theory. From the general definition of derived group, eq. \ceqref{liecm13},
and the characterization of the crossed module $\ZZ\msM_\varLambda$ given in the previous paragraph, 
it follows readily that 
\begin{equation}
\clO_\varLambda=\fke/\ZZ\fke_\varLambda[1]\times \msG/\mu\ZZ\msE_\varLambda,
\label{cmcentr1/1}
\end{equation}
where $\ZZ\fke_\varLambda$ is the Lie algebra of $\ZZ\msE_\varLambda$. 

%The associated central Weyl submodule $\ZZ\WW\msM_\varLambda=\ZZ\msM_\varLambda/\msM_\varlambda$ is defined.

In ordinary KKS theory, one considers mainly the coadjoint orbits of regular Lie algebra
elements for simplicity. When the relevant group is compact, the regular elements are those whose
stability group is a maximal torus. In derived KKS theory, regular elements are defined in a similar way
upon introducing suitable notions of compact crossed module and maximal toral crossed submodule
thereof. 

A Lie group crossed module $\msM=(\msE,\msG,\tau,\mu)$ is called compact if $\msG$ is a
compact group. A maximal toral crossed submodule $\msJ$ of a compact Lie group crossed module
$\msM$ consists of subgroups 
$\msT$ and $\msH$ of $\msG$ and $\msE$, respectively, with the following properties: $(i)$ 
$\msT$ is a maximal torus of $\msG$; % (in the usual Lie theoretic sense);
$(ii)$ for $A\in\msH$, $\tau(A)\in\msT$; $(iii)$ 
for $a\in\msT$ and $A\in\msH$, one has $\mu(a,A)=A$;
$(iv)$ for fixed $\msT$, $\msH$ is maximal with the above properties.  
In fact, it can be shown that $\msH=\mu\FF\msT$, where $\mu\FF\msT$ is the $\mu$--trivializer of
$\msT$, the subgroup of $\msE$ constituted by the elements $A\in\msE$ such that
$\mu(a,A)=A$ for $a\in\msT$. Therefore, $\msT$ determines $\msJ$ and so $\msJ$ may be referred to as
the maximal toral crossed submodule of $\msT$ in $\msM$.

%To stress this dependence when necessary, we shall write $\msJ=\MTOR_{\,\msM\!}\msT$.

A maximal toral \pagebreak crossed submodule $\msJ=(\msH,\msT)$ of $\msM$ has %certain
properties which make it akin to a familiar Lie theoretic maximal torus. 
First, $\msJ$ with the induced target and action structure maps 
is a crossed submodule of $\msM$ as indicated by its name.  
Second, by virtue of the Peiffer identity, % \ceqref{liecm2},
$\msH$ is automatically an Abelian subgroup of $\msE$, though not necessarily 
maximally Abelian, %\pagebreak
rendering $\msJ$ a fully Abelian object. Third, $\msJ$ is maximal, since
$\msT$ is and $\msH$ is fully determined by $\msT$. % as found above. 

Suppose so that the Lie group crossed module $\msM$ is compact.  
An element $\varLambda\in\fke$ is said to be regular if the centralizer $\ZZ\msM_\varLambda$
of $\varLambda$ %the associated 1--parameter crossed submodule $\msM_\varLambda$
is a maximal toral crossed submodule of $\msM$. 
In this case, from \ceqref{cmcentr1}, the derived coadjoint orbit of $\varLambda$ is
$\clO_\varLambda=\DD\msM/\DD\msJ$. By the special Abelian features of $\msJ$,
$\clO_\varLambda$ is expected to enjoy a particularly simple description. On account of 
\ceqref{cmcentr1/1}, we have $\clO_\varLambda=\fke/\fkh[1]\times\msG/\msT$. 

%In similar fashion, a cocharacter of $\msJ$ is a Lie group  
%crossed module morphism $X:\INN\mathsans{U}(1)\rightarrow\msJ$. 
%The component $f:\mathsans{U}(1)\rightarrow\msT$ of $X$ is then an ordinary cocharacter of
%the maximal torus $\msT$ of $\msG$. Further, $f$ is determined by the component $F:\mathsans{U}(1)\rightarrow\msH$
%as $f=\tau\circ F$. 
%%for $u\in\mathsans{U}(1)$ 
%%\begin{equation}
%%f(u)=\tau(F(u)).
%%\label{cocharacters1}
%%\end{equation}
%A cocharacter $X$ of $\msJ$ is completely characterized by the Lie differential 
%$\dot F:\fku(1)\rightarrow\fkh$ of $F$ by the relation $F(\ee^Z)=\ee^{\dot F(Z)}$
%%\begin{equation}
%%F(\ee^Z)=\ee^{\dot F(Z)}
%%\label{coweights5}
%%\end{equation}
%with $Z\in\fku(1)$. % and \ceqref{cocharacters1}.
%As $\fku(1)\simeq i\bbR$,
%$\dot F$ is uniquely determined by the element $\dot F(i)\in\fkh$. 
%Clearly $2\pi\dot F(i)$ lies in the kernel of the exponential map of $\msH$. 
%Further, $2\pi\dot F(i)\in\dot\tau^{-1}(L_{\msT\mathrm{int}})$,
%the inverse image in $\fkh$ by $\dot\tau$ of the integral lattice of $\msT$. 

%From ordinary character theory, $\dot F(i)$ belongs to the intersection 
%$\fkh\cap L_{\mathsans{U}\mathrm{int}}$ of $\fkh$ and the integral lattice
%$L_{\mathsans{U}\mathrm{int}}$ of any maximal torus $\mathsans{U}$ of $\msE$ containing 
%$\msH$. There exists in this way a one--to--one correspondence between cocharacters $F$
%of $\msJ$ and $\fkh\cap L_{\mathsans{U}\mathrm{int}}$. 

As an illustration %of the above notions
we present a few examples. 
With any pair of a Lie group $\msG$ and a normal subgroup $\msN$ of $\msG$,
there is associated a Lie group crossed module $\INN_\msG\msN=(\msN,\msG,\iota,\kappa)$,
where $\iota$ is the inclusion map of $\msN$ into $\msG$ and $\kappa$ is the left
conjugation action of $\msG$ on $\msN$. For $\varLambda\in\fkn$, the centralizer
crossed submodule of $\varLambda$ in $\INN_\msG\msN$ is the crossed module
$\INN_{\ZZ_{\,\msG\!}\varLambda}(\msN\cap\ZZ_{\,\msG\!}\varLambda)$, where $\ZZ_{\,\msG\!}\varLambda$
is the subgroup of $\msG$ of the elements $a\in\msG$ such that $\Ad a(\varLambda)=\varLambda$.
If $\msG$ is compact and
$\msT$ is a maximal torus of $\msG$, the maximal toral crossed submodule %$\MTOR_{\INN_\msG\msN}\msT$
of $\msT$ in $\INN_\msG\msN$ is the crossed module $\INN_\msT(\msN\cap\msT)$.
$\varLambda$ is regular precisely when $\ZZ_{\,\msG\!}\varLambda=\msT$ for some maximal torus $\msT$,
i.e. when $\varLambda$ is regular as an element of $\fkg$ in the usual sense.
In such a case, we have $\clO_\varLambda=\fkn/\fkn\cap\fkt[1]\times\msG/\msT$. 

With any central extension
$\!\mhfpt\xymatrix@C=1.3pc{1\ar[r]&\msC\ar[r]^-{\iota}&\msQ\ar[r]^-{\pi}&\msG\ar[r]&1}\!\mhfpt$
of Lie groups, there is associated a Lie group crossed module
$\msC(\!\xymatrix@C=1.3pc{\msQ\ar[r]^-{\pi}&\msG}\!)=(\msQ,\msG,\pi,\alpha)$, where the action $\alpha$
is given by $\alpha(a,A)=\sigma(a)A\sigma(a)^{-1}$ for $a\in\msG$, $A\in\msQ$ with 
$\sigma:\msG\rightarrow\msQ$ a section of the projection $\pi$, i.e. $\pi\circ\sigma=\id_\msG$,
whose choice is immaterial. If $\varLambda\in\fkq$, there is an induced Lie group central extension
$\!\mhfpt\xymatrix@C=1.3pc{1\ar[r]&\msC\ar[r]^-{\iota}
&\ZZ_{\,\msQ\!}\varLambda\ar[r]^-{\pi}&\sigma\ZZ_{\,\msG\!}\varLambda\ar[r]&1}\!\mhfpt$,
where $\ZZ_{\,\msQ\!}\varLambda$ and $\sigma\ZZ_{\,\msG\!}\varLambda$ are the subgroups of $\msQ$ and $\msG$
of the elements $A\in\msQ$ and $a\in\msG$ such that $\Ad A(\varLambda)=\varLambda$ and 
$\Ad\sigma(a)(\varLambda)=\varLambda$, respectively.
The centralizer crossed submodule of $\varLambda$ in $\msC(\!\xymatrix@C=1.3pc{\msQ\ar[r]^-{\pi}&\msG}\!)$
is so the crossed module
$\msC(\!\xymatrix@C=1.3pc{\ZZ_{\,\msQ\!}\varLambda\ar[r]^-{\pi}&\sigma\ZZ_{\,\msG\!}\varLambda}\!)$. 
If $\msG$ is compact and
$\msT$ is a maximal torus of $\msG$, there is an induced group sequence
$\!\!\xymatrix@C=1.3pc{1\ar[r]&\msC\ar[r]^-{\iota}&\sigma\FF_{\,\msQ\!}\msT\ar[r]^-{\pi}&\msT\ar[r]&1}\!\mhfpt$,
where $\sigma\FF_{\,\msQ\!}\msT$ is the subgroup of $\msQ$ of the elements $A\in\msA$ such that
$\sigma(a)A\sigma(a)^{-1}=A$ for $a\in\msT$. This sequence is generally not exact, as
$\pi$ may fail to be onto. It is if $\sigma$ can be chosen such that $\sigma:\msT\rightarrow\msQ$
is a group morphism for one and so all maximal tori $\msT$. In that case, the maximal toral crossed
submodule of $\msT$ in $\msC(\!\xymatrix@C=1.3pc{\msQ\ar[r]^-{\pi}&\msG}\!)$ is the crossed module
$\msC(\!\xymatrix@C=1.3pc{\sigma\FF_{\,\msQ\!}\msT\ar[r]^-{\pi}&\msT}\!)$.
$\varLambda$ is regular precisely when $\sigma\ZZ_{\,\msG\!}\varLambda=\msT$ for some maximal torus $\msT$.
This being so depends ultimately on the form of the projection $\pi$. If $\varLambda$ is regular, 
we have $\clO_\varLambda=\fkq/\sigma\FF_{\,\msQ\!}\fkt[1]\times\msG/\msT$, where
$\sigma\FF_{\,\msQ\!}\fkt$ is the Lie algebra of $\sigma\sigma\FF_{\,\msQ\!}\msT$. 

A Lie group crossed module $\msD(\rho)=(\msV,\msG,1_\msG,\rho)$ can be associated with the data consisting
of a Lie group $\msG$, a vector space $\msV$ regarded as an Abelian group, the trivial morphism $1_\msG$ of
$\msV$ into $\msG$ and a representation $\rho$ of $\msG$ in $\msV$. In this case, the centralizer crossed submodule
of an element $\varLambda\in\fkv$ in $\msD(\rho)$ is the crossed module $\msD(\rho\big|_{\INV_{\msG,\rho}\varLambda})$ 
where $\INV_{\msG,\rho}\varLambda$ is the invariance subgroup of $\varLambda$ in $\msG$. If $\msG$ is compact and $\msT$
is maximal torus of $\msG$, the maximal toral crossed submodule of $\msT$ in $\msD(\rho)$ is the crossed module
$\msD(\iota_{\msT,\msV_\msT})$, where $\msV_\msT$ is the subspace of $\msV$ spanned by the elements $X\in\msV$
such that $\rho(a)X=X$ for $a\in\msT$ and $\iota_{\msT,\msV_\msT}$ is the trivial representation of $\msT$ in $\msV_\msT$.
Thus, $\varLambda$ is regular only provided that the representation $\rho$ is trivial.
Clearly, the derived KKS theory of the crossed modules $\msD(\rho)$ is kind of trivial, making them less interesting
than the ones treated in the previous two paragraphs.

In the next several subsections, we shall work out the main elements of derived KKS theory from a general
standpoint. The model crossed modules treated above can be refereed to for illustration
of the abstract notions and exemplification of the most significant relations of the theory.
It turns out that in most case this is a
fairly straightforward task, which for this reason is left to the reader. In subsect. \cref{subsec:crlbreg},
which is the culmination of our formal development and contains the results of derived KKS theory most relevant
in paper II, we shall present in great detail further examples.

\subsection{\textcolor{blue}{\sffamily Operational description of derived homogeneous spaces % DM/DM'
}}\label{subsec:hkksop}

As in the standard theory expounded in sect. \cref{sec:clkks}, the fact that derived
coadjoint orbits are instances of derived homogeneous spaces allows for the application
of operational methods in derived KKS theory.
In this subsection patterned on subsect. \cref{subsec:ophomsp}, we take a general standpoint.
We consider a Lie group crossed module $\msM$ and a crossed submodule $\msM'$ of $\msM$ and
working in the derived framework
regard $\DD\msM$ as a principal $\DD\msM'$--bundle over the derived homogeneous 
space $\DD\msM/\DD\msM'$. We then offer an operational description of $\DD\msM$ as such,
using which the geometry of $\DD\msM/\DD\msM'$ can be studied as the basic geometry
of $\DD\msM$. 
Later, we shall adapt the resulting analysis to special choices of $\msM'$.

The basic data of the operation that we are going to construct are
the derived Lie algebra $\DD\fkm'$ of $\fkm'$ and the
internal function algebra $\iFun(T[1]\DD\msM)$ of $T[1]\DD\msM$
(cf. app. \cref{subsec:gradgeo}). 
Because of the graded nature of the derived group $\DD\msM'$, 
the ordinary algebra $\Fun(T[1]\DD\msM)$ cannot in fact be consistently employed.
%(cf. app. \cref{subsec:clkks}). 

%With the above in mind, we need to begin with
The detailed specification of the operation 
requires that appropriate coordinates
for the shifted tangent bundle $T[1]\DD\msM$ are used. 
By the isomorphism $T[1]\DD\msM\simeq\DD\msM\times\DD\fkm[1]$, it is natural to employ
coordinates adapted to the two Cartesian factors $\DD\msM$, $\DD\fkm[1]$, 
which we may think of as base and fiber coordinates of $T[1]\DD\msM$.
These are conveniently encoded in derived variables $\Gamma\in\DD\msM$, $\Sigma\in\DD\fkm[1]$
in an index free manner.

%To spell out the action of the operation's derivations on the function algebra
%$\iFun(T[1]\DD\msM)$ of $T[1]\DD\msM$, it is sufficient to specify it on suitable
%adapted coordinates of $T[1]\DD\msM$.
%of the clones $\DD\msM$, $\DD\fkm[1]$  (cf. app. \cref{subsec:gradgeo})

%To spell out the action of the operation's derivations on the function algebra
%$\iFun(T[1]\DD\msM)$ of $T[1]\DD\msM$, it is sufficient to specify it on suitable 
%adapted coordinates $\Gamma$, $\Sigma$ of $T[1]\DD\msM$. By the isomorphism
%$T[1]\DD\msM\simeq\DD\msM\times\DD\fkm[1]$, it is natural to choose $\Gamma$, $\Sigma$
%to be the standard projections $\DD\msM\times\DD\fkm[1]\rightarrow\DD\msM$, 
%$\DD\msM\times\DD\fkm[1]\rightarrow\DD\fkm[1]$, respectively, and think of them as
%generalized base and fiber coordinates of $T[1]\DD\msM$.

%To spell out the action of the operation's derivations on the function algebra
%$\iFun(T[1]\DD\msM)$ of $T[1]\DD\msM$, it is enough to specify it on a suitable set of
%adapted coordinates of $T[1]\DD\msM$. These consist of pairs constituted %formed
%by a derived Lie group valued map $\Gamma\in\iMap(T[1]\DD\msM,\DD\msM)$ 
%and a shifted derived Lie algebra valued map $\Sigma\in\iMap(T[1]\DD\msM,\DD\fkm[1])$ 
%satisfying an appropriate invertibility condition whose precise form will not matter.
%$\Gamma$ and $\Sigma$ can be thought of as base and fiber coordinates of $T[1]\DD\msM$.

The action of the de Rham derivation $d$ is expressed naturally via that of the corresponding
derived differential $\dd$ (cf. subsect. \cref{subsec:superfield}). As the derived coordinates
$\Gamma$, $\Sigma$ we use reflect 
the isomorphism $T[1]\DD\msM\simeq\DD\msM\times\DD\fkm[1]$, % in a derived setting, 
$\dd$ acts on $\Gamma$ as %takes the form 
\begin{equation}
\Gamma^{-1}\dd\Gamma=\Sigma,
\label{hkksop5}
\end{equation}
identifying $\Sigma$ as the derived Maurer--Cartan form associated with $\Gamma$. The action of
$\dd$ on $\Sigma$ is mandated by the requirement of nilpotence of $\dd$ and reduces to the derived 
Maurer--Cartan equation 
\begin{equation}
\dd\Sigma=-\tfrac{1}{2}[\Sigma,\Sigma].
\label{hkksop6}
\end{equation}
These identities are the derived counterpart of relations \ceqref{ophomsp1} of the standard theory.

%We can think of the $\Gamma$ as codifying coordinate choices of $\DD\msM$. These are however 
%more general than ordinary coordinates as they comprise 
%contributions that in ordinary parlance are describable as $0$-- and $1$--forms. 
%Correspondingly the $\Sigma$ are are more general than ordinary Maurer--Cartan elements.
%Note that for an ordinary Lie group $\msF$, we have
%$\iMap(T[1]\msF,\msF)=\Map(\msF,\msF)$ by grading considerations

%Our aim is describing derived Lie valued maps on $T[1]\DD\msM$ and the induced
%$\DD\msM'$--action on them by means of a suitable operation.
%We shall achieve this end by using certain derived Lie group valued 
%maps $\Gamma\in\iMap(T[1]\DD\msM,\DD\msM)$ 
%and the associated derived Maurer--Cartan
%maps $\Sigma\in\iMap(T[1]\DD\msM,\DD\fkm[1])$ defined by

The right $\DD\msM'$--action of $\DD\msM$ has an obvious expression in terms
of the derived coordinates $\Gamma$, $\Sigma$.  
On the base coordinate $\Gamma$, the action reads as 
\begin{equation}
\Gamma^{\msR\rmQ}=\Gamma\rmQ 
\label{hkksop7}
\end{equation}
with $\rmQ\in\DD\msM'$. 
The resulting action on the fiber coordinate $\Sigma$ is then determined by \ceqref{hkksop5}
and takes the form
\begin{equation}
\Sigma^{\msR\rmQ}=\Ad\rmQ^{-1}(\Sigma)+\rmQ^{-1}\dd\rmQ.
\label{hkksop8}
\end{equation}
These relations correspond to relations \ceqref{ophomsp2} in the standard theory.
Note however the appearance in \ceqref{hkksop8} of an inhomogeneous term $\rmQ^{-1}\dd\rmQ$
with no counterpart in \ceqref{ophomsp2} due to the special form of the derived differential $\dd$
%which is generically non vanishing in the derived setting
(cf. eq. \ceqref{superfield10}), which allows $\rmQ^{-1}\dd\rmQ$ to be non vanishing.

%The derived Lie group $\DD\msM'$ acts on $\DD\msM$ by right multiplication.
%Its coordinate expression
%by elements of the clone group $\DD\msM'$.
%The action is however mediated by
%$\DD\msM'$ of $\DD\msM'$.

In the operational set--up associated with the right $\DD\msM'$--action, $\Gamma$ behaves as 
\begin{align}
&\Gamma^{-1}j_{\msR\rmX}\Gamma=0,
\vphantom{\Big]}
\label{hkksop9}
\\
&\Gamma^{-1}l_{\msR\rmX}\Gamma=\rmX
\vphantom{\Big]}
\label{hkksop10}
\end{align}
with $\rmX\in\DD\fkm'$. %, the clone of derived Lie algebra $\DD\fkm'$.
%In more conventional geometric terms, \ceqref{hkksop9}, \ceqref{hkksop10} characterize 
%$\Gamma$ as a section of a certain associated $\DD\msM'$--bundle over
%$\DD\msM/\DD\msM'$. %Correspondingly, b
By virtue of \ceqref{hkksop5}, $\Sigma$ then satisfies %the relations
\begin{align}
&j_{\msR\rmX}\Sigma=\rmX,
\vphantom{\Big]}
\label{hkksop11}
\\
&l_{\msR\rmX}\Sigma=\dd\rmX-[\rmX,\Sigma].
\vphantom{\Big]}
\label{hkksop12}
\end{align}
These relations answer to relations \ceqref{ophomsp3} of the standard theory as appears from inspection. 
The appearance in \ceqref{hkksop12} of a non vanishing inhomogeneous term $\dd\rmX$ 
with no counterpart in \ceqref{ophomsp3} due to the special form of the derived differential $\dd$
is again to be noticed. 

For the sake of concreteness and completeness and later use, we express the above relations
in terms of the components $\gamma$, $\varGamma$, $\sigma$, $\varSigma$
of the derived coordinates $\Gamma$, $\Sigma$. This material constitutes a reference table, which  
the hurried reader can skip, if he/she wishes so, proceeding directly to the next subsection.

%$\gamma\in\iMap(T[1]\DD\msM,\msG)$, $\varGamma\in\iMap(T[1]\DD\msM,\fke[1])$ and
%$\sigma\in\iMap(T[1]\DD\msM,\fkg[1])$, $\varSigma\in\iMap(T[1]\DD\msM,\fke[2])$
% for the time being

The defining relation \ceqref{hkksop5} and the Maurer--Cartan equation \ceqref{hkksop6}
of the components $\gamma$, $\varGamma$, $\sigma$, $\varSigma$ of %the derived coordinates
$\Gamma$, $\Sigma$ read as 
\begin{align}
&\Ad\gamma^{-1}\mhfpt\left(d\gamma\gamma^{-1}+\dot\tau(\varGamma)\right)=\sigma,
\vphantom{\Big]}
\label{hkksop13}
\\
&\mu\sdot\left(\gamma^{-1},d\varGamma+\tfrac{1}{2}[\varGamma,\varGamma]\right)=\varSigma,
\vphantom{\Big]}
\label{hkksop14}
\\
%\end{align}
%\begin{align}
&d\sigma=-\tfrac{1}{2}[\sigma,\sigma]+\dot\tau(\varSigma), 
\vphantom{\Big]}
\label{hkksop15}
\\
&d\varSigma=-\sdot\mu\sdot(\sigma,\varSigma). 
\vphantom{\Big]}
\label{hkksop16}
\end{align}

The component expressions of the right $\DD\msM'$--action, given in eqs. \ceqref{hkksop7} and \ceqref{hkksop8},
involve besides the coordinate components $\gamma$, $\varGamma$, $\sigma$, $\varSigma$
the components $q$, $Q$  %.  %$q\in\msG'$, $Q\in\fke'[1]$
of the derived group parameter $\rmQ$. They take the form
\begin{align}
&\gamma^{\msR q,Q}=\gamma q,
\vphantom{\Big]}
\label{hkksop17}
\\
&\varGamma^{\msR q,Q}=\varGamma+\mu\sdot(\gamma,Q),
\vphantom{\Big]}
\label{hkksop18}
\\
%\end{align}
%for $\gamma$, $\varGamma$ and\hphantom{xxxxxxxxxxxxxx}
%\begin{align}
&\sigma^{\msR q,Q}=\Ad q^{-1}\left(\sigma+\dot\tau(Q)\right),
\vphantom{\Big]}
\label{hkksop19}
\\
&\varSigma^{\msR q,Q}=\mu\sdot\left(q^{-1},\varSigma+\sdot\mu\sdot(\sigma,Q)+\tfrac{1}{2}[Q,Q]\right).
\vphantom{\Big]}
\label{hkksop20}
\end{align}
%for $\sigma$, $\varSigma$. 

The component expressions of the structure relations
%\ceqref{hkksop9}, \ceqref{hkksop10} and \ceqref{hkksop11}, \ceqref{hkksop12}
\ceqref{hkksop9}--\ceqref{hkksop12}
of the associated right $\DD\fkm'$--operation
involve in addition to the coordinate components $\gamma$, $\varGamma$, $\sigma$, $\varSigma$
the components $x$, $X$ of the derived algebra parameter $\rmX$ in similar fashion 
and read explicitly as %, for $\gamma$, $\varGamma$ they take the form %They read as 
\begin{align}
&\gamma^{-1}j_{\msR x,X}\gamma=0,
\vphantom{\Big]}
\label{hkksop21}
\\
&\mu\sdot(\gamma^{-1},j_{\msR x,X}\varGamma)=0,
\vphantom{\Big]}
\label{hkksop22}
\\
&\gamma^{-1}l_{\msR x,X}\gamma=x,
\vphantom{\Big]}
\label{hkksop23}
\\
&\mu\sdot(\gamma^{-1},l_{\msR x,X}\varGamma)=X,
\vphantom{\Big]}
\label{hkksop24}
\\
%\end{align}
%while for $\sigma$, $\varSigma$ they read as \hphantom{xxxxxxxxxxxxxxxxx}
%\begin{align}
&j_{\msR x,X}\sigma=x,
\vphantom{\Big]}
\label{hkksop25}
\\
&j_{\msR x,X}\varSigma=X,
\vphantom{\Big]}
\label{hkksop26}
\\
&l_{\msR x,X}\sigma=-[x,\sigma]+\dot\tau(X),
\vphantom{\Big]}
\label{hkksop27}
\\
&l_{\msR x,X}\varSigma=-\sdot\mu\sdot(x,\varSigma)+\sdot\mu\sdot(\sigma,X).
\vphantom{\Big]}
\label{hkksop28}
\end{align}

\subsection{\textcolor{blue}{\sffamily Target kernel symmetry}}\label{subsec:tarker}

The target kernel symmetry of a derived homogeneous space
%$\DD\msM/\DD\msM'$ of the kind studied in subsect. \cref{subsec:hkksop} 
is the counterpart of the left symmetry of an ordinary homogeneous space 
described in subsect. \cref{subsec:ophomsp}.
%It is distinguished from this latter by the fact that $\msM_\tau$ is in general a proper
%crossed submodule of $\msM$.
Its properties however turn out to be a bit more involved.
The symmetry is the topic of this subsection.

Every Lie group crossed module $\msM=(\msE,\msG,\tau,\mu)$ is characterized as such 
by a canonical crossed submodule, the target kernel crossed module $\msM_\tau=(\ker\tau,\msG)$. 
With this there is associated a distinguished 
symmetry of the homogeneous space $\DD\msM/\DD\msM'$ of the kind studied 
in subsect. \cref{subsec:hkksop}.

The size and nature of the target kernel crossed module $\msM_\tau$ depends on crossed module $\msM$
to an important extent. For an illustration of this point, we shall refer here to the model crossed modules
described in detail in the last part of subsect. \cref{subsec:cmcentr}. 
If $\msM$ the crossed module $\INN_\msG\msN$ associated with a normal subgroup $\msN$ of the Lie group $\msG$,
$\ker\tau$ is the trivial group $1$; $\msM_\tau$ then reduces to the crossed module $(1,\msG)$,
which codifies the group $\msG$ as a crossed module. 
If $\msM$ is the crossed modules $\msC(\!\xymatrix@C=1.3pc{\msQ\ar[r]^-{\pi}&\msG}\!)$
yielded by a central extension $\!\mhfpt\xymatrix@C=1.3pc{1\ar[r]&\msC\ar[r]^-{\iota}&\msQ\ar[r]^-{\pi}&\msG\ar[r]&1}\!\mhfpt$,
$\ker\tau$ is just the central subgroup $\iota(\msC)$; $\msM_\tau$ is the identified with the crossed module
$(\iota(\msC),\msG)$ with trivial target and action maps. 
Finally, if $\msM$ is the crossed modules $\msD(\rho)$ corresponding to a representation $\rho$ of $\msG$
in a vector space $\msV$, $\ker\tau$ is the whole space $\msV$; $\msM_\tau$ therefore is the crossed module $(\msV,\msG)$, 
that is the whole crossed module $\msM$. 

As a manifold, $\DD\msM$ is endowed with a left $\DD\msM_\tau$--action, which we shall write in
right form for convenience. %In coordinates, it reads as
%The derived target kernel Lie group $\DD\msM_\tau$ acts on the derived Lie group $\DD\msM$.
This has a fairly simple expression in terms derived coordinates $\Gamma$, $\Sigma$.
On the base coordinate $\Gamma$, the action reads as 
\begin{equation}
\Gamma^{\msL\rmE}=\rmE^{-1}\Gamma
\label{tarker1}
\end{equation}
with $\rmE\in\DD\msM_\tau$. 
The action on the fiber coordinate $\Sigma$ is trivial
\begin{equation}
\Sigma^{\msL\rmE}=\Sigma. 
\label{tarker2}
\end{equation}
This follows readily from \ceqref{hkksop5} noting that  
$\msM_\tau$ is by construction the largest crossed submodule of $\msM$
such that $\rmE^{-1}\dd\rmE=0$ for $\rmE\in\DD\msM_\tau$, by \ceqref{superfield14}.
Comparison of \ceqref{tarker1}, \ceqref{tarker2} with \ceqref{ophomsp4} shows clearly that the target
kernel symmetry of the derived KKS theory answers to the left symmetry of the ordinary one
as we anticipated. The analogy is
not complete however: while in the derived theory the target kernel group $\DD\msM_\tau$ is a proper subgroup
of the ambient group $\DD\msM$ in general, in the ordinary one the left symmetry group and the ambient group
are the same, viz $\msG$. 

%involves again the associated clone Lie group structure
%(cf. subsect. \cref{subsec:clkks}). 

%Although we shall not regard $\DD\msM$ as a principal $\DD\msM_\tau$--bundle on $\DD\msM/\DD\msM_\tau$ here,
%In analogy to the ordinary case, 

The left $\DD\msM_\tau$--action \pagebreak is aptly described infinitesimally by an associated left $\DD\fkm_\tau$--operation.
The derivations $j_{\msL\rmH}$, $l_{\msL\rmH}$ thereof are indexed by a 
derived target kernel Lie algebra parameter $\rmH\in\DD\fkm_\tau{}$. They act as 
\begin{align}
&j_{\msL\rmH}\Gamma\Gamma^{-1}=0,
\vphantom{\Big]}
\label{tarker3}
\\
&l_{\msL\rmH}\Gamma\Gamma^{-1}=-\rmH \vphantom{\dot{\dot{\dot{f}}}}
\vphantom{\Big]}
\label{tarker4}
\end{align}
on the coordinate $\Gamma$ and 
\begin{align}
&j_{\msL\rmH}\Sigma=-\Ad\Gamma^{-1}(\rmH),
\vphantom{\Big]}
\label{tarker5}
\\
&l_{\msL\rmH}\Sigma=0
\vphantom{\Big]}
\label{tarker6}
\end{align}
on the coordinate $\Sigma$. Straightforward inspection reveals that
relations \ceqref{tarker3}--\ceqref{tarker6} of the derived theory are %manifestly
the counterpart of relations \ceqref{ophomsp5} of the ordinary one, as expected. 

The left $\DD\msM_\tau$--action manifestly commutes with the right $\DD\msM'$--action of eqs. \ceqref{hkksop7},
\ceqref{hkksop8}. It descends so %in this way
to a $\DD\msM_\tau$--action on the homogeneous space $\DD\msM/\DD\msM'$.
The commutativity of the $\DD\msM_\tau$-- and $\DD\msM'$--actions
is reflected by the graded commutativity of the derivations $j_{\msL \rmH}$, $l_{\msL \rmH}$
of eqs. \ceqref{tarker3}--\ceqref{tarker6} and the derivations $j_{\msR\rmX}$, $l_{\msR\rmX}$ of the right
$\DD\fkm'$--operation given by eqs. \ceqref{hkksop9}--\ceqref{hkksop12}. 

For completeness and later reference, we rewrite the above relations
in terms of the components $\gamma$, $\varGamma$, $\sigma$, $\varSigma$ 
of the derived coordinates $\Gamma$, $\Sigma$.
As in the analogous previous circumstances,
the uninterested reader can skip directly to the next subsection, if he/she wishes so.

The component expression of the left $\DD\msM_\tau$--action shown in eqs. \ceqref{tarker1}, 
\ceqref{tarker2} involves beside the coordinate components $\gamma$, $\varGamma$, $\sigma$, $\varSigma$
the components $e$, $E$ of the derived Lie group parameter $\rmE$.  
They take the form
%Eq. \ceqref{tarker1} reads explicitly as \hphantom{xxxxxxxxxx}
%The defining relation of the $\DD\msM_\tau$--action reads %in terms of $\gamma$, $\varGamma$ as
%in component form as 
\begin{align}
&\gamma^{\msL e,E}=e^{-1}\gamma, 
\vphantom{\Big]}
\label{tarker7}
\\
&\varGamma^{\msL e,E}=\mu\sdot(e^{-1},\varGamma-E),
\vphantom{\Big]}
\label{tarker8}
\\
%\end{align}
%The induced action on $\sigma$, $\varSigma$, eq.
%Eq. \ceqref{tarker2} takes similarly the form
%\begin{align}
&\sigma^{\msL e,E}=\sigma, 
\vphantom{\Big]}
\label{tarker9}
\\
&\varSigma^{\msL e,E}=\varSigma.
\vphantom{\Big]}
\label{tarker10}
\end{align}

The component expressions of the structure relations
%\ceqref{tarker3}, \ceqref{tarker4} and \ceqref{tarker5}, \ceqref{tarker6}
\ceqref{tarker3}--\ceqref{tarker6}
of the associated left $\DD\fkm_\tau$--operation
involve in addition to the %coordinate and Maurer--Cartan
components $\gamma$, $\varGamma$, $\sigma$, $\varSigma$
the components $h$, $H$ %$x\in\fkg^+$, $X\in\fke[1]^+$
of the derived Lie algebra parameter $\rmH$ in similar fashion.  %\in\DD\fkm'^+$.
Explicitly, they take the form %Thy read as \pagebreak 
\begin{align}
&j_{\msL h,H}\gamma\gamma^{-1}=0,
\vphantom{\Big]}
\label{tarker11}
\\
&j_{\msL h,H}\varGamma=0,
\vphantom{\Big]}
\label{tarker12}
\\
&l_{\msL h,H}\gamma\gamma^{-1}=-h,
\vphantom{\Big]}
\label{tarker13}
\\
&l_{\msL h,H}\varGamma=-H-\sdot\mu\sdot(h,\varGamma),
\vphantom{\Big]}
\label{tarker14}
\\
%\end{align}
%for $\gamma$, $\varGamma$ while they read as \hphantom{xxxxxxxxxxx}
%\begin{align}
&j_{\msL h,H}\sigma=-\Ad\gamma^{-1}(h),
\vphantom{\Big]}
\label{tarker15}
\\
&j_{\msL h,H}\varSigma=-\mu\sdot(\gamma^{-1},H+\sdot\mu\sdot(h,\varGamma)),
\vphantom{\Big]}
\label{tarker16}
\\
&l_{\msL h,H}\sigma=0,
\vphantom{\Big]}
\label{tarker17}
\\
&l_{\msL h,H}\varSigma=0.
\vphantom{\Big]}
\label{tarker18}  
\end{align}
%for $\sigma$, $\varSigma$. %the fiber coordinates. %in the Maurer--Cartan case.

%\vfil\eject

\subsection{\textcolor{blue}{\sffamily Derived automorphism symmetry
}}\label{subsec:derauto}

The automorphisms of the principal $\DD\msM'$--bundle $\DD\msM$
studied in subsect. \cref{subsec:hkksop} are the fiber preserving internal mappings
of $\DD\msM$ compatible with the right $\DD\msM'$--action. They are internal 
maps $\Psi\in\iMap(\DD\msM,\DD\msM')$ with certain covariance properties under the right $\DD\msM'$--action
and so naturally constitute a distinguished subgroup $\iAut_{\DD\msM'}(\DD\msM)$ of the infinite
dimensional Lie group $\iMap(\DD\msM,\DD\msM')$. The associated derived automorphism symmetry
is the subject of this subsection. 

The full content of the derived automorphism group $\iAut_{\DD\msM'}(\DD\msM)$ is specified by the requirement that 
automorphisms $\Psi\in\iAut_{\DD\msM'}(\DD\msM)$ transform 
under the right $\DD\msM'$--action of $\DD\msM$ according to 
\begin{equation}
\Psi^{\msR\rmQ}=\rmQ^{-1}\Psi\rmQ 
\vphantom{\Big]}
\label{hkksop31}
\end{equation}
for $\rmQ\in\DD\msM'$ (cf. subsect. \cref{subsec:hkksop}). 
\ceqref{hkksop31} replicates in derived theory
relation \ceqref{ophomsp6} of ordinary theory. It characterizes  
$\Psi$ as a section of an $\Ad_{\DD\msM'}$--bundle over
$\DD\msM/\DD\msM'$ associated to the $\DD\msM'$--bundle $\DD\msM$. 
In the corresponding right $\DD\fkm'$--operation, the $\Psi$ consequently
obey
\begin{align}
&j_{\msR\rmX}\Psi\Psi^{-1}=0,
\vphantom{\Big]}
\label{hkksop32}
\\
&l_{\msR\rmX}\Psi\Psi^{-1}=-\rmX+\Ad\Psi(\rmX)
\vphantom{\Big]}
\label{hkksop33}
\end{align}
with $\rmX\in\DD\fkm'$. In the above expressions and similar ones below, right composition of $\Psi$ with the bundle 
projection $T[1]\DD\msM\simeq\DD\msM\times\DD\fkm[1]\rightarrow\DD\msM$ is tacitly understood.
\ceqref{hkksop32}, \ceqref{hkksop33} are evidently the derived theory counterpart of the ordinary theory 
relations \ceqref{ophomsp7}. 

The derived automorphism group $\iAut_{\DD\msM'}(\DD\msM)$ acts on $\DD\msM$. % is given. 
An automorphism $\Psi\in\iAut_{\DD\msM'}(\DD\msM)$ acts on the base coordinate $\Gamma$ of $\DD\msM$ as %according to 
\begin{equation}
{}^\Psi\Gamma=\Gamma\Psi^{-1}. 
\label{hkksop29}
\end{equation}
This action yields a corresponding action on the fiber coordinate $\Sigma$ of $\DD\msM$, viz 
\begin{equation}
{}^\Psi\Sigma=\Ad\Psi(\Sigma)-\dd\Psi\Psi^{-1},
\label{hkksop30}
\end{equation}
on account of relation \ceqref{hkksop5}. It is immediate to see from \ceqref{hkksop7} that
the right $\DD\msM'$--action on $\iAut_{\DD\msM'}(\DD\msM)$ given in eq. \ceqref{hkksop31} 
is determined by the compatibility requirement that
$({}^\Psi\Gamma)^{\msR\rmQ}={}^{\Psi^{\msR\rmQ}}\Gamma^{\msR\rmQ}$ with $\rmQ\in\DD\msM'$ 
for the action \ceqref{hkksop29}. 
The formal correspondence of relations \ceqref{hkksop29}, \ceqref{hkksop29} and \ceqref{ophomsp8} in the derived and
ordinary theory is again evident.

The left target kernel $\DD\msM_\tau$--action also extends to the derived automorphism group $\iAut_{\DD\msM'}(\DD\msM)$
(cf. subsect. \cref{subsec:tarker}). The extension is trivial: $\Psi$ is left invariant,
\begin{equation}
\Psi^{\msL\rmE}=\Psi.
\label{tarker20}
\end{equation}
for $\rmE\in\DD\msM_\tau$. In the left $\DD\fkm_\tau$--operation yielded by the $\DD\msM_\tau$--action one has so
\begin{align}
&j_{\msL\rmH}\Psi\Psi^{-1}=0,
\vphantom{\Big]}
\label{tarker21}
\\
&l_{\msL\rmH}\Psi\Psi^{-1}=0
\vphantom{\Big]}
\label{tarker22}
\end{align}
with $\rmH\in\DD\fkm_\tau{}$. Relations \ceqref{tarker20}--\ceqref{tarker22} obviously reproduce in the derived
theory relations \ceqref{ophomsp9}, \ceqref{ophomsp10} of the ordinary theory. \vfil\eject

The $\DD\msM_\tau$--action \ceqref{tarker20} is determined by the requirement of compatibility %\pagebreak
with the automorphism action \ceqref{hkksop29} on the base coordinate $\Gamma$ 
demanding that $({}^\Psi\Gamma)^{\msL\rmE}={}^{\Psi^{\msL\rmE}}\Gamma^{\msL\rmE}$
%\begin{equation}
%({}^\Psi\Gamma)^{\msL\rmE}={}^{\Psi^{\msL\rmE}}\Gamma^{\msL\rmE}
%\label{tarker19}
%\end{equation}
with $\rmE\in\DD\msM_\tau{}$ for all automorphisms
$\Psi\in\iAut_{\DD\msM'}(\DD\msM)$, analogously to the right $\DD\msM'$--action. 

Again, for the sake of concreteness and completeness and later use, we provide a
reference list of the component expression of the above relations. 
%via %in terms of
%the components $\psi$, $\varPsi$ %$\psi\in\iMap(T[1]\DD\msM,\msG')$, $\varPsi\iMap(T[1]\DD\msM,\fke'[1])$
%of the automorphisms $\Psi$ and
%the components $\gamma$, $\varGamma$ and $\sigma$, $\varSigma$ of the %base and fiber
%coordinate maps $\Gamma$ and $\Sigma$.
The hurried reader can once more skip to the next subsection.

The component expression of the right $\DD\msM'$--action on automorphisms,
given by eq. \ceqref{hkksop31}, involves in addition to the components $\psi$, $\varPsi$
of the automorphism $\Psi$ the components $q$, $Q$ %$q\in\msG'$, $Q\in\fke'[1]$
of the derived group parameter $\rmQ$, %. It reads as 
\begin{align}
&\psi^{\msR q,Q}=q^{-1}\psi q,
\vphantom{\Big]}
\label{hkksop38}
\\
&\varPsi^{\msR q,Q}=\mu\sdot(q^{-1},\varPsi-Q+\mu\sdot(\psi,Q)).
\vphantom{\Big]}
\label{hkksop39}
\end{align}

%Similarly, t
The component expressions of the structure relations \ceqref{hkksop32}, \ceqref{hkksop33}
of the right $\DD\fkm'$--operation involves correspondingly the components $x$, $X$ 
of the derived algebra parameter $\rmX$ and read as 
\begin{align}
&j_{\msR x,X} \psi\psi^{-1}=0, 
\vphantom{\Big]}
\label{hkksop40}
\\
&j_{\msR x,X} \varPsi=0, 
\vphantom{\Big]}
\label{hkksop41}
\\
&l_{\msR x,X} \psi\psi^{-1}=-x+\Ad \psi(x),
\vphantom{\Big]} 
\label{hkksop42}
\\
&l_{\msR x,X} \varPsi=-\sdot\mu\sdot\hfpt(x,\varPsi)-X+\mu\sdot\hfpt(\psi,X). 
\vphantom{\Big]}
\label{hkksop43}
\end{align}

Relations \ceqref{hkksop29}, \ceqref{hkksop30} describing 
the automorphism action in coordinate form read componentwise as %\hphantom{xxxxxxxxx}
\begin{align}
&{}^{\psi,\varPsi}\gamma=\gamma\psi^{-1},
\vphantom{\Big]}
\label{hkksop34}
\\
&{}^{\psi,\varPsi}\varGamma=\varGamma-\mu\sdot\hfpt(\gamma \psi^{-1},\varPsi),
\vphantom{\Big]}
\label{hkksop35}
\\
%\end{align}
%The ensuing relation \ceqref{hkksop30} similarly takes the form
%\begin{align}
&{}^{\psi,\varPsi}\sigma=\Ad\psi(\sigma)-d\psi\psi^{-1}-\dot\tau(\varPsi),
\vphantom{\Big]}
\label{hkksop36}
\\
&{}^{\psi,\varPsi}\varSigma=\mu\sdot\hfpt(\psi,\varSigma)%-\sdot\mu\sdot(\Ad\psi(\sigma),\varPsi)
-d\varPsi-\tfrac{1}{2}[\varPsi,\varPsi]-\sdot\mu\sdot(\Ad\psi(\sigma)-d\psi\psi^{-1}-\dot\tau(\varPsi),\varPsi). 
\vphantom{\Big]}
\label{hkksop37}
\end{align}

%\vfil\eject

%Again, for completeness and later reference,
%we rewrite the above relations also in terms of the components $\psi$, $\varPsi$
%of the automorphism $\Psi$. 
%The uninterested reader can skip directly to the next subsection.

The component \pagebreak expression of the left $\DD\msM_\tau$--action relation \ceqref{tarker20} reads as
follows \hphantom{xxxxxxxxxxxxxxxxx}
%involves beside $\psi$, $\varPsi$ the components $e$, $E$ %$q\in\msG'$, $Q\in\fke'[1]$
%of the derived Lie group parameter $\rmE$ and reads 
\begin{align}
&\psi^{\msL e,E}=\psi,
\vphantom{\Big]}
\label{tarker23}
\\
&\varPsi^{\msL e,E}=\varPsi.
\vphantom{\Big]}
\label{tarker24}
\end{align}
The component expression of the structure relations \ceqref{tarker21}, \ceqref{tarker22}
in the associated $\DD\fkm_\tau$--operation takes consequently the form 
\begin{align}
&j_{\msL h,H}\psi\psi^{-1}=0, 
\vphantom{\Big]}
\label{tarker25}
\\
&j_{\msL h,H}\varPsi=0, 
\vphantom{\Big]}
\label{tarker26}
\\
&l_{\msL h,H}\psi\psi^{-1}=0,
\vphantom{\Big]}
\label{tarker27}
\\
&l_{\msL h,H}\varPsi=0. 
\vphantom{\Big]}
\label{tarker28}  
\end{align}

%\vfil\eject

\subsection{\textcolor{blue}{\sffamily The derived unitary line bundle of a character}}\label{subsec:crline}

In this subsection, the operational framework of subsect. \cref{subsec:hkksop}.
describing a derived homogeneous space $\DD\msM/\DD\msM'$ is adapted to the
case occurring in the derived KKS theory of regular coadjoint orbits 
(cf. subsect. \cref{subsec:cmcentr})
where $\msM$ is compact and $\msM'$ is a maximal toral crossed submodule $\msJ$ of $\msM$.
On a regular derived homogeneous space
further geometrical objects can be considered such as derived unitary line bundles, 
their sections and their unitary connections. 

%constructions can be realized. In particular, we shall introduce
%the derived unitary line bundle of a character. %We begin with the definition of this notion. 

With any Lie group $\msL$, there is associated its inner automorphism %Lie group
crossed module $\INN\msL=(\msL,\msL,\id_{\msL},\Ad_{\msL})$,
where $\Ad_{\msL}$ denotes the conjugation action of $\msL$ on itself.
Similarly, with any Lie algebra $\fkl$, there is associated its inner
derivation %Lie algebra
crossed module $\INN\fkl=(\fkl,\fkl,\id_{\fkl},\ad_{\fkl})$,
where $\ad_{\fkl}$ denotes the adjoint action of $\fkl$ on itself. If $\fkl$ is the Lie algebra of 
$\msL$, then $\INN\fkl$ is the Lie algebra crossed module of $\INN\msL$. 

The inner automorphism crossed module $\INN\msU(1)$ %=(\msU(1),\msU(1),\id_{\msU(1)},\mu_{\msU(1)})$
of the Lie group $\msU(1)$ and the Lie algebra crossed module
$\INN\fku(1)$ %=(\fku(1),\fku(1),t_{\fku(1)},m_{\fku(1)})$
of the Lie algebra $\fku(1)$ feature in the theory of crossed module characters outlined 
momentarily.  For these, the conjugation action $\Ad_{\msU(1)}$ and the adjoint action $\ad_{\fku(1)}$
appearing in their definition are trivial due to the Abelian nature 
of $\msU(1)$ and $\fku(1)$, respectively. 

Let $\msJ$ be a maximal toral crossed submodule of the Lie group crossed module 
$\msM$. A character of $\msJ$ is a Lie group crossed module morphism
$\beta:\msJ\rightarrow\INN\mathsans{U}(1)$ (cf. subsect. \cref{subsec:liecrmod}).
The component
$\xi:\msT\rightarrow\mathsans{U}(1)$ of $\beta$ is then an ordinary character of
the maximal torus $\msT$ of $\msG$. Further, the component $\varXi:\msH\rightarrow\mathsans{U}(1)$ 
is determined by $\xi$ as $\varXi=\xi\circ\tau$. 
%for $A\in\msH$
%\begin{equation}
%\varXi(A)=\xi(\tau(A)).
%\label{coweights0}
%\end{equation}
There exists in this way a one--to--one correspondence between the character set 
of $\msJ$ and that of $\msT$.

Having defined the notion of character, we consider next
a compact Lie group crossed module $\msM=(\msE,\msG,\tau,\mu)$
(cf. subsect. \cref{subsec:cmcentr}) and
a maximal toral crossed submodule $\msJ=(\msH,\msT)$ of $\msM$ equipped with a character
$\beta=(\varXi,\xi)$. 

Consider the complex vector space %$\DD\bbC=\iMap(\bbR[-1],\bbC)$. This is simply
\begin{equation}
\DD\bbC\simeq\bbC\oplus\bbC[1].
\label{crline1}
\end{equation}
The coordinate $\rmZ\in\DD\bbC$ has therefore a component expansion of the form 
%An element $\rmZ\in\DD\bbC$ is viewed as an 
%internal function $\rmZ\in\iMap(\bbR[-1],\bbC)$ as %\linebreak \vskip-1cm through
\begin{equation}
\rmZ(\alpha)=z+\alpha Z
\label{crline1/1}
\end{equation}
with $\alpha\in\bbR[-1]$, where $z\in\bbC$, $Z\in\bbC[1]$.
Conjugation in $\DD\bbC$ is defined such that $\rmZ^*$ has components $z^*$, $Z^*$. 
%by component conjugation. 
$\DD\bbC$ has further a  natural field structure with 
\begin{equation}
\rmW\rmZ(\alpha)=wz+\alpha(zW+wZ), \qquad \rmZ^{-1}(\alpha)=z^{-1}-\alpha z^{-2}Z.
\label{crline1/2}
\end{equation}
For this reason, we shall call $\DD\bbC$ the derived complex line though it is in fact
2--dimensional. %, indeed viewable also as the shifted tangent bundle $T[1]\bbC$. 

%There is a natural multiplicative action of $\DD\INN\msU(1)$ on $\DD\bbC$
%via an embedding $\iota:\DD\INN\msU(1)\rightarrow\DD\bbC$ defined in terms
%of the coordinate $\rmP\in\DD\INN\msU(1)$ by
%There is also an induced multiplicative action of $\DD\INN\fku(1)$ on $\DD\bbC$
%obtained via the associated embedding $\dot\iota:\DD\INN\fku(1)\rightarrow\DD\bbC$ with
%By the coordinate $\rmP\in\DD\INN\msU(1)$
%of $\DD\INN\msU(1)$ % defined by 

The derived Lie group $\DD\INN\msU(1)$ acts multiplicatively on $\DD\bbC$ via an
embedding $\iota:\DD\INN\msU(1)\rightarrow\DD\bbC$. The coordinate expression of $\iota$ reads as 
\begin{equation}
\iota(\rmP)(\alpha)=p+\alpha pP,
\label{crline1/3}
\end{equation}
where $\rmP\in\DD\INN\msU(1)$. 
Likewise, the derived $\DD\INN\fku(1)$ acts multiplicatively on $\DD\bbC$ via the associated embedding
$\dot\iota:\DD\INN\fku(1)\rightarrow\DD\bbC$ given by 
\begin{equation}
\dot\iota(\rmY)(\alpha)=y+\alpha Y 
\label{crline1/4}
\end{equation}
in coordinate form, where $\rmY\in\DD\INN\fku(1)$. % of $\INN\fku(1)$.
In what follows, $\iota$ and $\dot\iota$ will be tacitly understood for brevity.

\vfil\eject

Via the character $\beta$, $\DD\bbC$ carries the left action of $\DD\msJ$ defined by
\begin{equation}
\rho_\beta(\rmQ)(Z)=\DD\beta(\rmQ)^{-1}\rmZ
\label{crline2/1}
\end{equation}
with $\rmQ\in\DD\msJ$, where $\DD\beta$ is defined in subsect. \cref{subsec:dergralg}.
Writing $\rmZ$ and $\rmQ$ in the forms \ceqref{crline1/1} and \ceqref{superfield1}
respectively, $\rho_\beta(\rmQ)(\rmZ)$ can be expressed in components as 
\begin{equation}
\rho_\beta(\rmQ)(\rmZ)(\alpha)=\xi(q)^{-1}z+\alpha\xi(q)^{-1}(Z-\dot\varXi(Q)z),
\label{crline2}
\end{equation}
where $\varXi=\xi\circ\tau$. 

We introduce next the vector bundle on $\DD\msM/\DD\msJ$ defined as
\begin{equation}
\clL_\beta=\DD\msM\times_\beta\DD\bbC=\DD\msM\times\DD\bbC/\sim_\beta,
\label{crline3}
\end{equation}
where $\sim_\beta$ is the equivalence relation
\begin{equation}
(\Gamma,\rmZ)\sim_\beta(\Gamma\rmQ,\rho_\beta(\rmQ^{-1})(\rmZ))
\label{crline4}
\end{equation}
in coordinate form with $\rmQ\in\DD\msJ$. $\clL_\beta$ is the derived unitary 
line bundle associated with the character $\beta$. $\clL_\beta$ is termed in this way by its analogy 
to the ordinary unitary line bundle of a character (cf. subsect. \cref{subsec:stkkspreq}),
though $\clL_\beta$ is strictly speaking a rank 2 vector bundle and so not a line bundle in the usual sense.
%We shall regard $\clL_\beta$ as the derived prequantum line bundle. 

The definitions of and results about sections and connections of $\clL_\beta$ provided next elegantly 
replicate in derived theory the analogous notions of ordinary one reviewed in subsect.
\ceqref{subsec:stkkspreq}, showing the effectiveness of the derived approach. 

A derived degree $p$ $\bbC$--valued function is defined naturally as an internal  map $\rmS\in\iMap(T[1]\DD\msM,\DD\bbC[p])$.
It has a component expansions of the form 
\begin{equation}
\rmS(\alpha)=s+(-1)^p\alpha S
\label{wcrline1}
\end{equation}
with $\alpha\in\bbR[-1]$, where
$s\in\iMap(T[1]\DD\msM,\bbC[p])$, $S\in\iMap(T[1]\DD\msM,\bbC[p+1])$.
A derived differential $\dd:\iMap(T[1]\DD\msM,\DD\bbC[p])
\rightarrow\iMap(T[1]\DD\msM,\DD\bbC[p+1])$ is defined through the component expression
\begin{equation}
\dd\rmS(\alpha)=ds+(-1)^pS+(-1)^{p+1}\alpha dS.
\label{wcrline2}
\end{equation}
$\dd$ is nilpotent as is easily checked.

\vfil\eject

In the operational framework we are employing, 
the space $\DD\Omega^p(\clL_\beta)$ of derived $p$--form sections of
the derived line bundle $\clL_\beta$ can be identified with the subspace %$\iMap_\beta(T[1]\DD\msM,\DD\bbC[p])$ 
of the space of derived degree $p$ $\bbC$--valued functions $\iMap(T[1]\DD\msM,\DD\bbC[p])$ spanned by
the elements which are basic with respect to the action %\pagebreak
$\rho_\beta$ defining $\clL_\beta$. A derived function 
$\rmS\in\iMap(T[1]\DD\msM,\DD\bbC[p])$ is basic if it satisfies the horintality and covariance conditions
%explicitly reading as \hphantom{xxxxxxxxxx}
\begin{align}
&j_{\msR\rmX}\rmS=0,
\vphantom{\Big]}
\label{crline5/1}
\\
&l_{\msR\rmX}\rmS=\DD\dot\beta(\rmX)\rmS
\vphantom{\Big]}
\label{crline7/1}
\end{align}
for $\rmX\in\DD\fkj$. Compare the above derived relations with the corresponding ones
of the ordinary case in eq. \ceqref{stkkspreq3}. 
%(see the discussion at the end of subsect. \cref{subsec:opers}). \vspace{.33mm}

The unitary connections of the derived line bundle
$\clL_\beta$ can also be described in the operational formalism. 
A derived unitary connection of $\clL_\beta$ is an internal derived
function $\rmA\in\iMap(T[1]\DD\msM,\DD\INN\fku(1)[1])$ satisfying for $\rmX\in\DD\fkj$ %the conditions 
\begin{align}
&j_{\msR\rmX}\rmA=-\DD\dot\beta(\rmX),
\vphantom{\Big]}
\label{crline5/2}
\\
&l_{\msR\rmX}\rmA=-\dd\DD\dot\beta(\rmX),
\vphantom{\Big]}
\label{crline7/2}
\end{align}
where $\dd$ is the derived differential of $\DD\msM$. 
The curvature of $\rmA$ is the derived function $\rmB\in\iMap(T[1]\DD\msM,\DD\INN\fku(1)[2])$ defined by 
\begin{equation}
\rmB=\dd\rmA.
\label{ncrline1}
\end{equation}
By construction, $\rmB$ obeys the derived Bianchi identity
\begin{equation}
\dd\rmB=0.
\label{ncrline1/1}
\end{equation}
Further, from  \ceqref{crline5/2}, \ceqref{crline7/2}, $\rmB$ satisfies 
\begin{align}
&j_{\msR\rmX}\rmB=0
\vphantom{\Big]}
\label{crline5/3}
\\
&l_{\msR\rmX}\rmB=0.
\vphantom{\Big]}
\label{crline7/3}
\end{align} 
\ceqref{crline5/3}--\ceqref{crline7/3} indicate that %$\rmB\in\DD\Omega^2(\clL_1)$ %$\rmB\in \iMap_1(T[1]\DD\msM,\DD\bbC[2])$
$\rmB$ is a $2$--form section of the derived line bundle $\clL_1$ 
of the trivial character $1:\msJ\rightarrow\Inn\mathsans{U}(1)\vphantom{\ul{\ul{\ul{g}}}}$,
\pagebreak which we may describe as an $\Inn\mathsans{U}(1)$--valued 
derived 2--form on $\DD\msM/\DD\msJ$. %$\rmB\in\rmA^2(\DD\msM/\DD\msJ)$.
The operational conditions \ceqref{crline5/2}, \ceqref{crline7/2} specifying a connection
extend the ordinary requisites \ceqref{stkkspreq4}.
%As in similar instances encountered earlier, t
The right hand side of \ceqref{crline7/2}
does not vanish again by the special nature of the derived differential $\dd$ (cf. eq. \ceqref{superfield10}).
Conversely, the definition \ceqref{ncrline1}, the 
Bianchi identity \ceqref{ncrline1/1} and the operational 
relations \ceqref{crline5/3}, \ceqref{crline7/3} specifying  a connection's curvature 
exactly match their ordinary counterparts of eqs. \ceqref{stkkspreq5} and \ceqref{stkkspreq6}.
The unitary connections of $\clL_\beta$ form an affine space $\iConn(\clL_\beta)$.

When a connection $\rmA$ of $\clL_\beta$ is assigned, the derived covariant differential of
%a basic function $\mrS\in\iMap_\beta(T[1]\DD\msM,\DD\bbC[p])$ 
a $p$--form section $\rmS$ of $\clL_\beta$ can be defined,  
\begin{equation}
\dd_\rmA\rmS=\dd\rmS+\rmA\rm S. 
\label{ncrline4}
\end{equation}
By virtue of \ceqref{crline5/1}, \ceqref{crline7/1} and 
\ceqref{crline5/2}, \ceqref{crline7/2}, $\dd_\rmA\rmS$ is a $p+1$--form section of $\clL_\beta$
as required. The derived Ricci identity holds,  
\begin{equation}
\dd_\rmA\dd_\rmA\rmS=\rmB\rmS.
\label{ncrline5}
\end{equation}
%holds. %, where $B$ is the curvature of $A$ given by \ceqref{ncrline1}.

In the operational framework, it is further possible to characterize the derived 
gauge transformations of the derived line bundle
$\clL_\beta$. A gauge transformation of $\clL_\beta$ can be defined 
as an internal derived function $\rmU\in\iMap(T[1]\DD\msM,\DD\INN\msU(1))$ satisfying 
the horizontality and conjugation covariance conditions
\begin{align}
&j_{\msR\rmX}\rmU\rmU^{-1}=0,
\vphantom{\Big]}
\label{xgcrline5/2}
\\
&l_{\msR\rmX}\rmU\rmU^{-1}=0
\vphantom{\Big]}
\label{xgcrline7/2}
\end{align}
for $\rmX\in\DD\fkj$. It is useful to compare the derived conditions \ceqref{xgcrline5/2}, \ceqref{xgcrline7/2} with the
corresponding relations \ceqref{stkkspreq7} of the ordinary theory.
Derived gauge transformations constitute a distinguished subgroup $\iGau(\clL_\beta)$ of the mapping group
$\iMap(T[1]\DD\msM,\DD\INN\msU(1))$. 

A gauge transformation $\rmU$ acts on a degree $p$ section $\rmS$ of $\clL_\beta$ as
\begin{equation}
{}^\rmU\rmS=\rmU\rmS.
\label{gggcrline3}
\end{equation}
$\rmU$ also acts on a connection $\rmA$ of $\clL_\beta$ as 
\begin{equation}
{}^\rmU\rmA=\rmA-\dd\rmU\rmU^{-1}.
\label{gggcrline1}
\end{equation}
The curvature $\rmB$ of $\rmA$ is then gauge invariant
\begin{equation}
{}^\rmU\rmB=\rmB.
%\vphantom{\Big]}
\label{gggcrline2}
\end{equation}
By construction, the action is such that 
\begin{equation}
\dd_{{}^\rmU\rmA}{}^\rmU\rmS={}^\rmU(\dd_\rmA\rmS)=\rmU\dd_\rmA\rmS
\label{gggcrline4}
\end{equation}
as usual. The derived gauge transformation properties of sections, connections and curvature thereof
of the bundle $\clL_\beta$ shown in
eqs. \ceqref{gggcrline3}--\ceqref{gggcrline2} are totally analogous to their ordinary counterparts
\ceqref{stkkspreq8}--\ceqref{stkkspreq10}.

As in earlier similar instances, 
for concreteness and completeness, we express the above relations
in terms of the components of the fields involved. This will also exemplify how
of the derived framework effectively encodes non trivial higher gauge theoretic relations. 
The uninterested reader can skip directly to the paragraph following eq. \ceqref{xglcrline18/1}. 

Consider a $p$--form section $\rmS$ of $\clL_\beta$ with components $s$, $S$. In terms of these latter,
the horizontality and covariance conditions \ceqref{crline5/1}, \ceqref{crline7/1} take the form 
\begin{align}
&j_{\msR x,X}s=0,
\vphantom{\Big]}
\label{crline5}
\\
&j_{\msR x,X}S=0,
\vphantom{\Big]}
\label{crline6}
\\
&l_{\msR x,X}s=\dot\xi(x)s,
\vphantom{\Big]}
\label{crline7}
\\
&l_{\msR x,X}S=\dot\xi(x)S+(-1)^p\dot\varXi(X)s,
\vphantom{\Big]}
\label{crline8}
\end{align} 
where $x$, $X$ are the components of the Lie algebra element $\rmX$.
%$x\in\fkg$, $X\in\fke[1]$ are the components of $\rmX\in\DD\fkj$.  

A connection $\rmA$ of $\clL_\beta$ can similarly be expressed 
through its components $a$, $A$. By \ceqref{crline5/2}, \ceqref{crline7/2}, 
$a$, $A$ obey 
\begin{align}
&j_{\msR x,X}a=-\dot\xi(x),
\vphantom{\Big]}
\label{crline11}
\\
&j_{\msR x,X}A=-\dot\varXi(X),
\vphantom{\Big]}
\label{crline12}
\\
&l_{\msR x,X}a=-\dot\varXi(X),
\vphantom{\Big]}
\label{crline13}
\\
&l_{\msR x,X}A=0.
\vphantom{\Big]}
\label{crline14}
\end{align}
%with $x\in\fkg$, $X\in\fke[1]$.  %$\rmX\in\DD\fkj$. 
By virtue of \ceqref{ncrline1} \pagebreak  
the components of the curvature $\rmB$ of $\rmA$, $b$, $B$, 
are expressible in terms of $a$, $A$ according to 
\begin{align}
&b=da-A,
\vphantom{\Big]}
\label{crline17}
\\
&B=dA,
\vphantom{\Big]}
\label{crline18}
\end{align}
where $d$ denotes the de Rham differential of $\DD\msM$. 
The Bianchi identity \ceqref{ncrline1/1} obeyed by $\rmB$ translates into the relations 
\begin{align}
&db+B=0,
\vphantom{\Big]}
\label{crline17/1}
\\
&dB=0,
\vphantom{\Big]}
\label{crline18/1}
\end{align}
By \ceqref{crline5/3}, \ceqref{crline7/3}, $b$, $B$ satisfy furthermore 
\begin{align}
&j_{\msR x,X}b=0,
\vphantom{\Big]}
\label{crline19}
\\
&j_{\msR x,X}B=0,
\vphantom{\Big]}
\label{crline20}
\\
&l_{\msR x,X}b=0, 
\vphantom{\Big]}
\label{crline21}
\\
&l_{\msR x,X}B=0.
\vphantom{\Big]}
\label{crline22}
\end{align}

We denote by $d_{a,A}s$, $d_{a,A}S$ the components of the covariant differential
$\dd_\rmA\rmS$ of a $p$--form section $\rmS$ with respect to an assigned connection
$\rmA$ of $\clL_\beta$. Using \ceqref{ncrline4}, $d_{a,A}s$, $d_{a,A}S$
are found to be given by 
\begin{align}
&d_{a,A}s=ds+a s+(-1)^pS,
\vphantom{\Big]}
\label{crline9}
\\
&d_{a,A}S=dS+a S+(-1)^pA s. 
\vphantom{\Big]}
\label{crline10}
\end{align}
Note that $d_{a,A}s$, $d_{a,A}S$ are not separately linear is $s$, $S$. The component
notation used here, though suggestive, must therefore be used with some care.
The Ricci identity \ceqref{ncrline5} can correspondingly be written as  
\begin{align}
&d_{a,A}d_{a,A}s=bs,
\vphantom{\Big]}
\label{crline15}
\\
&d_{a,A}d_{a,A}S=bS+(-1)^pBs. 
\vphantom{\Big]}
\label{crline16}
\end{align}

A gauge transformation $\rmU$ of $\clL_\beta$ can similarly be expressed 
through its components $u$, $U$. By \ceqref{xgcrline5/2}, \ceqref{xgcrline7/2}, $u$, $U$ obey  \pagebreak 
\begin{align}
&j_{\msR x,X}uu^{-1}=0,
\vphantom{\Big]}
\label{glcrline11}
\\
&j_{\msR x,X}U=0,
\vphantom{\Big]}
\label{glcrline12}
\\
&l_{\msR x,X}uu^{-1}=0,
\vphantom{\Big]}
\label{glcrline13}
\\
&l_{\msR x,X}U=0.
\vphantom{\Big]}
\label{glcrline14}
\end{align} 
%with $x\in\fkg$, $X\in\fke[1]$. 

On account of \ceqref{gggcrline3}, the action of a gauge transformation $\rmU$ on a $p$--form
section $\rmS$ of $\clL_\beta$ takes in components the form 
\begin{align}
&{}^{u,U}s=us,
\vphantom{\Big]}
\label{yglcrline17}
\\
&{}^{u,U}S=u(S+(-1)^pUs).
\vphantom{\Big]}
\label{yglcrline18}
\end{align}
By \ceqref{gggcrline1}, the action of $\rmU$ on a 
connection $\rmA$ $\clL_\beta$ reads in component form as 
\begin{align}
&{}^{u,U}a=a-duu^{-1}-U,
\vphantom{\Big]}
\label{xglcrline17}
\\
&{}^{u,U}A=A-dU.
\vphantom{\Big]}
\label{xglcrline18}
\end{align}
Similarly, by \ceqref{gggcrline2}, the action of $\rmU$ on the curvature $\rmB$ of $\rmA$ is componentwise
%components $b$, $B$ of the curvature $\rmB$ of $\rmA$ reads in components as 
\begin{align}
&{}^{u,U}b=b,
\vphantom{\Big]}
\label{xglcrline17/1}
\\
&{}^{u,U}B=B. 
\vphantom{\Big]}
\label{xglcrline18/1}
\end{align}

We conclude this subsection with an examination of basic differential topological issues 
concerning  derived unitary line bundles. Is there a topological classification of
such bundles based on a cohomological characterization of their curvature
analogous to that of ordinary line bundle? This is a far reaching question whose full answer
lies beyond the scope of the present work. We shall limit ourselves 
to the following considerations.

We start with the following premises. First, the derived line bundle $\clL_\beta$ of a character
$\beta$ of $\msJ$ is a graded vector bundle on the graded manifold $\DD\msM/\DD\msJ$.
Second, a unitary connection $\rmA$ of $\clL_\beta$ is a non ordinary $\DD\Inn\fku(1)[1]$--valued internal
map on the graded manifold $T[1]\DD\msM$ with special properties. Similarly, the
curvature $\rmB$ of $\rmA$ is a non ordinary $\DD\INN\fku(1)[2]$--valued internal map on
$T[1]\DD\msM$.  The bundle theoretic \pagebreak set--up we are concerned with here, therefore, is a non standard 
one. This renders the application of standard differential topological results problematic.
We may try to tackle the issue anyway from the derived standpoint on which our whole approach is based.

The characters $\beta$ of the maximal toral crossed submodule $\msJ$ are in one--to--one correspondence
with the characters $\xi$ of the maximal torus $\msT$ of $\msG$. These constitute a lattice, the dual
lattice $\Lambda_\msG{}^*$ of the integer lattice $\Lambda_\msG$ of $\fkt$. The derived line bundles
$\clL_\beta$ of the characters $\beta$ thus form a discrete family of vector bundles
on $\DD\msM/\DD\msJ$ organized as a kind of lattice isomorphic in some sense to $\Lambda_\msG{}^*$. 

Below, we shall consider derived cohomology, that is the cohomology of the complex
of $\Inn\mathsans{U}(1)$--valued forms on $\DD\msM/\DD\msJ$ and the derived
differential $\dd$, and so the terms 'closed' and 'exact' will tacitly refer to such
complex.  We saw earlier that, due to \ceqref{crline5/3}, \ceqref{crline7/3}, the curvature $\rmB$
of a unitary connection $\rmA$ of a derived line bundle $\clL_\beta$ is an $\Inn\mathsans{U}(1)$--valued
derived 2--form. By \ceqref{ncrline1/1}, $\rmB$ is further closed. 
If $\rmA$, $\rmA'$ are connections of $\clL_\beta$, then their difference
$\rmA'-\rmA$ is a derived 1--form on account of
\ceqref{crline5/2}, \ceqref{crline7/2} and consequently, by \ceqref{ncrline1},
the difference of the curvatures $\rmB$, $\rmB'$ of
$\rmA$, $\rmA'$, $\rmB'-\rmB=\dd(\rmA'-\rmA)$, is an exact derived 2--form. 
Hence, the degree 2 derived cohomology class $\rmc(\clL_\beta)=[-i\rmB/2\pi]_\dd$ does not depend
on the connection $\rmA$ underlying $\rmB$ and so constitutes a differential topological
property of the line bundle $\clL_\beta$, the derived Chern class of $\clL_\beta$,
much in analogy to the ordinary line bundle theory. 

The lattice nature of the family of derived line bundles $\clL_\beta$ for varying character $\beta$
indicates their derived Chern classes $\rmc(\clL_\beta)$ belong to some kind of lattice in the 2nd derived
cohomology. We do not know whether there is a way of characterizing this cohomological lattice in terms
of some form integrality of derived periods, in analogy to the ordinary theory. 

The derived closedness condition of the curvature $\rmB$ of a derived line bundle $\clL_\beta$ 
can be written through its components $b$, $B$ and reads as in \ceqref{crline17/1},
\ceqref{crline18/1}. In customary de Rham cohomology, $B$ is so exact whilst $b$ is
not even closed. Hence, the derived Chern class $\rmc(\clL_\beta)$ cannot be straightforwardly framed
in de Rham cohomology.

A relationship of the theory of derived line bundles with connection
to the theory of twisted Hermitian line bundles of
bundle gerbes proposed by C. Rogers in ref. \ccite{Rogers:2011zc} is conceivable %possible
but remains to be elucidated. The point is that our approach to derived line bundles,
based as it is on an operational set--up, is essentially a total space one while 
Rogers's is a base space one. Clarifying the nature of the relations between the two
standpoints, if a relation does exist, would require further work.

%\vspace{-.4mm}

\subsection{\textcolor{blue}{\sffamily Derived Poisson structure of a derived line bundle}}\label{subsec:quapoi}

%\vspace{-.2mm}

In subsect. \cref{subsec:crline}, 
we built the derived unitary line bundle $\clL_\beta$ of a given character $\beta$
on the homogeneous space $\DD\msM/\DD\msJ$. 
By \ceqref{ncrline1}--\ceqref{crline7/3}, the curvature $\rmB$ of a unitary connection $\rmA$ of $\clL_\beta$
is a $\dd$--closed $\INN\fku(1)$--valued derived 2--form. $-i\rmB$ can so be thought of as a derived presymplectic
structure on $\DD\msM/\DD\msJ$. One can construct an associated derived Poisson structure 
duplicating in the derived framework the  standard theory reviewed in subsect. \cref {subsec:stkkssympl}
as we shall show below.
%will be described presently. 
%Moreover, as shown above, when $\rmA$ is invariant $\rmB$ is invariant as well.
%We view $\rmB$ as a derived presymplectic structure.
%a degree 2 $\dd$--closed basic derived function on $\DD\msM$.

%There is a canonical  $\iVect(\DD\msM)$--operation on the algebra $\iFun(T[1]\DD\msM)$, where  
%$\iVect(\DD\msM)$ denotes the Lie algebra of vector fields on $\DD\msM$. It consists of the de Rham
%and the contraction and Lie derivative vector fields of $T[1]\DD\msM$, $d$, $j_V$, $l_V$
%with $V\in\iVect(\DD\msM)$, obeying familiar relations analogous to 
%\ceqref{hkksop1}--\ceqref{hkksop4}. This operation is the largest that $\DD\msM$ supports. 
%The vector fields $j_\rmX$, $l_\rmX$ with $\rmX\in\DD\fkj$ of the $\DD\fkj$--operation 
%are just the vector fields $j_{\msR\rmX}$, $l_{\msR\rmX}$, 
%where $S_{\msR\rmX}$ is the vertical vector field of the $\DD\msJ$--action corresponding to $\rmX$.

Below, we shall systematically use  %rely heavily on
the internal Cartan calculus of $\DD\msM$ (cf. subsect. \cref{subsec:clkks}). 
%employ the form of Cartan calculus appropriate for the graded manifold $\DD\msM$. 
This features a set of graded derivations of $\iFun(T[1]\DD\msM)$ including 
the degree $-1$ contractions
$j_V$ and Lie derivatives $l_V$ along the vector fields $V\in\iVect(\DD\msM)$, where $\iVect(\DD\msM)$
is the Lie algebra of internal vector fields of $\DD\msM$, and the de Rham differential $d$ and
obeying relations \ceqref{hkksop1}--\ceqref{hkksop4}. 
The derivations $j_{\msR\rmX}$, $l_{\msR\rmX}$ with $\rmX\in\DD\fkj$ of the right $\DD\msJ$--operation 
are just % the derivations
$j_{S_{\msR\rmX}}$, $l_{S_{\msR\rmX}}$, 
where $S_{\msR\rmX}$ is the vertical vector field of the $\DD\msJ$--action corresponding to $\rmX$.

%For a fixed derived connection $\rmA$ of $\clL_\beta$, 
Next, we consider the set $\iVect_{\msi\rmA}(\DD\msM)$ 
of all vector fields $V\in\iVect(\DD\msM)$ leaving $\rmB$ invariant, that is satisfying \hphantom{xxxxxxxxxxxxx}
\begin{align}
l_V\rmB=0, 
\label{phiinv1}
\end{align}
%We consider also 
and the set $\iVect_{\msk\rmA}(\DD\msM)$ of all vector fields $V\in\iVect(\DD\msM)$ obeying 
%leaving $\rmB$ invariant, that is satisfying \hphantom{xxxxxxxxxx}
\begin{align}
j_V\rmB=0.
\label{phiinv2}
\end{align}
It is straightforward to verify that $\iVect_{\msi\rmA}(\DD\msM)$ is a Lie subalgebra of $\iVect(\DD\msM)$
and that $\iVect_{\msk\rmA}(\DD\msM)$ is a Lie ideal of $\iVect_{\msi\rmA}(\DD\msM)$. We can thus construct the
quotient Lie algebra $\iVect_{\msq\rmA}(\DD\msM)=\iVect_{\msi\rmA}(\DD\msM)/\iVect_{\msk\rmA}(\DD\msM)$.

Let $\iVect_\msv(\DD\msM)$ be the Lie subalgebra of $\iVect(\DD\msM)$ of the
vertical vector fields $S_{\msR\rmX}$ with $\rmX\in\DD\fkj$. $\iVect_\msv(\DD\msM)$ is a Lie subalgebra
of $\iVect_{\msi\rmA}(\DD\msM)$, as $l_{\msR\rmX}\rmB=0$ for $\rmX\in\DD\fkj$ by \ceqref{crline7/3}.
The Lie derivatives $l_{S_{\msR\rmX}}$ are thus derivations of the Lie algebra
$\iVect_{\msi\rmA}(\DD\msM)$ preserving the Lie ideal $\iVect_{\msk\rmA}(\DD\msM)$. 
They so induce derivations $l_{\msq S_{\msR\rmX}}$ of the Lie algebra $\iVect_{\msq\rmA}(\DD\msM)$ defined
in the previous paragraph. 
$\iVect_{\msb\rmA}(\DD\msM)=\bigcap_{\rmX\in\DD\fkj}\ker l_{\msq S_{\msR\rmX}}$ is consequently
a Lie subalgebra of $\iVect_{\msq\rmA}(\DD\msM)$.
Since we aim eventually to construct a structure on the derived homogeneous space
$\DD\msM/\DD\msJ$, $\iVect_{\msb\rmA}(\DD\msM)$ is the relevant vector field algebra. %For this reason, 
%We describe its content %more
Explicitly, $\iVect_{\msb\rmA}(\DD\msM)$ consists of the vector fields
$V\in\iVect(\DD\msM)$ obeying \ceqref{phiinv1} and defined modulo vector fields $V'\in\iVect(\DD\msM)$ 
obeying \ceqref{phiinv2} with the property that $l_{S_{\msR\rmX}}V$ obeys \ceqref{phiinv2} 
for all $\rmX\in\DD\fkj$.

%\vspace{-.33mm}

The derived real line is the real vector space $\DD\bbR=\bbR\oplus\bbR[1]$. $\DD\bbR$ 
has properties completely analogous to those of the derived complex line $\DD\bbC$ studied earlier
in subsect. \cref{subsec:crline},
in particular it is a field (cf. eqs. \ceqref{crline1/1}, \ceqref{crline1/1}).
The space $\iMap(T[1]\DD\msM,\DD\bbR[p])$ of degree $p$ derived $\bbR$--valued functions is
then available.
It has a component reduction analogous to that of its complex counterpart
(cf. eq. \ceqref{wcrline1}) and is acted upon by the derived differential $\dd$
(cf. eq. \ceqref{wcrline2}).  
Below, we shall concentrate on the space $\iDFnc(\DD\msM)=\iMap(T[1]\DD\msM,\DD\bbR[0])$ of derived functions
of $\DD\msM$. An element $\rmF\in\iDFnc(\DD\msM)$ is therefore a derived field: 
$\rmF(\alpha)=f+\alpha F$ with $f\in\iMap(T[1]\DD\msM,\bbR[0])$, $F\in\iMap(T[1]\DD\msM,\bbR[1])$.
$\iDFnc(\DD\msM)$ has a natural algebra structure induced by the field structure of $\DD\bbR$. 
There exists a special mapping $\varpi:\iDFnc(\DD\msM)\rightarrow\iDFnc(\DD\msM)$ %the map
defined componentwise 
as $\varpi\rmF(\alpha)=f$. $\varpi$ is an algebra morphism. Its range $\iDFnc_\varpi(\DD\msM)$ is therefore
a subalgebra of $\iDFnc(\DD\msM)$, which by construction is %, $\iDFnc_\varpi(\DD\msM)$ is %evidently
isomorphic to $\iMap(T[1]\DD\msM,\bbR[0])$. In what follows, the functions of $\iDFnc_\varpi(\DD\msM)$ will be called short
and $\iDFnc_\varpi(\DD\msM)$ will be referred to as the short subalgebra.

%Its elements have a component structure analogous to that of their complex counterpart
% and are acted upon by the derived differential $\dd$
%(cf. eqs. \ceqref{wcrline1}, \ceqref{wcrline2}).  

By the isomorphism $\iDFnc_\varpi(\DD\msM)\simeq\iMap(T[1]\DD\msM,\bbR[0])$ noticed above
and by the further isomorphism $\iMap(T[1]\DD\msM,\bbR[0])\simeq\iMap(\DD\msM,\bbR[0])$,  
for any $\rmF\in\iDFnc_\varpi(\DD\msM)$ and $V\in\iVect(\DD\msM)$, the product $\rmF V=f\rmV\in\iVect(\DD\msM)$
is defined, rendering $\iVect(\DD\msM)$ a $\iDFnc_\varpi(\DD\msM)$--module.
This restricted module structure is responsible for %many of
the special features of the derived Poisson set--up.
%\vfil\eject % theoretic construction carried out below. 

Again, as we aim eventually to obtain a structure on $\DD\msM/\DD\msJ$, we restrict the
range of derived functions we consider to the subset
$\iDFnc_\msb(\DD\msM)$ of $\iDFnc(\DD\msM)$ of the basic ones. This is 
formed by the derived functions $\rmF\in\iDFnc(\DD\msM)$ obeying 
\begin{align}
&j_{\msR\rmX}\rmF=0,
\vphantom{\Big]}
\label{phiinv-1}
\\
&l_{\msR\rmX}\rmF=0
\vphantom{\Big]}
\label{phiinv0}
\end{align}
for $\rmX\in\DD\fkj$. \ceqref{phiinv0} answers to relation \ceqref{stkkssympl1}
of the standard theory, where a corresponding restriction is imposed.
The counterpart of \ceqref{phiinv-1} does no appear in
this latter as it is automatically fulfilled for grading reasons. 
$\iDFnc_\msb(\DD\msM)$ constitutes a subalgebra of $\iDFnc(\DD\msM)$,
as the $j_{\msR\rmX}$, $l_{\msR\rmX}$ are derivations. Since $\varpi j_{\msR\rmX}=j_{\msR\rmX}\varpi$,
$\varpi l_{\msR\rmX}=l_{\msR\rmX}\varpi$, $\iDFnc_\msb(\DD\msM)$ is invariant under $\varpi$ and consequent\-ly
$\iDFnc_{\varpi\msb}(\DD\msM)=\varpi\iDFnc_\msb(\DD\msM)$ is the short subalgebra of $\iDFnc_\msb(\DD\msM)$.
%, its short subalgebra. 

% algebra $\iDFnc_\msb(\DD\msM)$.
%We consider so the space $\iDFnc_\msb(\DD\msM)$ of basic derived functions of $\DD\msM$, 
%the set of all $\rmF\in\iDFnc(\DD\msM)$ such that $j_{\msR\rmX}\rmF=0$,

%Note that if $V\in\iVect_{\msi\rmA}(\DD\msM)$,
%$\rmF V\not\in\iVect_{\msi\rmA}(\DD\msM)$ in general. If $V\in\iVect_{\msk\rmA}(\DD\msM)$, conversely,
%$\rmF V\in\iVect_{\msk\rmA}(\DD\msM)$ too.

%\vspace{-.33mm}

Suppose that $\rmF\in\iDFnc_\msb(\DD\msM)$ and $P_\rmF\in\iVect(\DD\msM)$ obey the equation
%\vspace{.25mm}
\begin{align}
\dd\rmF-ij_{P_\rmF}\rmB=0. 
\label{phiinv3}
\end{align}
%\vspace{.25mm}
Then, as is straightforward enough to check, 
$P_\rmF\in\iVect_{\msi\rmA}(\DD\msM)$, $P_\rmF$ is determined by eq. \ceqref{phiinv3}
mod $\iVect_{\msk\rmA}(\DD\msM)$ and $l_{S_{\msR\rmX}}P_\rmF\in\iVect_{\msk\rmA}(\DD\msM)$ for all $\rmX\in\DD\fkj$. 
Thus, $P_\rmF\in\iVect_{\msb\rmA}(\DD\msM)$ and as such $P_\rmF$ is uniquely determined by $\rmF$.
If the vector field $P_\rmF$ exists, $\rmF$ is said to be a Hamiltonian derived function
and $P_\rmF$ is called the Hamiltonian vector field of $\rmF$ as
\ceqref{phiinv3} is clearly a derived extension of \ceqref{stkkssympl2}, the relation
defining Hamiltonian functions and vector fields in the standard theory. 
We denote by $\iDFnc_\rmA(\DD\msM)$ the set of the Hamiltonian functions of $\iDFnc_\msb(\DD\msM)$ and by
$\iVect_{\rmA}(\DD\msM)$ that of the Hamiltonian vector fields of $\iVect_{\msb\rmA}(\DD\msM)$. 
We denote similarly by $\iDFnc_{\varpi\rmA}(\DD\msM)$ the set of the short functions of $\iDFnc_\rmA(\DD\msM)$ 
and by $\iVect_{\varpi\rmA}(\DD\msM)$ that of the associated vector fields of $\iVect_{\rmA}(\DD\msM)$. 

%We denote similarly by $\iDFnc_{\varpi\rmA}(\DD\msM)$ the set of the Hamiltonian elements of $\iDFnc_{\varpi\msb}(\DD\msM)$
%and by $\iVect_{\varpi\rmA}(\DD\msM)$ that of the associated Hamiltonian elements of $\iVect_{\msb\rmA}(\DD\msM)$. 

$\iDFnc_\rmA(\DD\msM)$ is only a vector subspace of $\iDFnc_\msb(\DD\msM)$, while $\iDFnc_{\varpi\rmA}(\DD\msM)$ is a
subalgebra of $\iDFnc_{\varpi\msb}(\DD\msM)$. The reason for this somewhat surprising 
difference in their algebraic properties can be ultimately traced back to the fact that $\iVect(\DD\msM)$
is not a $\iDFnc(\DD\msM)$--module but only a $\iDFnc_\varpi(\DD\msM)$--module.
We shall come back to this point momentarily. \vfil\eject

It can also be seen that  $\iVect_{\rmA}(\DD\msM)$
is a Lie subalgebra of $\iVect_{\msb\rmA}(\DD\msM)$ and $\iVect_{\varpi\rmA}(\DD\msM)$ is a Lie ideal of 
$\iVect_{\rmA}(\DD\msM)$. %\vfil\eject

Mimicking a similar construction of Poisson theory, we may define a bracket
$\{\cdot,\cdot\}_\rmA:\iDFnc_\rmA(\DD\msM)\times\iDFnc_\rmA(\DD\msM)\rightarrow\iDFnc_\rmA(\DD\msM)$
as follows. It can be shown that if $\rmF,\rmG$ are Hamiltonian derived functions and $P_\rmF,P_\rmG$
are the associated Hamiltonian vector fields, then $-ij_{P_\rmG}j_{P_\rmF}\rmB$ is also a Hamiltonian
function and $[P_\rmF,P_\rmG]$ is the associated Hamiltonian vector field.
Taking relation \ceqref{stkkssympl3} of the standard theory as a model, we then define the derived Poisson bracket
%may then define
%a bracket $\{\cdot,\cdot\}_\rmA:\iDFnc_\rmA(\DD\msM)\times\iDFnc_\rmA(\DD\msM)\rightarrow\iDFnc_\rmA(\DD\msM)$ by 
\begin{align}
\{\rmF,\rmG\}_\rmA=-ij_{P_\rmG}j_{P_\rmF}\rmB.
\label{phiinv4}
\end{align}
This bracket is bilinear and antisymmetric. Unlike its ordinary counterpart, however, it fails to satisfy the
Jacobi identity. One has indeed
\begin{align}
\{\rmF,\{\rmG,\rmH\}_\rmA\}_\rmA+\{\rmG,\{\rmH,\rmF\}_\rmA\}_\rmA+\{\rmH,\{\rmF,\rmG\}_\rmA\}_\rmA
=\langle\rmF,\rmG,\rmH\rangle_\rmA
\label{phiinv5}
\end{align}
for $\rmF,\rmG,\rmH\in\iDFnc_\rmA(\DD\msM)$, 
the Jacobiator in the right hand side being given by 
\begin{align}
\langle\rmF,\rmG,\rmH\rangle_\rmA=i\dd j_{P_\rmH}j_{P_\rmG}j_{P_\rmF}\rmB. 
\label{phiinv6}
\end{align}
Note that, even though the derived Lie group $\DD\msM$ is a non negatively graded manifold
and $-i\rmB$ is a degree $2$ derived function, $\langle\rmF,\rmG,\rmH\rangle_\rmA$ 
is generally non vanishing because the degree $3$ component $B$ of $\rmB$ is not
necessarily annihilated by $j_{P_\rmH}j_{P_\rmG}j_{P_\rmF}$. By virtue of \ceqref{phiinv3} and $\dd$--exactness,
however, the Hamiltonian vector field $P_{\langle\rmF,\rmG,\rmH\rangle_\rmA}$ of 
$\langle\rmF,\rmG,\rmH\rangle_\rmA$ vanishes and so $\langle\rmF,\rmG,\rmH\rangle_\rmA$  
is central, %so that %\hphantom{xxxxxxxxxxxxxx} 
\begin{align}
\{\langle\rmF,\rmG,\rmH\rangle_\rmA,\cdot\}_\rmA=0.
\label{phiinv7}
\end{align}
%for $\rmI\in\iDFnc_\rmA(\DD\msM)$.
$\iDFnc_\rmA(\DD\msM)$ equipped with the bracket $\{\cdot,\cdot\}_\rmA$ can
so be described %characterized
as a twisted Lie algebra. The connection $\rmA$ is called simple if the Jacobiator vanishes. % identically.

It is remarkable that the twisted Lie bracket $\{\cdot,\cdot\}_\rmA$ just introduced 
restricts to a honest Poisson bracket on the short subalgebra $\iDFnc_{\varpi\rmA}(\DD\msM)\subset\iDFnc_\rmA(\DD\msM)$. 
%As we found above, $\iDFnc_{\varpi\rmA}(\DD\msM)$ unlike $\iDFnc_\rmA(\DD\msM)$ is an algebra. 
Indeed, on $\iDFnc_{\varpi\rmA}(\DD\msM)$ the bracket $\{\cdot,\cdot\}_\rmA$ turns out to be Leibniz in both arguments
and the Jacobiator $\langle\cdot,\cdot,\cdot\rangle_\rmA$ to vanish identically. 
As a Lie algebra, $\iDFnc_{\varpi\rmA}(\DD\msM)$ is a Lie ideal of $\iDFnc_\rmA(\DD\msM)$. With this we mean 
that the Lie bracket $\{\rmF,\rmG\}_\rmA$ is short when at least one of its two arguments is and that
the Jacobiator $\langle\rmF,\rmG,\rmH\rangle_\rmA$ vanishes when %ever
at least one of its three arguments is short.

The function space $\iDFnc_\rmA(\DD\msM)$ together with 
the short function algebra $\iDFnc_{\varpi\rmA}(\DD\msM)\subset\iDFnc_\rmA(\DD\msM)$
and the bracket $\{\cdot,\cdot\}_\rmA$ %which we have built above
are the basic elements of the derived Poisson structure associated with the unitary connection 
$\rmA$ of $\clL_\beta$. It possibly exemplifies a higher form of Poisson geometrical structure, though its precise
framing in Poisson theory remains to be elucidated.

%bracket $\{\cdot,\cdot\}_\rmA$ which we have built above
%are the constitutive elements of the derived Poisson structure associated with the character
%$\beta$ of $\msJ$ of and the connection $\rmA$ of the derived line bundle $\clL_\beta$. 

A remarkable property of the derived Poisson structure of $\rmA$ is its gauge invariance. 
This follows immediately from the gauge invariance of $\rmB$ itself (cf. eq. \ceqref{gggcrline2}.

Next, we express the most relevant relations obtained above in components to make the relationship
of derived to ordinary Poisson theory clearer. 
Consider a derived functions $\rmF\in\iDFnc_\msb(\DD\msM)$. The basicness conditions \ceqref{phiinv-1},
\ceqref{phiinv0} %, $j_{\msR\rmX}\rmF=0$, $l_{\msR\rmX}\rmF=0$ with $\rmX\in\DD\fkj$
read in components as
$j_{\msR x,X}f=0$, $l_{\msR x,X}f=0$,  $j_{\msR x,X}F=0$, $l_{\msR x,X}F=0$.  
Further, when $\rmF\in\iDFnc_\rmA(\DD\msM)$, the Hamiltonian relation \ceqref{phiinv3} takes the form 
\begin{align}
&df+F-ij_{P_{f,F}}b=0,
\vphantom{\Big]}
\label{phiinv8}
\\
&dF-ij_{P_{f,F}}B=0.
\vphantom{\Big]}
\label{phiinv9}
\end{align}

For $\rmF,\rmG\in\iDFnc_\rmA(\DD\msM)$, the Lie bracket $\{\rmF,\rmG\}_\rmA$ has a component expression  of the form 
$\{\rmF,\rmG\}_\rmA(\alpha)=\{f,g\}_{a,A}+\alpha\{F,G\}_{a,A}$. 
%$\{\rmF,\rmG\}_\rmA(\alpha)=\{f,g\}_{a,A}+\alpha\{f,G\}_{a,A}
%=\{f,g\}_{a,A}+\alpha\{F,g\}_{a,A}$. 
%\begin{equation}
%\{\rmF,\rmG\}_\rmA(\alpha)=\{f,g\}_{a,A}+\alpha\{f,G\}_{a,A}.
%\label{}  %phiinv10}
%\end{equation} 
The notation used here is merely suggestive and should be handled with care, %\pagebreak
since in general $\{f,g\}_{a,A}$, $\{F,G\}_{a,A}$ 
%$\{f,g\}_{a,A}$, $\{f,G\}_{a,A}$, $\{F,g\}_{a,A}$ 
both depend on $f,F$, $g,G$, as the derived expression \ceqref{phiinv4} gives 
\begin{align}
&\{f,g\}_{a,A}=-ij_{P_{g,G}}j_{P_{f,F}}b,
\vphantom{\Big]}
\label{phiinv10}
\\
&\{F,G\}_{a,A}=-ij_{P_{g,G}}j_{P_{f,F}}B.
\vphantom{\Big]}
\label{phiinv11}
\end{align}
When $\rmF\in\iDFnc_{\varpi\rmA}(\DD\msM)$, we have that $\{\rmF,\rmG\}_\rmA\in\iDFnc_{\varpi\rmA}(\DD\msM)$
and therefore $\{F,G\}_{a,A}=0$. %When $\rmF,\rmG\in\iDFnc_{\varpi\rmA}(\DD\msM)$,
The Poisson bracket $\{\rmF,\rmG\}_\rmA$ reduces in this way to the degree 0 component $\{f,g\}_{a,A}$ only.

For $\rmF,\rmG,\rmH\in\iDFnc_\rmA(\DD\msM)$, the Jacobiator $\langle\rmF,\rmG,\rmH\rangle_\rmA$ has the component structure
$\langle\rmF,\rmG,\rmH\rangle_\rmA(\alpha)=\langle f,g,h\rangle_{a,A}+\alpha\langle F,G,H\rangle_{a,A}$, 
where again the notation used is merely suggestive \pagebreak for reasons already explained. The Jacobiator components 
$\langle f,g,h\rangle_{a,A}$, $\langle F,G,H\rangle_{a,A}$ are given by 
\begin{align}
&\langle f,g,h\rangle_{a,A}=-ij_{P_{h,H}}j_{P_{g,G}}j_{P_{f,F}}B,
\vphantom{\Big]}
\label{phiinv20}
\\
&\langle F,G,H\rangle_{a,A}=idj_{P_{h,H}}j_{P_{g,G}}j_{P_{f,F}}B.
\vphantom{\Big]}
\label{phiinv21}
\end{align}
Notice that they depend only on the degree 3 components $B$ of $\rmB$. 
A term of the form $idj_{P_{h,H}}j_{P_{g,G}}j_{P_{f,F}}b$ does not appear in the right hand side of
\ceqref{phiinv20}, since it vanishes identically for grading reasons. 

%To recap, with the derived curvature $\rmB$ of an invariant connection $\rmA$ there is associated a
%twisted Lie algebra structure on a suitable vector space $\iDFnc_{\rmA}(\DD\msM)$ of Hamiltonian basic derived functions
%that subsumes a genuine Poisson structure on the subalgebra $\iDFnc_{\varpi\rmA}(\DD\msM)$ of short
%Hamiltonian functions. We call this structure, the quasi Poisson structure of $\rmB$.  

The notion of derived Poisson structure worked out above is reminiscent of that
of pre--2--symplectic Lie 2--algebra proposed by Rogers in \ccite{Rogers:2010nw}. 
%As we review next, t
Roughly, the latter is obtained from the former 
by replacing the derived gauge field and curvature
$\rmA$ and $\rmB$ by their degree 2 and 3 components $A$ and $B$, respectively.

%A related construction was originally worked out by Rogers in ref. \ccite{Rogers:2010nw}.
%It proceeds roughly as that of the derived Poisson structure detailed above
%with the derived gauge field and curvature $\rmA$ and $\rmB$ replaced by their degree 2 and 3
%components $A$ and $B$, respectively.

We consider the set $\iVect_{\msi A}(\DD\msM)$ of all vector
fields $V\in\iVect(\DD\msM)$ satisfying %\hphantom{xxxxxxxxxx}
\begin{align}
l_VB=0
\label{phiinv12}
\end{align}
and the set $\iVect_{\msk A}(\DD\msM)$ of all vector
fields $V\in\iVect(\DD\msM)$ obeying 
%leaving $\rmB$ invariant, that is satisfying \hphantom{xxxxxxxxxx}
\begin{align}
j_VB=0.
\label{phiinv13}
\end{align}
Again, $\iVect_{\msi A}(\DD\msM)$ is a Lie subalgebra of $\iVect(\DD\msM)$
and $\iVect_{\msk A}(\DD\msM)$ is a Lie ideal of $\iVect_{\msi A}(\DD\msM)$. 
It is therefore possible to construct 
the quotient Lie algebra $\iVect_{\msq A}(\DD\msM)=\iVect_{\msi A}(\DD\msM)/\iVect_{\msk A}(\DD\msM)$. 
The Lie derivatives $l_{S_{\msR\rmX}}$ with $\rmX\in\DD\fkj$ induce
derivations $l_{\msq S_{\msR\rmX}}$ of $\iVect_{\msq A}(\DD\msM)$ and hence
$\iVect_{\msb A}(\DD\msM)$ $=\bigcap_{\rmX\in\DD\fkj}\ker l_{\msq S_{\msR\rmX}}$ is 
a Lie subalgebra of $\iVect_{\msq A}(\DD\msM)$. 

The relevant function space is again the basic space $\iDFnc_\msb(\DD\msM)$
decomposed as the direct sum of its degree 0 and 1 components $\iDFnc_{\msb 0}(\DD\msM)$,
$\iDFnc_{\msb 1}(\DD\msM)$. The Hamiltonian functions are the elements of $F\in\iDFnc_{\msb 1}(\DD\msM)$
for which there exists a vector field $P_F\in\iVect_{\msb A}$, necessarily unique, such that 
\begin{equation}
dF-ij_{P_F}B=0.
\label{phiinv14}  
\end{equation}
We denote by $\iDFnc^*\!\!\!{}_{A 1}(\DD\msM)$ the subspace of $\iDFnc_{\msb 1}(\DD\msM)$
they form and by $\iDFnc^*\!\!\!{}_A(\DD\msM)=\iDFnc_{\msb 0}(\DD\msM)\oplus
\iDFnc^*\!\!\!{}_{A 1}(\DD\msM)$ the corresponding subspace of $\iDFnc_\msb(\DD\msM)$. 

We finally introduce a unary, a binary and a trinary bracket
$\{\cdot\}_A$, $\{\cdot,\cdot\}_A$, $\{\cdot,\cdot,\cdot\}_A$
on $\iDFnc^*\!\!\!{}_A(\DD\msM)$ whose only non zero instances are 
\begin{align}
&\{f\}_A=df,
\vphantom{\Big]}
\label{phiinv15}
\\
&\{F,G\}_A=-ij_{P_G}j_{P_F}B,
\vphantom{\Big]}
\label{phiinv16}
  \\
&\{F,G,H\}_A=ij_{P_H}j_{P_G}j_{P_F}B
\vphantom{\Big]}
\label{phiinv17}
\end{align}
with $f\in\iDFnc_{\msb 0}(\DD\msM)$, $F,G,H\iDFnc^*\!\!\!{}_{A 1}(\DD\msM)$. 
By means of a $-1$ degree shift placing $\iDFnc_{\msb 0}(\DD\msM)$,
$\iDFnc^*\!\!\!{}_{A 1}(\DD\msM)$ respectively in degree $-1$,
$0$, the graded vector space $\iDFnc^*\!\!\!{}_A(\DD\msM)$ equipped with the brackets
$\{\cdot\}_A$, $\{\cdot,\cdot\}_A$, $\{\cdot,\cdot,\cdot\}_A$
is a semistrict Lie 2--algebra \ccite{Baez:2003fs}. 

By \ceqref{phiinv9}, we have a mapping $\lambda:\iDFnc_{\rmA}(\DD\msM)\rightarrow
\iDFnc^*\!\!\!{}_{A 1}(\DD\msM)$ given by $\lambda(\rmF)=F$ defining a subspace %. Denote by
$\iDFnc^\lambda\!\!\!\!{}_\rmA(\DD\msM)=\iDFnc_{\msb 0}(\DD\msM)\oplus
\lambda(\iDFnc_\rmA(\DD\msM))$ of $\iDFnc^*\!\!\!{}_A(\DD\msM)$. 
$\iDFnc^\lambda\!\!\!\!{}_\rmA(\DD\msM)$ is a Lie 2--subalgebra of $\iDFnc^*\!\!\!{}_A(\DD\msM)$
\footnote{$\vphantom{\dot{\dot{\dot{a}}}}$  Note here that for $f\in\iDFnc_{\msb 0}(\DD\msM)$
one has $\{f\}_A\in\lambda(\iDFnc_\rmA(\DD\msM))$, as $df=\lambda(\rmQ_f)$ 
where $\rmQ_f\in\iDFnc_{\rmA}(\DD\msM)$ with $\rmQ_f(\alpha)=-f+\alpha df$.}. Further, 
\begin{align}
&\lambda(\{\rmF,\rmG\}_\rmA)=\{F,G\}_{a,A}=\{F,G\}_A,
\vphantom{\Big]}
\label{phiinv18}
\\
&\lambda(\langle\rmF,\rmG,\rmH\rangle_\rmA)=\{\{F,G,H\}_A\}_A.
\vphantom{\Big]}
\label{phiinv19}  
\end{align}
This clarifies the relationship between the two formulations.

%\vfil\eject  

\subsection{\textcolor{blue}{\sffamily Hamiltonian nature of the target kernel symmetry}}\label{subsec:twoplct}

As detailed in subsect. \cref{subsec:tarker}, the derived group $\DD\msM$ is characterized by the left action of 
the derived target kernel group $\DD\msM_\tau$. The left $\DD\msM_\tau$--action commutes with the right $\DD\msJ$--action.
It therefore descends on one on the regular homogeneous manifold $\DD\msM/\DD\msJ$ and, for each character $\beta$,
the derived unitary line bundle $\clL_\beta$. 
In this subsection, we examine the left target kernel symmetry from the standpoint of the derived
Poisson theory of subsect. \cref{subsec:quapoi} and show its Hamiltonian nature. Our analysis
is patterned again on the corresponding one of the standard formulation,
reviewed in subsect. \cref{subsec:stkkssympl}, where the left symmetry can be similarly
shown to be Hamiltonian. 

At the infinitesimal level, the left $\DD\msM_\tau$--action is codified by the associated left 
$\DD\fkm_\tau$--operation. % (cf. subsect. \cref{subsec:tarker}).
%A $p$--form section $\rmS$ of
%$\clL_\beta$ is left invariant if 
%\begin{equation}
%l_{\msL\rmH}\rmS=0. 
%\label{tkcrline1}
%\end{equation}
%for $\rmH\in\DD\fkm_\tau$. Similarly, a
A derived unitary connection $\rmA$ of $\clL_\beta$ is left invariant if 
\begin{equation}
l_{\msL\rmH}\rmA=0.
\label{tkcrline2}
\end{equation}
This property implies the left invariance of the associated curvature $\rmB$, 
\begin{equation}
l_{\msL\rmH}\rmB=0
\label{tkcrline3}
\end{equation}
on account of \ceqref{ncrline1}. \ceqref{tkcrline2}, \ceqref{tkcrline3} answer to
and have the same meaning as that of the
invariance relations \ceqref{stkkssympl4}, \ceqref{stkkssympl5} of the standard theory. 

A derived gauge transformation $\rmU$ of $\clL_\beta$ is invariant if 
\begin{equation}
l_{\msL\rmH}\rmU\rmU^{-1}=0
\label{tkcrline4}
\end{equation}
%Clearly, if $\rmS$ is an invariant $p$--form section and $\rmU$ is an invariant gauge transformation,
%then ${}^\rmU\rmS$ is also an invariant $p$--form section. Similarly, if
just as in  the standard theory, see eq. \ceqref{stkkssympl9}. 
If $\rmA$ is an invariant connection
and $\rmU$ is an invariant gauge transformation, then ${}^\rmU\rmA$ is also an invariant connection. 

%are just the vector fields $j_{S_{\msL\rmH}}$, $l_{S_{\msL\rmH}}$
%where $S_{\msL\rmH}$ is the vertical vector field of the $\DD\msM_\tau$--action corresponding to $\rmH$.

%We resume from now on our earlier notation for the $\DD\msM_\tau$--action on $\DD\msM$ 
%according to which $j_{S_{\msL\rmH}}=j_{\msL\rmH}$, $l_{S_{\msL\rmH}}=l_{\rmX}$ 
%for $\rmH\in\DD\fkm_\tau$, 

%We now equip the derived line bundle $\clL_\beta$ with an invariant connection $\rmA$.
%Then, the curvature $\rmB$ of $\rmA$ is invariant as well

As the curvature $\rmB$ of an invariant unitary connection
$\rmA$ is also invariant, the derived Poisson structure associated with $\rmA$ is expected to have
special properties with regard to the $\DD\msM_\tau$ target kernel symmetry. In particular,
the natural question arises about whether the $\DD\msM_\tau$--action is Hamiltonian.
It turns out that it is, as we show next.
We recall here that 
the derivations $j_{\msL\rmH}$, $l_{\msL\rmH}$ with $\rmH\in\DD\fkm_\tau$ are just the Cartan calculus derivations
$j_{S_{\msL\rmH}}$, $l_{S_{\msL\rmH}}$ considered in subsect. \cref{subsec:quapoi}, where 
$S_{\msL\rmH}$ is the vertical vector field of the left $\DD\msM_\tau$--action corresponding to $\rmH$.

The expression of the moment map of the left $\DD\msM_\tau$--action is analogous to that of the moment
map of the left $\msG$--action in the standard theory in eq. \ceqref{stkkssympl6}.
It is the degree 0 derived function valued map %mapping
$\rmQ_\rmA:\DD\fkm_\tau\rightarrow\iDFnc_\msb(\DD\msM)$ given by %
%The degree 0 moment map of the left $\DD\msM_\tau$--action can be explicitly  %precisely %explicitly written
%put down. Its expression is analogous to that of the moment map of the left $\msG$--action
%of the standard theory given in \ceqref{stkkssympl6}. 
%It is the derived function valued map %mapping
%$\rmQ_\rmA:\DD\fkm_\tau\rightarrow\iDFnc_\msb(\DD\msM)$ given by %
\begin{equation}
\rmQ_\rmA(\rmH)=-ij_{\msL\rmH}\rmA
\label{tkcrline5}
\end{equation}
with $\rmH\in\DD\fkm_\tau$. By \ceqref{crline5/2}, \ceqref{crline7/2}
$\rmQ_\rmA(\rmH)$ satisfies the basicness conditions \ceqref{phiinv-1}, \ceqref{phiinv0}. Thus, 
$\rmQ_\rmA(\rmH)\in\iDFnc_\msb(\DD\msM)$ as indicated and so $\rmQ_\rmA(\rmH)$ defines a derived function 
on $\DD\msM/\DD\msJ$ as required.
Further, by the invariance of $\rmA$, $\rmQ_\rmA$ obeys the moment map equation 
\begin{equation}
\dd\rmQ_\rmA(\rmH)-ij_{\msL\rmH}\rmB=0.
\label{tkcrline6}
\end{equation}
$\rmQ_\rmA(\rmH)$ is therefore a Hamiltonian derived function 
with Hamiltonian vector field $P_{\rmQ_\rmA(\rmH)}=S_{\msL\rmH}$.
%, so that $\rmQ_\rmA(\rmH)\in\iDFnc_\rmA(\DD\msM)$.
%in the derived Poisson structure of $\rmA$.

%$j_{\msR\rmX}\rmQ_\rmA(H)=0$, $l_{\msR\rmX}\rmQ_\rmA(H)=0$ for $\rmX\in\DD\fkj$

$\rmQ_\rmA$ obeys the derived equivariance identity 
\begin{equation}
l_{\msL\rmH}\rmQ_\rmA(\rmK)-\rmQ_\rmA([\rmH,\rmK])=0
\label{tkcrline7}
\end{equation}
for $\rmH,\rmK\in\DD\fkm_\tau$. \ceqref{tkcrline7} formally reproduces 
the corresponding relation of the standard theory, eq. \ceqref{stkkssympl8}, 
only in part. In fact, in the present derived Poisson set--up, 
$l_{\msL\rmH}\rmQ_\rmA(\rmK)\neq\{\rmQ_\rmA(\rmH),\rmQ_\rmA(\rmK)\}_\rmA$, 
in general in contrast to ordinary Poisson theory. The defect map
$\rmC_\rmA:\DD\msM_\tau\times\DD\msM_\tau\rightarrow\iDFnc_\msb(\DD\msM)$ which 
measures the failure for such an equality to hold is given by \hphantom{xxxxxxxxx}
\begin{equation}
\rmC_\rmA(\rmH,\rmK)=-i\dd j_{\msL\rmK}j_{\msL\rmH}\rmA
\label{tkcrline18}
\end{equation}
with $\rmH,\rmK\in\DD\fkm_\tau$. From \ceqref{crline5/2}, \ceqref{crline7/2} again, 
$\rmC_\rmA(\rmH,\rmK)$ satisfies the basicness requirements \ceqref{phiinv-1}, \ceqref{phiinv0} 
and therefore $\rmC_\rmA(\rmH,\rmK)\in\iDFnc_\msb(\DD\msM)$ defining a function on $\DD\msM/\DD\msJ$.
The defect relation 
\begin{equation}
l_{\msL\rmH}\rmQ_\rmA(\rmK)=\{\rmQ_\rmA(\rmH),\rmQ_\rmA(\rmK)\}_\rmA-\rmC_\rmA(\rmH,\rmK)
%\rmQ_\rmA([\rmH,\rmK])
\label{tkcrline17}
\end{equation}
holds by construction. 
Through \ceqref{tkcrline17}, we can reexpress relation \ceqref{tkcrline7} suggestively in the form  
\begin{equation}
\{\rmQ_\rmA(\rmH),\rmQ_\rmA(\rmK)\}_\rmA-\rmQ_\rmA([\rmH,\rmK])=\rmC_\rmA(\rmH,\rmK).
\label{tkcrline20}
\end{equation}
Analogously to the Poisson Jacobiator encountered  in subsect. \cref{subsec:quapoi}, $\rmC_\rmA(\rmH,\rmK)$
is generally non vanishing because the degree $2$ component $A$ of $\rmA$ is not
necessarily annihilated by $j_{\msL\rmK}j_{\msL\rmH}$. By \ceqref{phiinv3} and $\dd$--exactness,
however, $\rmC_\rmA(\rmH,\rmK)$ is a Hamiltonian derived function and its
Hamiltonian vector field $P_{\rmC_\rmA(\rmH,\rmK)}$ vanishes. Hence, $\rmC_\rmA(\rmH,\rmK)$ is central, \hphantom{xxxxxxx}
%the Hamiltonian vector field $P_{\rmC_\rmA(\rmH,\rmK)}$ of 
%$\rmC_\rmA(\rmH,\rmK)$ vanishes and so $\rmC_\rmA(\rmH,\rmK)$ is central, 
\begin{align}
\{\rmC_\rmA(\rmH,\rmK),\cdot\}_\rmA=0.
\label{tkcrline19}
\end{align}
An invariant connection $\rmA$ is called strict if $\rmC_\rmA=0$. % for all for $\rmH,\rmK$.
For a strict connection, the equivariance relation \ceqref{tkcrline20} takes a more familiar
form analogous to that relation \ceqref{stkkssympl8} does in ordinary Poisson theory. 

%$j_\rmX\rmC_\rmA(H,K)=0$, $l_\rmX\rmC_\rmA(H,K)=0$ for $\rmX\in\DD\fkj$ 

By \ceqref{gggcrline1} and \ceqref{tkcrline5}, \pagebreak under an invariant gauge transformation $\rmU$, the moment map
$\rmQ_\rmA$ varies by a $\dd$--exact term
\begin{equation}
{}^\rmU\rmQ_\rmA(\rmH)=\rmQ_{{}^\rmU\rmA}(\rmH)=\rmQ_\rmA(\rmH)-i\dd(j_{\msL\rmH}\rmU\rmU^{-1}),
\label{tkcrline8}
\end{equation}
thanks to the invariance relation \ceqref{tkcrline4}. 
It is immediately checked that the moment map properties \ceqref{tkcrline6}, \ceqref{tkcrline7} 
are compatible with the gauge transformation action \ceqref{tkcrline8}.

It is interesting to examine the form the above relations take in components.
By the defining relation \ceqref{tkcrline5}, the moment map components read as 
\begin{align}
&q_{a,A}(h,H)=-ij_{\msL h,H}a,
\vphantom{\Big]}
\label{tkcrline9}
\\
&Q_{a,A}(h,H)=-ij_{\msL h,H}A.
\vphantom{\Big]}
\label{tkcrline10}
\end{align} 
The moment map equation \ceqref{tkcrline6} yields the pair of equations 
\begin{align}
&dq_{a,A}(h,H)+Q_{a,A}(h,H)-ij_{\msL h,H}b=0,
\vphantom{\Big]}
\label{tkcrline11}
\\
&dQ_{a,A}(h,H)-ij_{\msL h,H}B=0.
\vphantom{\Big]}
\label{tkcrline12}
\end{align} 

By \ceqref{tkcrline18}, the defect map components are given by 
\begin{align}
&c_{a,A}(h,H,k,K)=ij_{\msL k,K}j_{\msL h,H}A,
\vphantom{\Big]}
\label{tkcrline28}
\\
&c_{a,A}(h,H,k,K)=-idj_{\msL k,K}j_{\msL h,H}A.
\vphantom{\Big]}
\label{tkcrline29}
\end{align}
Notice that they depend only on the degree 2 components $A$ of $\rmA$. 
A term of the form $-idj_{\msL k,K}j_{\msL h,H}a$ does not appear in the right hand side of
\ceqref{tkcrline27}, since it vanishes identically for grading reasons. 
The defect equivariance relation \ceqref{tkcrline20} assume componentwise the shape
\begin{align}
&\{q_{a,A}(h,H),q_{a,A}(k,K)\}_{a,A}
-q_{a,A}([h,k],\sdot\mu\sdot(h,K)-\sdot\mu\sdot(k,H))
\vphantom{\Big]}
\label{tkcrline13}
\\
&\hspace{9cm}=c_{a,A}(h,H,k,K),
\vphantom{\Big]}
\nonumber
\\
&\{Q_{a,A}(h,H),Q_{a,A}(k,K)\}_{a,A}
-Q_{a,A}([h,k],\sdot\mu\sdot(h,K)-\sdot\mu\sdot(k,H))
\vphantom{\Big]}
\label{tkcrline14}
\\
&\hspace{9cm}=C_{a,A}(h,H,k,K), 
\vphantom{\Big]}
\nonumber
\end{align} 
where the derived Poisson bracket components are defined as in eqs. \ceqref{phiinv10}, \ceqref{phiinv11}.
%$\vphantom{\ul{\ul{\ul{g}}}}$\vfil\eject

From \ceqref{tkcrline8}, the gauge variation of the moment map components are 
\begin{align}
&{}^{u,U}q_{a,A}(h,H)=q_{a,A}(h,H)-id(j_{\msL h,H}uu^{-1})+ij_{\msL h,H}U,
\vphantom{\Big]}
\label{tkcrline15}
\\
&{}^{u,U}Q_{a,A}(h,H)=Q_{a,A}(h,H)-idj_{\msL h,H}U. 
\vphantom{\Big]}
\label{tkcrline16}
\end{align} 

%The above properties characterize $\rmQ$ as a homotopy moment map 
%for the $\DD\msM_\tau$--action with respect the pre--2--plectic structure
%$B$ \ccite{Callies:2013jbu}.

%All the above relations can be expressed also in terms of the components $s$, $S$ of a %n invariant
%$p$--form section $\rm S$, $a$, $A$ and $b$, $B$ of a %n invariant
%a connection $\rmA$ and its curvature $\rmB$ and $u$, $U$ of a gauge transformation $\rmU$  
%and the components $h$, $H$ of the derived Lie algebra parameter $\rmH\in\DD\msM_\tau$. %\pagebreak 
%Being simple invariance relations, this is a straightforward task, which we leave to reader. 

The above notion of derived moment map %illustrated above bears an apparent formal resemblance with that
is related to that of homotopy moment map worked out by Callies {\it et al.} \ccite{Callies:2013jbu}. 
The relationship between the two %forms of moment maps
can be understood in terms of that
intercurring between the derived Poisson and pre--2--symplectic structures described at the end of 
subsect. \cref{subsec:quapoi}. Again, as we outline next, the latter is obtained from the former roughly
by replacing the derived gauge field and curvature
$\rmA$ and $\rmB$ by their degree 2 and 3 components $A$ and $B$, respectively.

Similarly to the derived case, we assume that the degree 2 component $A$ of the relevant connection
$\rmA$ is invariant so that 
\begin{equation}
l_{\msL h,H}A=0
\label{tkcrline21}
\end{equation}
for $h\in\fkg$, $H\in\ker\dot\tau[1]$. 
As a consequence, by \ceqref{crline18},
also the degree 3 component $B$ of the associated curvature $\rmB$ is 
\begin{equation}
l_{\msL h,H}B=0.
\label{tkcrline22}
\end{equation}

The homotopy moment map $Q_A:\DD\fkm_\tau\rightarrow\iDFnc_{\msb 1}(\DD\msM)$ is defined as
\begin{equation}
Q_A(h,H)=-ij_{\msL h,H}A
\label{tkcrline23}
\end{equation}
with $h\in\fkg$, $H\in\ker\dot\tau[1]$ analogously to \cref{tkcrline5}. % as \hphantom{xxxxxxxxxx}
The basicness of $Q_A(h,H)$ is checked using relations
\ceqref{crline12}, \ceqref{crline14}. $Q_A(h,H)$ is %turns out to be
Hamiltonian with Hamiltonian vector field $S_{\msL h,H}$ in the pre--2--plectic
Lie 2--algebra structure of $\rmA$ as
\begin{equation}
dQ_A(h,H)-ij_{\msL h,H}B=0,
\label{tkcrline24}
\end{equation}
so that $Q_A(h,H)$ is a Hamiltonian function.  
%$Q_A(h,H)\in\iDFnc^*\!\!\!{}_A(\DD\msM)$. 

The defect map $R_A:\DD\msM_\tau\times\DD\msM_\tau\rightarrow\iDFnc_{\msb 0}(\DD\msM)$
associated with the mo\-ment map $Q_A$ is given by 
\begin{equation}
R_A(h,H,k,K)=-ij_{\msL k,K}j_{\msL h,H}A.
\label{tkcrline25}
\end{equation}
%\vfil\eject\noindent
The basicness of $R_A(h,H,k,K)$ is checked easily using again relations
\ceqref{crline12}, \ceqref{crline14}. The moment map obeys further the relations 
\begin{align}
&\{Q_A(h,H),Q_A(k,K)\}_A
-Q_A([h,k],\sdot\mu\sdot(h,K)-\sdot\mu\sdot(k,H))\vphantom{\Big]}
\label{tkcrline26}
\\
&\hspace{8cm}=dR_A(h,H,k,K),
\vphantom{\Big]}
\nonumber
\\
&\{Q_A(h,H),Q_A(k,K),Q_A(l,L)\}_A
=R_A(h,H,[k,l],\sdot\mu\sdot(k,L)-\sdot\mu\sdot(l,K))
\vphantom{\Big]}
\label{tkcrline27}
\\
&\hspace{.5cm}+R_A(k,K,[l,h],\sdot\mu\sdot(l,H)-\sdot\mu\sdot(h,L))
+R_A(l,L,[h,k],\sdot\mu\sdot(h,K)-\sdot\mu\sdot(k,H)).
\vphantom{\Big]}
\nonumber
\end{align}
Relations \ceqref{tkcrline26}, \ceqref{tkcrline27} characterize the homotopy moment map $Q_A$
in the pre--2--plectic Lie 2--algebra set--up \ccite{Callies:2013jbu}. 

In subsect. \cref{subsec:quapoi}, we exhibited a mapping
$\lambda:\iDFnc_{\rmA}(\DD\msM)\rightarrow \iDFnc^*\!\!\!{}_{A 1}(\DD\msM)$
connecting the Hamiltonian derived  and pre--2--plectic function spaces preserving the bracket structure
in the appropriate sense (cf eqs. \ceqref{phiinv18}, \ceqref{phiinv19}).
$\lambda$ also relates the derived and homotopy moment map data in a consistent manner,
\begin{align}
&\lambda(\rmQ_\rmA(\rmH))=Q_{a,A}(h,H)=Q_A(h,H),
\vphantom{\Big]}
\label{tkcrline30}
\\
&\lambda(C_\rmA(\rmH,\rmK))=\{R_A(h,H,k,K)\}_A.
\vphantom{\Big]}
\label{tkcrline31}  
\end{align}
This clarifies again the relationship between our constructions and those of ref. \ccite{Callies:2013jbu} .

%\vfil\eject

\subsection{\textcolor{blue}{\sffamily Derived prequantization}}\label{subsec:prequcrkks}

In derived KKS theory, prequantization is implemented along lines analogous to those of ordinary
KKS prequantization as reviewed in subsect. \cref{subsec:prequstkks}. 
Derived prequantization is however more involved than its ordinary counterpart
for a number of reasons. In this subsection, we illustrate its construction
and the problems which affect it. 

The foundation on which derived prequantization rests is the derived Poisson structure of the regular derived homogeneous
space $\DD\msM/\DD\msJ$ associated with a unitary connection $\rmA$ of the derived line bundle $\clL_\beta$
of a character $\beta$ of $\msJ$ introduced and studied in subsect. \cref{subsec:quapoi}.
This however is not a genuine Poisson structure. The derived Poisson bracket $\{\cdot,\cdot\}_\rmA$ has a
generally non trivial Jacobiator and the Hamiltonian derived 
function space $\iDFnc_\rmA(\DD\msM)$ is not an algebra. It is therefore necessary to impose a suitable 
condition on $\rmA$ and to restrict the range of Hamiltonian functions to an appropriate subspace
$\iDFnc_{\rmA\msh}(\DD\msM)$ of the space $\iDFnc_\rmA(\DD\msM)$. Upon doing so, 
$\{\cdot,\cdot\}_\rmA$ becomes a genuine Lie bracket as required by prequantization. It can then be
shown that with any function $\rmF\in\iDFnc_{\rmA\msh}(\DD\msM)$ there is associated an endomorphism $\widehat{F}$
of a certain subspace $\DD\Omega^0{}_\msh(\clL_\beta)$ of the space $\DD\Omega^0(\clL_\beta)$
of 0--form sections of $\clL_\beta$ such that for $\rmF,\rmG\in\iDFnc_{\rmA\msh}(\DD\msM)$, one has  
$[\widehat{\rmF},\widehat{\rmG}]=i\widehat{\{\rmF,\rmG\}}_\rmA$. $\clL_\beta$ gets so interpreted
as the derived prequantum line bundle. 

In this way, a derived KKS prequantization is defined associating an operator with any suitably restricted
Hamiltonian derived function such that the resulting operator commutator structure is fully compatible
with the derived Poisson bracket structure. However, there apparently is no natural
derived prequantum Hilbert space structure on $\DD\msM/\DD\msJ$ with respect to
which the operators yielded by derived prequantization are formally Hermitian. This seems
to be in line with the findings of several higher prequantization schemes available in the literature, see in
particular refs. \ccite{Hawkins:2006jbu,Fiorenza:2013kqa,Fiorenza:2013lpq}. %,Fiorenza:2013kqa,Fiorenza:2013frs}.

Derived prequantization requires the underlying derived presymplectic structure to be
symplectic as in the ordinary setting. We thus assume that the curvature $\rmB$
of the connection $\rmA$ is non singular. In this way, if $V\in\iVect(\DD\msM)$ and $j_V\rmB=0$, then
$V=S_{\msR\rmX}$ for some $\rmX\in\DD\fkj$ pointwise on $\DD\msM$.

For any Hamiltonian derived function $\rmF\in\iDFnc_\rmA(\DD\msM)$ and derived 
0--form section $\rmS\in\DD\Omega^0(\clL_\beta)$, we set \hphantom{xxxxxxx}
\begin{equation}
\widetilde{\rmF}\rmS=ij_{P_\rmF}\dd_\rmA\rmS+\rmF\rmS, 
\label{prequcrkks1}
\end{equation}
where $P_\rmF$ is the Hamiltonian vector field of $\rmF$ (cf. eq. \ceqref{phiinv3}).  
It can be readily verified that $\widetilde{\rmF}\rmS\in\DD\Omega^0(\clL_\beta)$
using \ceqref{crline5/1}, \ceqref{crline7/1} and \ceqref{phiinv-1}, \ceqref{phiinv0}.
$\widetilde{\rmF}$ is clearly an endomorphism of the vector space $\DD\Omega^0(\clL_\beta)$.
%Hence, $\widetilde{\rmF}\in\End(\DD\Omega^0(\clL_\beta))$.
By the structural analogy of eq. \ceqref{prequcrkks1} to relation \ceqref{prequstkks1} 
providing the prequantization of Hamiltonian functions in ordinary KKS theory, 
$\widetilde{\rmF}$ can naively be thought of as the derived prequantization of $\rmF$.
Unfortunately,$\vphantom{\ul{\ul{\ul{\ul{\ul{g}}}}}}$
we have $[\widetilde{\rmF},\widetilde{\rmG}]\neq i\widetilde{\{\rmF,\rmG\}}_\rmA$
for $\rmF,\rmG\in\iDFnc_\rmA(\DD\msM)$ in general
\footnote{$\vphantom{\dot{\dot{\dot{a}}}}$ We have in fact that
\begin{align}
&[\widetilde{\rmF},\widetilde{\rmG}]=i\widetilde{\{\rmF,\rmG\}}_\rmA-\dd_\rmA j_{P_\rmG}j_{P_\rmF}\dd_\rmA
\nonumber
\\
&\qquad\qquad\qquad
+((j_{P_\rmG}j_{P_\rmF}\rmA)+i(j_{P_\rmF}\rmG)-i(j_{P_\rmG}\rmF))\dd_\rmA+(j_{P_\rmG}\rmB)j_{P_\rmF}-(j_{P_\rmF}\rmB)j_{P_\rmG}. 
\end{align}
The terms in the right hand side beyond the first one do not vanish in general. 
A relation analogous to the above one holds also in ordinary KKS theory, where however 
the extra terms can be shown to vanish identically for grading reasons.}.
Our simpleminded derived %\pagebreak 
prequantization approach requires therefore appropriate modifications in order to be viable. 

We assume in what follows that the connection $\rmA$ of $\clL_\beta$ of 1--form type, that is
with the property that $j_Vj_W\rmA=0$ for any two vector fields $V,W\in\iVect(\DD\msM)$. %In such a case, t
The curvature $\rmB$ of $\rmA$ is then of 2--form type, i.e. one has $j_Vj_Wj_Z\rmB=0$ for 
any $V,W,Z\in\iVect(\DD\msM)$. Furthermore, $\rmA$ is both simple and strict (cf. subsects.
\cref{subsec:quapoi}, \cref{subsec:twoplct}). 

A derived function $\rmF\in\iDFnc_\msb(\DD\msM)$ is said to be pure if
$j_V\rmF=0$ for any vector field $V\in\iVect(\DD\msM)$. The pure derived functions 
constitute a subalgebra $\iDFnc_{\msb\msh}(\DD\msM)$ of $\iDFnc_\msb(\DD\msM)$. Similarly,  
a 0--form section $\rmS\in\DD\Omega^0(\clL_\beta)$ is said to be pure if
$j_V\rmS=0$ for any vector field $V\in\iVect(\DD\msM)$. The pure 0--from sections 
form a subspace $\DD\Omega^0{}_\msh(\clL_\beta)$ of $\DD\Omega^0(\clL_\beta)$.

Let $\iDFnc_{\rmA\msh}(\DD\msM)=\iDFnc_\rmA(\DD\msM)\cap\iDFnc_{\msb\msh}(\DD\msM)$ be the space
of pure Ham\-iltonian derived functions. For any function pair $\rmF,\rmG\in\iDFnc_{\rmA\msh}(\DD\msM)$, one has
$\{\rmF,\rmG\}_\rmA\in\iDFnc_{\rmA\msh}(\DD\msM)$. Further, the restriction of
the twisted Lie bracket $\{\cdot,\cdot\}_\rmA$
to $\iDFnc_{\rmA\msh}(\DD\msM)$ is a genuine Lie bracket, since the connection $\rmA$ is simple and so
the Jacobiator $\langle\cdot,\cdot,\cdot\rangle_\rmA$ of $\{\cdot,\cdot\}_\rmA$ vanishes identically
(cf. eqs. \ceqref{phiinv5}, \ceqref{phiinv6}).
Note that, like $\iDFnc_\rmA(\DD\msM)$, $\iDFnc_{\rmA\msh}(\DD\msM)$ depends on the connection
$\rmA$ and is a vector space but not an algebra and so $\{\cdot,\cdot\}_\rmA$ is a Lie but not a Poisson bracket. 

It can be readily shown %straightforwardly demonstrated
that $\widetilde{\rmF}\rmS\in\DD\Omega^0{}_\msh(\clL_\beta)$
for any $\rmF\in\iDFnc_{\rmA\msh}(\DD\msM)$ and $\rmS\in\DD\Omega^0{}_\msh(\clL_\beta)$ 
as a consequence of $\rmA$ being of 1--form type. 
Let $\widehat{\rmF}$ be the restriction of $\widetilde{\rmF}$ to $\DD\Omega^0{}_\msh(\clL_\beta)$. 
Then, $\widehat{\rmF}$ is an endomorphism of the vector space $\DD\Omega^0{}_\msh(\clL_\beta)$.
Furthermore, the commutation relation 
\begin{equation}
[\widehat{\rmF},\widehat{\rmG}]=i\widehat{\{\rmF,\rmG\}}_\rmA\vphantom{\bigg]}
\label{prequcrkks2}
\end{equation}
holds for any two functions $\rmF,\rmG\in\iDFnc_{\rmA\msh}(\DD\msM)$. Note that 
\ceqref{prequcrkks2} is consistent with the Jacobi property
of the endomorphism commutator because $\{\cdot,\cdot\}_\rmA$
is a Lie bracket on $\iDFnc_{\rmA\msh}(\DD\msM)$. \ceqref{prequcrkks2}
reproduces the commutation relation \ceqref{prequstkks2} of the ordinary theory. 

The target kernel symmetry moment maps are pure Hamiltonian derived functions.
By \ceqref{tkcrline5} %, \ceqref{tkcrline6}
and the 1--form property of the connection $\rmA$ it is indeed evident that
$\rmQ_\rmA(\rmH)\in\iDFnc_{\msb\msh}(\DD\msM)$ for $\rmH\in\DD\fkm_\tau$.
Since $\rmQ_\rmA(\rmH)\in\iDFnc_\rmA(\DD\msM)$, $\rmQ_\rmA(\rmH)\in\iDFnc_{\rmA\msh}(\DD\msM)$ 
and so the operators $\widehat{\rmQ}_\rmA(\rmH)$
is defined. From \ceqref{tkcrline5}, \ceqref{tkcrline6} and \ceqref{prequcrkks1},
its expression is very simple
\begin{equation}
\widehat{\rmQ}_\rmA(\rmH)=il_{\msL\rmH}.
\label{prequcrkks3}
\end{equation}
The commutation relation \ceqref{prequcrkks2} takes for the $\widehat{\rmQ}_\rmA(\rmH)$
the expected form 
\begin{equation}
[\widehat{\rmQ}_\rmA(\rmH),\widehat{\rmQ}_\rmA(\rmK)]=i\widehat{\rmQ}_\rmA([\rmH,\rmK]).
\label{prequcrkks4}
\end{equation}
with $\rmH,\rmK\in\DD\fkm_\tau$. \ceqref{prequcrkks3}, \ceqref{prequcrkks4}
replicate in derived KKS theory the ordinary KKS relations 
\ceqref{prequstkks3}, \ceqref{prequstkks4}. 

We conclude this subsection explaining why no natural derived prequantum Hilbert
space structure on $\DD\msM/\DD\msJ$ can be defined in derived KKS theory
with respect to which the operators yielded by derived prequantization are formally Hermitian.
Duplicating the standard expression of the prequantum Hilbert inner product of ordinary
KKS theory given in eq. \ceqref{prequstkks5} in our derived setting furnishes
the following tentative expression for the derived prequantum Hilbert inner product
of two sections  $\rmS,\rmT\in\DD\Omega^0{}_\msh(\clL_\beta)$:
\begin{equation}
\langle\rmS,\rmT\rangle=\frac{1}{n!}\int_{T[1](\DD\msM/\DD\msJ)}\varrho_{\DD\msM/\DD\msJ}(-iB)^n \rmS^*\rmT
\qquad \text{(tentative)}.
\label{prequcrkks5}
\end{equation}
There are a minor and a major problem with this formula: the existence of an appropriate value
of the half dimension $n$ and the well--definedness of integration, respectively.
If $\msM=(\msE,\msG)$ and $\msJ=(\msH,\msT)$, the degree of the Berezinian $\varrho_{\DD\msM/\DD\msJ}$ is
$d=-\dim\msG+\dim\msT+\dim\msE-\dim\msH$. To have a degree 0 integrand, we should so set $n=-d/2$,
as $\rmB$ has degree 2. However, unlike ordinary KKS theory,
$-d/2$ is not guaranteed to be a positive integer as far as we can see. Further,
it is known that on a non negatively graded manifold of degree 1 such as $\DD\msM/\DD\msJ$,
integration is defined and convergent only if the integrand is a Dirac type distribution
in the coordinates of positive even degree (in supergeometry forms of this kind
are called integral \ccite{bernstein:1977ifs}). Unfortunately, the integrand
in the right hand side of eq. \ceqref{prequcrkks5} does not have
this property. In conclusion, the prequantum Hilbert structure
of ordinary KKS theory does not have a viable derived analog.

%\vfil\eject 

\subsection{\textcolor{blue}{\sffamily Paths to derived quantization}}\label{subsec:derqu}

The analysis of subsect. \cref{subsec:prequcrkks} indicates that conventionally designed
derived KKS prequantization falls short of meeting its prerequisites in spite of reproducing
some of the basic features of ordinary KKS prequantization.
The non existence of a derived prequantum Hilbert structure precludes the successful
completion of the geometric quantization program of derived KKS theory
as ordinarily envisaged.
At this point, it is therefore important to explore the range of 
quantization options at our disposal. 

In the incomplete prequantization framework of subsect. \cref{subsec:prequcrkks}, we assumed that
the underlying derived connection $\rmA$ was of 1--form type and we restricted to the space of pure Hamiltonian
derived functions $\iDFnc_{\rmA\msh}(\DD\msM)$ in order to dispose of the Jacobiator 
$\langle\cdot,\cdot,\cdot\rangle_\rmA$ and so deal with a genuine Lie algebra structure
$\{\cdot,\cdot\}_\rmA$, a minimal requirement for any operator/wave--function type quantization scheme such as
geometric quantization. If we do not impose such restriction on the connection and the Hamiltonian functions,
we face a twisted Lie algebra structure $\{\cdot,\cdot\}_\rmA$. Multiple advancements in quantum mechanics
and string theory have revealed that twisted Poisson structure featuring non trivial Jacobiators are physically relevant
and that their quantization must be tackled. (See \ccite{Szabo:2019hhg} for a updated review and extensive referencing.)
Since operator/wave--function quantization is not viable in this case, alternative quantization schemes
must be sought.

To the best of our knowledge, there exist three main approaches to quantisation of a twisted Poisson manifold.
The first approach, developed in \ccite{Mylonas:2012pg}, is an extension of the standard
framework of deformation quantisation. The basic idea informing it is that the non trivial
Poisson Jacobiator leads to a non associative star product algebra, which is explicitly constructed.
The method suffers the usual problem of being based on purely formal power series in $\hbar$. 
The second approach, originally proposed is \ccite{Kupriyanov:2018xji},
is an adaptation of the classic technique of symplectic realisation of Poisson theory 
and consists in embedding the relevant twisted Poisson manifold into a symplectic
manifold of twice the dimension. In this way, standard operator/wave--function based geometric
quantisation is again possible. Ineliminable spurious extra variables with no evident
interpretation must however be introduced in order to absorb the Jacobiator into the
higher dimensional symplectic structure. In \ccite{Bunk:2018qvk}, a third approach was suggested,
which can be described as higher geometric quantization. Its basic principle is replacing the usual quantum
line bundle and the connection with curvature equal to the symplectic form by a quantum bundle gerbe
and a connection with curvature equal to the Jacobiator form. Correspondingly, 
the customary quantum Hilbert space of sections of the line bundle and its operator algebra
get substituted by a 2--Hilbert space of sections of the gerbe, a categorified Hilbert space structured
as a monoidal category, and its endofunctor category. 
Although this approach provides a higher quantum Hilbert space formulation naturally incorporating
non associativity, its interpretation in more conventional physical terms remains to be elucidated.

The natural question arises about to what extent one or the other of the three approaches to
quantization of twisted Poisson manifolds outlined in the previous paragraph 
can be utilized for the quantization of the derived Poisson manifold
$\DD\msM/\DD\msJ$. An exhaustive answer to it would require an in--depth analysis
that lies beyond the scope of the present work. Here, we shall limit ourselves
to the following considerations. 

The three methods have been formulated for ordinary manifolds. $\DD\msM/\DD\msJ$ is instead a 
genuinely graded manifold. Hence, derived KKS quantization would require
a graded geometric extension of such approaches, which to the best of our knowledge is presently lacking.
As the findings of subsect. \cref{subsec:prequcrkks} show, such extension might not
be straightforward or even possible.

The Hamiltonian derived function space $\iDFnc_\rmA(\DD\msM)$ does not contain only functions
meant as 0--forms on $\DD\msM/\DD\msJ$ but also 1--forms. In fact, it is a subspace of 
$\iDFnc(\DD\msM)=\iMap(T[1]\DD\msM,\DD\bbR[0])$ consisting of inhomogeneous forms of form
degree up to $1$ obeying the basicness conditions \ceqref{phiinv-1}, \ceqref{phiinv0}
and the Ham\-iltonian condition \ceqref{phiinv3}. Furthermore, it is not an algebra, but only a vector space.  
This makes the first two approaches apparently hardly workable leaving utilizing the third as an open option.

On the whole, in fact, of the three approaches the third one appears to be the most likely adaptable
to the derived KKS framework. This indeed involves in an essential way derived line bundles with connection.
Geometric objects of this kind, as observed at the end of subsect. \cref{subsec:crline}, 
are likely related to Roger's twisted line bundles of bundle gerbes \ccite{Rogers:2011zc}.
We shall not however attempt to follow this strategy. 

The issues affecting derived KKS geometric quantization exposed in subsect. \cref{subsec:prequcrkks} 
do not by themselves imply that geometric quantization is outright impossible in the derived KKS set--up,
but only that the problem of geometric quantization in that context cannot 
be coped with by a straightforward extension or adaptation to the derived setting
of the basic techniques used to solve the corresponding problem in the ordinary KKS set--up. 

In paper II, we shall provide an indirect solution of the problem of geometric quantization
of derived KKS theory.
Specifically, we shall elaborate a derived 2--dimensional TCO sigma model and present 
substantial evidence that such model is the appropriate derived counterpart of the
ordinary 1--dimensional TCO quantum mechanical model.
Since that latter provides a quantization scheme for ordinary KKS theory
equivalent to geometric quantization, it is conceivable that the former may furnish
a quantization framework of derived KKS theory taking the place of geometric quantization. 

The derived TCO model is based on derived KKS geometry much as the ordinary TCO model is
on ordinary KKS geometry. This justifies the construction of the derived KKS theory
that we have carried out in this section and shall complete in the next subsection.

%Before proceeding further in our construction of the derived theory, it is worthy exploring
%whether there are other quantization options. 

%\vfil\eject 

\subsection{\textcolor{blue}{\sffamily Derived KKS theory in the regular case}}\label{subsec:crlbreg}

%In subsects. \cref{subsec:crline}--\cref{subsec:twoplct},
%we studied derived line bundles and their connections and gauge transformations
%the associated derived Poisson structures and the Hamiltonian target kernel symmetry, 
%but we did not provide relevant concrete examples of these objects. This we shall do in the present subsection
%paving the way to a better understanding of the quantization of the derived KKS set--up. 

%The geometric set--up we work with is that considered in subsect. \cref{subsec:crline} consisting of 
%a Lie group crossed module $\msM=(\msE,\msG,\tau,\mu)$ with $\msG$ a compact Lie group
%and a maximal toral crossed submodule $\msJ=(\msH,\msT)$ of $\msM$ equipped with a character
%$\beta=(\xi,\varXi)$. We require in addition that $\beta$ satisfies the admissibility condition
%stated next. 

%We have that $-i\dot\xi\in\fkt^*$. Thus, there exists an element $\varLambda

In subsects. \cref{subsec:crline}--\cref{subsec:twoplct},
we studied the derived unitary line bundles and their connections %and gauge transformations
and the associated derived Poisson structures on a regular derived homogeneous space,
finding that the target kernel symmetry is Hamiltonian.
As in ordinary KKS theory, a full fledged derived KKS theory takes shape when
the underlying homogeneous space is a coadjoint orbit.
In this subsection, we expound its construction in full detail.

The results presented below provide a concrete illustration of 
the abstract theory of subsects. \cref{subsec:crline}--\cref{subsec:twoplct}
by applying it to the description of regular derived coadjoint orbits and their derived Poisson structures.
Most importantly, however, 
they prepare the ground and provide the necessary geometric underpinning
for the quantization of such orbits and the construction of the associated derived TCO
sigma model presented in paper II.

%This we shall show in the present subsection
%preparing the ground for quantization of the derived KKS set--up and the construction
%of the derived TCO model presented in paper II. 

The derived presymplectic structure underlying a derived Poisson structure
stems from the curvature of a unitary connection of a derived line bundle
and so satisfies a quantization condition. This property plays an important role
in derived KKS theory, as in the ordinary theory, since it limits the range of quantizable coadjoint orbits
as derived symplectic manifolds. 

%The derived presymplectic structures underlying the derived Poisson structures studied in
%subsect. \cref{subsec:quapoi} stem from the curvature of unitary connections of derived line bundles
%and so satisfy a quantization condition. This property plays an important role
%in derived KKS theory, as in the ordinary theory, since it limits the range of quantizable coadjoint orbits
%as derived symplectic manifolds. 

The data of the construction we are presenting are (cf. subsect. \cref{subsec:cmcentr}):  %the following: %$(i)$
a compact Lie group crossed module $\msM=(\msE,\msG,\tau,\mu)$, %  %$(ii)$
a maximal toral crossed sub\-module $\msJ=(\msH,\msT)$ of $\msM$, %$(iii)$
an invariant pairing $\langle\cdot,\cdot\rangle$ of $\msM$ and %$(iv)$
an element $\varLambda\in\fke$ satisfying the following two admissibility conditions.
First, $\msJ=\ZZ\msM_\varLambda$, where $\msM_\varLambda$ is the
1--parameter crossed submodule of $\msM$ generated by $\varLambda$ and 
$\ZZ\msM_\varLambda$ is its centralizer crossed module; second, the map
$\xi_\varLambda:\msT\rightarrow \msU(1)$ defined by
\begin{equation}
\xi_\varLambda(\ee^x)=\ee^{i\langle x,\varLambda\rangle}
\label{crlbreg0}
\end{equation}
with $x\in\fkt$ is a character of the maximal torus $\msT$ of $\msG$. The first admissibility 
requirement implies that $\varLambda$ is a regular element of $\fke$ and that the derived coadjoint orbit
of $\varLambda$ is $\clO_\varLambda=\DD\msM/\DD\msJ$.
The second  entails that the restriction of the mapping $x\rightarrow \langle x,\varLambda\rangle/2\pi$ to the integer
lattice $\Lambda_\msG$ of $\msT$ belongs to the dual integral lattice $\Lambda_\msG{}^*$ of $\Lambda_\msG$, 
an integrality property.

By equipping $\msM$ with an invariant pairing, we are tacitly assuming that the crossed module $\msM$ is balanced
(cf. subsect. \cref{subsec:liecrmod}). The theory developed hitherto does not require such restriction
at any point. Demanding it, however, involves only a seeming loss of generality. Let us discuss this point in
some depth.

Any Lie group crossed module $\msM$ can always be extended to a balanced crossed module $\msM^c$. In fact, if
$\msW=(\msI,\msK)$ is a trivial Abelian crossed module 
\footnote{$\vphantom{\dot{\dot{\dot{a}}}}$ %To the reader's benefit,
A trivial Abelian Lie group crossed module is a Lie group crossed module $\msW=(\msI,\msK,\upsilon,\nu)$, where 
$\msI$, $\msK$ are Abelian Lie groups and the target and action maps $\upsilon$ and $\nu$ are trivial:
$\upsilon(Z)=1_\msK$ and $\nu(z,Z)=Z$ for $z\in\msK$, $Z\in\msI$.}
such that $\dim\msE+\dim\msI=\dim\msG+\dim\msK$, then the product crossed module
\footnote{$\vphantom{\dot{\dot{\dot{a}}}}$ 
Let $\msM_1=(\msE_1,\msG_1,\tau_1,\mu_1)$, $\msM_2=(\msE_2,\msG_2,\tau_2,\mu_2)$ be Lie group crossed modules.
The product of $\msM_1$, $\msM_2$ is the Lie group crossed module $\msM_1\times\msM_2=
(\msE_1\times\msE_2,\msG_1\times\msG_2,\tau_1\times\tau_2,\mu_1\times\mu_2)$ with target and action maps 
$\tau_1\times\tau_2$, $\mu_1\times\mu_2$ given by $\tau_1\times\tau_2(A_1\times A_2)=\tau_1(A_1)\times \tau_2(A_2)$
and $\mu_1\times\mu_2(a_1\times a_2,A_1\times A_2)=\mu_1(a_1,A_1)\times\mu_2(a_2,A_2)$ for $a_1\in\msG_1$,
$a_2\in\msG_2$, $A_1\in\msE_1$, $A_2\in\msE_2$.
The product crossed modules $\msM_1\times 1_2=(\msE_1\times1_{\msE_2},\msG_1\times1_{\msG_2})$,
$1_1\times\msM_2=(1_{\msE_1}\times\msE_2,1_{\msG_1}\times\msG_2)$, where $1_1=(1_{\msE_1},1_{\msG_1})$,
$1_2=(1_{\msE_2},1_{\msG_2})$ are trivial Abelian crossed modules, are crossed submodules of 
$\msM_1\times\msM_2$ isomorphic to $\msM_1$, $\msM_2$, respectively.}
$\msM^c=\msM\times\msW$ is a balanced crossed module containing $\msM$ as a submodule.
$\msM^c$ depends on the choice of $\msW$ and so is not uniquely defined. There is however a
minimal choice of $\msW$ and so of $\msM^c$ for which either $\msI$ or $\msK$ are the trivial
group $1$. The following discussion does not assume however that $\msM^c$ is minimal. 

An element $\varLambda\in\fke$ can be viewed as an element $\varLambda\in\fke\oplus\fkn$,
which we denote by the same symbol. The 1--parameter submodules $\msM_\varLambda$ and $\msM^c{}_\varLambda$
of $\varLambda$ in $\msM$ and $\msM^c$, respectively,  
are trivially related as $\msM^c{}_\varLambda=\msM_\varLambda\times 1_\msW$, where $1_\msW=(1_\msI,1_\msK)$
is a trivial Abelian crossed module. Correspondingly, the centralizer crossed submodules 
$\ZZ\msM_\varLambda$ and $\ZZ\msM^c{}_\varLambda$ of $\varLambda$ in $\msM$ and $\msM^c$ are simply related
as $\ZZ\msM^c{}_\varLambda=\ZZ\msM_\varLambda\times\msW$.

The derived Lie groups of $\msM^c$ and $\ZZ\msM^c{}_\varLambda$ factorize as 
$\DD\msM^c=\DD\msM\times\DD\msW$, $\DD\ZZ\msM^c{}_\varLambda=\DD\ZZ\msM_\varLambda\times\DD\msW$
by the definition of $\msM^c$ and the factorization of $\ZZ\msM^c{}_\varLambda$
shown in the previous paragraph
\footnote{$\vphantom{\dot{\dot{\dot{a}}}}$
For any two Lie group crossed modules $\msM_1$, $\msM_2$ with product $\msM_1\times\msM_2$,
the derived Lie groups $\DD\msM_1$, $\DD\msM_2$, $\DD(\msM_1\times\msM_2)$ stand in the relation  
$\DD(\msM_1\times\msM_2)=\DD\msM_1\times\DD\msM_2$, as is straightforward to verify. }. As a consequence, 
the derived coadjoint orbit $\clO_\varLambda$ of $\varLambda$ has the 
homogeneous space realization $\clO_\varLambda=\DD\msM^c/\DD\ZZ\msM^c{}_\varLambda$ alternative to
the basic realization $\clO_\varLambda=\DD\msM/\DD\ZZ\msM_\varLambda$ given in 
\ceqref{cmcentr1}. 

If the crossed module $\msM$ is compact and the trivial Abelian crossed module $\msW$ is taken to be
compact, the balanced crossed module $\msM^c$ is compact as well. If the group $\msK$ is further connected,
then for any maximal toral crossed submodule $\msJ$ of $\msM$
the crossed module $\msJ^c=\msJ\times\msW$ is a maximal toral crossed submodule of $\msM^c$. 
In such a case, as $\ZZ\msM^c{}_\varLambda=\ZZ\msM_\varLambda\times\msW$, if $\varLambda$ is regular
with respect to the crossed module structure of $\msM$, it is also regular with respect to that of $\msM^c$.

The above discussion shows that we are allowed to substitute the crossed module $\msM$ with a balanced extension
$\msM^c$ without changing the geometry the derived KKS theory describes.
%From now on, so, we tacitly assume that the crossed module $\msM$ is balanced.

%%%%%%%%%%%%%

The inclusion of an invariant pairing in our construction constitutes a fresh element
of the theory not considered up to this point. Given its crucial ilportance,
the natural question arises about the existence of an invariant pairing on a balanced crossed
module $\msM$. We cannot solve this issue in full generality. Existence can nevertheless be proven
under fairly broad assumptions.

To state our result, we need to introduce a few basic notions of crossed module theory.
A Lie group crossed module $\msM$ (not necessarily balanced) is said to be inert on the target
kernel if $\mu\sdot(a,X)=X$ for $a\in\msG$ and $X\in\ker\dot\tau$. $\msM$ is called
inert on the target cokernel if $\Ad a(x)=x$ for $a\in\msG$ and $x\in\fkg/\im\dot\tau$.
It is not difficult to verify that if $\msM$ is inert on the target kernel, respectively cokernel,
then any balanced extension $\msM^c$ of $\msM$ also is. 

If $\msM$ is a balanced crossed module that is compact and inert on the target kernel and cokernel,
then $\msM$ admits and invariant pairing $\langle\cdot,\cdot\rangle$. 
The key point of the proof is the existence of a left and right invariant normalized
Haar measure $\omega_\msG$ on the target group $\msG$ ensured by the compactness of this latter.
By the classic method of averaging on $\msG$, $\omega_\msG$ allows one to construct an invariant pairing
from a given non invariant one
\footnote{$\vphantom{\dot{\dot{\dot{a}}}}$ 
In outline the proof runs as follows.
The singular value decomposition theorem furnishes a simple
expression of $\dot\tau$: $\dot\tau=\sum_it_iu_i\otimes U_i{}^t$, where the scalars $t_i\geq 0$ and 
the vectors $u_i\in\fkg$ and $U_i\in\fke$ constitute orthonormal bases of $\fkg$ and $\fke$
with respect to $\msG$--invariant inner products
$(\cdot,\cdot)_\fkg$ and  $(\cdot,\cdot)_\fke$ of $\fkg$ and $\fke$, 
the $\msG$--actions being $\Ad$ and $\mu\sdot$ for $\fkg$ and $\fke$, respectively.
Using the $u_i$ and $U_i$, we can construct 
a non singular bilinear map $\langle\cdot,\cdot\rangle_{\hfpt 0}:\fkg\times\fke\rightarrow\mathbb{R}$ by
setting $\langle x,X\rangle_{\hfpt 0}=\sum_i(x,u_i)_\fkg(U_i,X)_\fke$ with $x\in\fkg$, $X\fke$.
$\langle\cdot,\cdot\rangle_{\hfpt 0}$ enjoys the symmetry property \ceqref{crmodinv1/rep} but not the invariance
property \ceqref{crmodinv14/rep} in general. This can be enforced by averaging on $\msG$, that is by setting
$\langle x,X\rangle=\int_\msG d\omega_\msG(g)\hfpt\langle \Ad g(x),\mu\sdot(g,X)\rangle_{\hfpt 0}$.
The property of inertness on the target kernel and cokernel are sufficient to ensure the non singularity
of $\langle\cdot,\cdot\rangle$.}.
We observe however that the above conditions on $\msM$ are only sufficient for the occurrence of
an invariant pairing, which may exist even when such conditions are not met.

As an illustration, we examine again the model crossed modules introduced in subsect. \cref{subsec:cmcentr}.
Consider the Lie group crossed module $\INN_\msG\msN=(\msN,\msG,\iota,\kappa)$ associated with a normal subgroup
$\msN$ of a Lie group $\msG$. Since $\dim\msN\leq\dim\msG$, $\INN_\msG\msN$ is not balanced
in general. A minimal choice of a balanced extension of $\INN_\msG\msN$ would be $\INN_\msG\msN^c=(\msN\times\msI,\msG)$,
where $\msI$ is an Abelian group such that $\dim\msI=\dim\msG-\dim\msN$.  
Since $\ker\dot\iota=0_\fkn$ in the present case, $\INN_\msG\msN$  and thus $\INN_\msG\msN^c$ 
are trivially inert on the target kernel. Conversely, $\INN_\msG\msN$ and $\INN_\msG\msN^c$ are not inert on the target
cokernel in general, because the adjoint action of $\msG$ on $\fkg/\fkn$ needs not be trivial. They are only if
$[\fkg,\fkg]\subset\fkn$. In such a case, provided $\msG$ is compact, an invariant pairing can be constructed
on $\INN_\msG\msN^c$. 

Consider next the Lie group crossed module $\msC(\!\xymatrix@C=1.3pc{\msQ\ar[r]^-{\pi}&\msG}\!)=(\msQ,\msG,\pi,\alpha)$
associated with a central extension 
$\!\mhfpt\xymatrix@C=1.3pc{1\ar[r]&\msC\ar[r]^-{\iota}&\msQ\ar[r]^-{\pi}&\msG\ar[r]&1}\!\mhfpt$.
Since $\dim\msQ\geq\dim\msG$, $\msC(\!\xymatrix@C=1.3pc{\msQ\ar[r]^-{\pi}&\msG}\!)$ is not balanced
in general. A minimal choice of a balanced extension of $\msC(\!\xymatrix@C=1.3pc{\msQ\ar[r]^-{\pi}&\msG}\!)$
would be $\msC(\!\xymatrix@C=1.3pc{\msQ\ar[r]^-{\pi}&\msG}\!)^c=(\msQ,\msG\times\msK)$,
where $\msK$ is an Abelian group such that $\dim\msK=\dim\msQ-\dim\msG$, e.g. $\msK=\msC$. Since
$\dot\iota(\fkc)$ is central in $\fkq$ and $\alpha\sdot(a,\cdot)=\Ad\sigma(a)$ for a section $\sigma:\msG\rightarrow\msQ$
of $\pi$, $\msC(\!\xymatrix@C=1.3pc{\msQ\ar[r]^-{\pi}&\msG}\!)$ and thus
$\msC(\!\xymatrix@C=1.3pc{\msQ\ar[r]^-{\pi}&\msG}\!)^c$ are inert on the target kernel.
$\msC(\!\xymatrix@C=1.3pc{\msQ\ar[r]^-{\pi}&\msG}\!)$ and $\msC(\!\xymatrix@C=1.3pc{\msQ\ar[r]^-{\pi}&\msG}\!)^c$
are trivially inert on the target cokernel since $\im\dot\pi=\fkg$. In this way, when $\msG$ is compact,
the existence of an invariant pairing on $\msC(\!\xymatrix@C=1.3pc{\msQ\ar[r]^-{\pi}&\msG}\!)^c$
is ensured.

Consider finally the Lie group crossed modules $\msD(\rho)=(\msV,\msG,1_\msG,\rho)$, where $\rho$ is a representation
of Lie group $\msG$ on the vector space $\msV$. $\msD(\rho)$ needs not be balanced in general. A minimal balanced
extension of $\msD(\rho)$ is of the form $\msD(\rho)^c=(\msV\times\msI,\msG)$ or
$\msD(\rho)^c=(\msV,\msG\times\msK)$ depending on whether $\dim\msV\leq\dim\msG$ or $\dim\msV\geq\dim\msG$
for Abelian groups $\msI$ or $\msK$ of suitable dimensions, respectively. Since $\ker\dot1_\msG=\msV$, $\msD(\rho)$
and $\msD(\rho)^c$ are inert on the target kernel only if $\rho$ is trivial. Since $\im\dot1_\msG=0_\fkg$,
$\msD(\rho)$ and $\msD(\rho)^c$ are inert on the target cokernel only if $\fkg$ is Abelian. For $\msG$ compact,
under such restrictions, then the existence of an invariant pairing on $\msD(\rho)^c$ is ensured. Clearly the
set-up emerging here is rather trivial.

%%%%%%%%%%%%%%%

The character \ceqref{crlbreg0} enters the derived KKS theory much as the character \ceqref{stkksorb1}
enters the standard one. Indeed, as explained in subsect. \cref{subsec:crline}, $\xi_\varLambda$ defines
a character of $\msJ$, $\beta_\varLambda=(\varXi_\varLambda,\xi_\varLambda)$ where
$\varXi_\varLambda=\xi_\varLambda\circ\tau$, and therefore a derived unitary line bundle
$\clL_\varLambda:=\clL_{\beta_\varLambda}$ on $\DD\msM/\DD\msJ$ 
through the geometrical construction we detailed  there. 

%The character $\xi_\varLambda$ of $\msT$ mentioned in the previous paragraph
% and focus on a related canonical group of gauge transformations. 

We are going to equip the derived line bundle $\clL_\varLambda$ with a canonical unitary connection $\rmA_\varLambda$
built upon the above data only. This connection is analogous to the canonical connection
\ceqref{stkksorb2} of a regular dual integral lattice element of ordinary KKS theory 
and is indeed the appropriate derived enhancement of this latter.
In the framework of subsect. \cref{subsec:hkksop}, the components $a_\varLambda$, $A_\varLambda$
of $\rmA_\varLambda$ are expressed in terms of the Maurer--Cartan
elements $\sigma$, $\varSigma$ of $\DD\msM$ and read as
%the maps $a_\varLambda\in\iMap(T[1]\DD\msM,\fku(1))[1])$,
%$A_\varLambda\in\iMap(T[1]\DD\msM,\fku(1)[2])$ given by 
\begin{align}
&a_\varLambda=-i\langle\sigma,\varLambda\rangle,
\vphantom{\Big]}
\label{crlbreg1}
\\
&A_\varLambda=-i\langle\dot\tau(\varLambda),\varSigma\rangle. %\vphantom{\ul{\ul{\ul{\ul{g}}}}}
\vphantom{\Big]}
\label{crlbreg2}
\end{align}
Employing relations \ceqref{hkksop25}--\ceqref{hkksop28}, it is readily checked that
$a_\varLambda$, $A_\varLambda$ obey relations \ceqref{crline11}--\ceqref{crline14}
%with respect to the character components $\xi_\varLambda$, $\varXi_\varLambda$
and so are the components of $\clL_\varLambda$. %\vfil\eject

%We work directly in components for convenience.
%in the framework of subsect. \cref{subsec:hkksop} as appropriate
%expressing the main entities. Using the Maurer--Cartan
%elements $\sigma$, $\varSigma$ of $\DD\msM$, the components $a_\varLambda$, $A_\varLambda$
%of $\rmA_\varLambda$ are

The curvature $\rmB_\varLambda$ of the connection $\rmA_\varLambda$ is the object that interests us most
because of its eventual relation to the derived KKS symplectic structure.
It is as expected a derived enhancement of the curvature of the ordinary KKS theory
canonical connection shown in \ceqref{stkksorb3}. 
The components $b_\varLambda$, $B_\varLambda$ of $\rmB_\lambda$ can be readily obtained
from the components $a_\varLambda$, $A_\varLambda$ of $\rmA_\varLambda$ given
in \ceqref{crlbreg1}, \ceqref{crlbreg2} em\-ploying relations \ceqref{crline17}, \ceqref{crline18}, % read as 
\begin{align}
&b_\varLambda=\frac{i}{2}\langle[\sigma,\sigma],\varLambda\rangle,
\vphantom{\Big]}
\label{crlbreg3}
\\
%\end{align}
%\begin{align}
&B_\varLambda=i\langle\dot\tau(\varLambda),\sdot\mu\sdot(\sigma,\varSigma)\rangle. 
\vphantom{\Big]}
\label{crlbreg4}
\end{align}
It can be checked exploiting identities \ceqref{hkksop15}, \ceqref{hkksop16} that 
$b_\varLambda$, $B_\varLambda$ obey the Bianchi identities \ceqref{crline17/1}, \ceqref{crline18/1}
as required. It can also be verified using relations \ceqref{hkksop25}--\ceqref{hkksop28}
that they obey relations \ceqref{crline19}--\ceqref{crline22}.
%with respect to $\xi_\varLambda$, $\varXi_\varLambda$.

The analysis of subsect. \cref{subsec:quapoi} shows that $-i\rmB_\varLambda$ constitutes a derived
presymplectic structure on $\DD\msM/\DD\msJ$.
As $j_V\rmB=0$ for a vector field $V\in\iVect(\DD\msM)$ only if $V=S_{\msR\rmX}$ for some $\rmX\in\DD\fkj$
pointwise on $\DD\msM$, $-i\rmB_\varLambda$  turns out to be non singular and so
a derived symplectic structure, the derived KKS $\varLambda$--structure. 
With it, there is associated a derived Poisson structure with brackets
$\{\cdot,\cdot\}_\varLambda=\{\cdot,\cdot\}_{\rmA_\varLambda}$.

%The analysis of subsect. \cref{subsec:quapoi} shows that $-i\rmB_\varLambda$ constitutes a derived
%presymplectic structure on $\DD\msM/\DD\msJ$, the derived KKS $\varLambda$--structure. 
%In contrast to its ordinary counterpart, \pagebreak $-i\rmB_\varLambda$ is not symplectic in general. 
%It is only if $\ker\dot\tau\neq 0$. In fact, by \ceqref{crlbreg4}, if $\ker\dot\tau\neq 0$ one has
%$j_V\rmB=0$ for any non zero vector field $V\in\iVect(\DD\msM)$
%such that $j_V\sigma=0$ and $j_V\varSigma\in\ker\dot\tau[1]$, e.g. $V=S_{\msL 0,H}$ with
%$H\in\ker\dot\tau[1]$ (cf. eqs. \ceqref{tarker15}, \ceqref{tarker16}). 
%In spite of this, the $\varLambda$--structure keeps its relevance. With it,
%there is associated a derived Poisson structure with brackets
%$\{\cdot,\cdot\}_\varLambda=\{\cdot,\cdot\}_{\rmA_\varLambda}$. 

Just as the standard KKS set--up of subsect. \cref{subsec:stkksorb} is fully left invariant, the derived KKS
set--up constructed above is invariant under the 
left target kernel symmetry dealt with in subsect. \cref{subsec:tarker}. 
The connection $\rmA_\varLambda$ defined componentwise by eqs. \ceqref{crlbreg1}, \ceqref{crlbreg2}
is invariant as a consequence of the invariance of the Maurer--Cartan map $\Sigma$. Indeed, relations
\ceqref{tarker17}, \ceqref{tarker18} evidently imply that $\rmA$ satisfies the invariance condition \ceqref{tkcrline2}.
In this way, also the associated curvature $\rmB_\varLambda$, given componentwise by eqs.
\ceqref{crlbreg3}, \ceqref{crlbreg4},
satisfies \ceqref{tkcrline3} and so is invariant. 

For reasons detailed in subsect. \cref{subsec:twoplct} in full generality, in derived KKS theory 
the target kernel $\DD\msM_\tau$--symmetry action on $\DD\msM/\DD\msJ$ is Hamiltonian with respect to the
derived KKS $\varLambda$--structure similarly to its counterpart
in ordinary KKS theory. The components of the associated derived moment map $\rmQ_\rmA$ are %given by %is available
%\vspace{1.125mm}
\begin{align}
&q_\varLambda(h,H)=\left\langle h,\mu\sdot(\gamma,\varLambda)\right\rangle,
\vphantom{\Big]}
\label{crlbreg11}
\\
&Q_\varLambda(h,H)=\left\langle h,[\varGamma,\mu\sdot(\gamma,\varLambda)]\right\rangle.
\vphantom{\Big]}  
\label{crlbreg12}
\end{align}
%\vspace{1.125mm}
This moment map manifestly is the proper derived counterpart of the ordinary KKS theory moment map
\ceqref{stkksorb4}. 
Using \ceqref{hkksop13}, \ceqref{hkksop14} and \ceqref{tarker11}--\ceqref{tarker14}, it is straightforward to 
verify relations \ceqref{tkcrline11}, \ceqref{tkcrline12}.
The defect map components \ceqref{tkcrline28}, \ceqref{tkcrline29} vanish
as a consequence of %\ceqref{tarker11}, \ceqref{tarker12} and
\ceqref{tarker15}, \ceqref{tarker16} and so the connection $\rmA_\varLambda$ turns out to be strict. 
The Poisson bracket relations \ceqref{tkcrline13}, \ceqref{tkcrline14} thus hold strictly, i.e. 
with vanishing right hand sides. 

The structure described above enjoys natural invariance properties under the derived automorphism
group action of the principal $\DD\msJ$--bundle $\DD\msM$ 
similarly again to the ordinary KKS theory surveyed in subsect. \cref{subsec:stkksorb}.
Indeed, we can associate with any automorphism $\Psi$ of $\DD\msM$ a gauge transformation
$\rmU_\varLambda$ of the derived line bundle $\clL_\varLambda$.
This gauge transformation is analogous to the gauge transformation \ceqref{scrlbreg5}
one associates with an automorphism of ordinary KKS theory and is indeed again a derived extension of this latter.
In the  description of automorphisms expounded in subsect. \cref{subsec:derauto},
the components $u_\varLambda$, $U_\varLambda$ of $\rmU_\varLambda$ are expressed in terms of 
the components $\psi$, $\varPsi$ of $\Psi$ as follows, %according to
\vspace{.75mm}
\begin{align}
&u_\varLambda=\exp\left(-i\int_{1_\msG}^{\,\cdot}\langle d\psi\psi^{-1},\varLambda\rangle\right),
\vphantom{\Big]}
\label{crlbreg5}
\\
&U_\varLambda=-i\langle\dot\tau(\varLambda),\varPsi\rangle.
\vphantom{\Big]_{\ul{g}}}
\label{crlbreg6}
\end{align}
\vspace{.75mm}
Since $d\psi\psi^{-1}\in\iMap(T[1]\DD\msM,\fkt[1])$ can be equated to an element 
of $\Omega^1(\msG,\fkt)$ 
with periods in the lattice $\Lambda_\msG$, $u_\varLambda$ is singlevalued as required
by virtue of the integrali\-ty property of $\varLambda$. The choice of $1_\msG$ as base point of
integration is conventional but natural. We are assuming here that $\msG$ is connected. 
Employing relations \ceqref{hkksop40}--\ceqref{hkksop43}, it is readily checked that
$u_\varLambda$, $U_\varLambda$ obey relations \ceqref{glcrline11}--\ceqref{glcrline14}
%with respect to the character components $\xi_\varLambda$, $\varXi_\varLambda$
and thus are the components of a gauge transformation of  $\clL_\varLambda$. %the derived line bundle

%$\Psi\in\iAut_{\DD\msJ}(\DD\msM)$ a gauge transformation $\rmU_\varLambda\in\iGau(\clL_\varLambda)$

% elements $u_\varLambda\in\iMap(T[1]\DD\msM,\msU(1))$, $U_\varLambda\in\iMap(T[1]\DD\msM,\fku(1)[1])$ by

%We now consider the automorphism action on $\DD\msM$.
%We work again in the framework subsect. \cref{subsec:hkksop} expressing again the main entities
%directly in components for convenience.  
% takes in the set--up under consideration. 
%We work again directly in components. As we explained in subsect. \cref{subsec:hkksop}, %the derived Lie group
%$\DD\msM$ is a principal $\DD\msJ$--bundle over $\DD\msM/\DD\msJ$ described through an appropriate 
%$\DD\fkj$--operation by generalized coordinate 
%and Maurer--Cartan maps $\gamma$, $\varGamma$ and $\sigma$, $\varSigma$ related as in
%\ceqref{hkksop13}, \ceqref{hkksop14}, satisfying \ceqref{hkksop15}, \ceqref{hkksop16} and obeying
%\ceqref{hkksop21}--\ceqref{hkksop24} and \ceqref{hkksop25}--\ceqref{hkksop28}.

The gauge transformation $\rmU_\varLambda$ %defined componentwise by \ceqref{crlbreg5}, \ceqref{crlbreg6}
is invariant as a consequence of the invariance of automorphism map $\Psi$, as relation 
\ceqref{tarker27}, \ceqref{tarker28} entail that $\rmU_\varLambda$ obeys the invariance condition \ceqref{tkcrline4}.

A mapping $\Psi\rightarrow \rmU_\varLambda$ 
from the automorphism group $\iAut_{\DD\msJ}(\DD\msM)$ of $\DD\msM$ to
the gauge transformation group $\iGau(\clL_\varLambda)$ of $\clL_\varLambda$
is established by \ceqref{crlbreg5}, \ceqref{crlbreg6}. 
As is easily checked, this mapping is a group morphism as a consequence
of the identity of $\ZZ\msM_\varLambda$ and $\msJ$. 

For an automorphism $\Psi\in\iAut_{\DD\msJ}(\DD\msM)$, let ${}^\Psi\rmA_\varLambda$ be the connection 
whose components ${}^{\psi,\varPsi}a_\varLambda$,
${}^{\psi,\varPsi}A_\varLambda$ are given by \ceqref{crlbreg1}, \ceqref{crlbreg2}
with $\sigma$, $\varSigma$ replaced by their transforms ${}^{\psi,\varPsi}\sigma$, ${}^{\psi,\varPsi}\varSigma$
(cf. eqs. \ceqref{hkksop36}, \ceqref{hkksop37}).
Then, ${}^{\psi,\varPsi}a_\varLambda={}^{u_\varLambda,U_\varLambda}a_\varLambda$,
${}^{\psi,\varPsi}A_\varLambda={}^{u_\varLambda,U_\varLambda}A_\varLambda$, 
%\begin{align}
%&{}^{\psi,\varPsi}a_\varLambda={}^{u_\varLambda,U_\varLambda}a_\varLambda,
%\vphantom{\Big]}
%\label{crlbreg7}
%\\
%&{}^{\psi,\varPsi}A_\varLambda={}^{u_\varLambda,U_\varLambda}A_\varLambda,
%\vphantom{\Big]}
%\label{crlbreg8}
%\end{align}
%where $u_\varLambda$, $U_\varLambda$ are given by \ceqref{crlbreg5}, \ceqref{crlbreg6} and 
where ${}^{u_\varLambda,U_\varLambda}a_\varLambda$, ${}^{u_\varLambda,U_\varLambda}A_\varLambda$ are given
in terms of $u_\varLambda$, $U_\varLambda$ by \ceqref{xglcrline17}, \ceqref{xglcrline18}. 
Consequently, the curvature $\rmB_\varLambda$ is automorphism invariant, 
${}^{\psi,\varPsi}b_\varLambda=b_\varLambda$, ${}^{\psi,\varPsi}B_\varLambda=B_\varLambda$.
The moment map $\rmQ_\varLambda$ is also invariant,
${}^{\psi,\varPsi}q_\varLambda=q_\varLambda$, ${}^{\psi,\varPsi}Q_\varLambda=Q_\varLambda$.
%the right hand sides of these relations are expressed according to 
%\begin{align}
%&{}^{\psi,\varPsi}b_\varLambda={}^{u_\varLambda,U_\varLambda}b_\varLambda,
%\vphantom{\Big]}
%\label{crlbreg9}
%\\
%&{}^{\psi,\varPsi}B_\varLambda={}^{u_\varLambda,U_\varLambda}B_\varLambda, %\pagebreak
%\vphantom{\Big]}
%\label{crlbreg10}
%\end{align}
%where ${}^{u_\varLambda,U_\varLambda}b_\varLambda$, ${}^{u_\varLambda,U_\varLambda}B_\varLambda$
%are given in terms of $u_\varLambda$, $U_\varLambda$ by 
%the right hand side of these relations is expressed according to
%\ceqref{xglcrline17/1}, \ceqref{xglcrline18/1}.

%As explained in subsect. \cref{subsec:prequcrkks}, 
%derived prequantization requires the derived KKS $\varLambda$--structure to be truly
%symplectic. This requirement is met only when the underling crossed module
%$\msM$ is quasi injective, i.e. such that $\ker\tau$ a discrete subgroup of $\msE$
%and so that $\ker\dot\tau=0$. It is not clear whether such a restriction can be
%dropped through a more extensive reformulation of the theory, which is left to future work.

We concluded this subsection by noting that the connection $\rmA_\varLambda$ is of 1--form type
as is immediate to realize by inspection of \ceqref{crlbreg1}, \ceqref{crlbreg2}.
Since its curvature $\rmB_\varLambda$ is also non singular as already
noticed, the incomplete derived prequantization procedure
described in subsect. \cref{subsec:prequcrkks} can be implemented in the present case.

%\vfil\eject

\subsection{\textcolor{blue}{\sffamily Conclusions}}\label{subsec:hkksend}

In this section, we have gone a long way toward a complete and satisfactory
formulation of higher KKS theory. We have done so in the derived framework,
that seems to be ideally suited for this purpose. 
As we have seen, derived prequantization
suffers certain limitation which we have detailed. We believe that the reason for this is ultimately that
the appropriate geometric quantization of derived KKS theory, whatever form it takes,  cannot have 
some kind of quantum mechanical model, albeit exotic, as its end result but a two
dimensional quantum field theory. This seems to give further support to standard expectations %to the extent
that derived KKS theory can be regarded as some kind of categorification of the ordinary theory. 
The TCO model elaborated in paper II, whose geometric foundation
is provided precisely by derived KKS theory, is an attempt to concretize the above intuitions.

\vfil\eject

\vfil\eject

\begin{appendix}
%\appendix

\section{\textcolor{blue}{\sffamily Appendixes}}\label{app:app}

The following appendixes review certain nonstandard notions whose knowledge is tacitly assumed 
throughout the paper. %They do not have any
Our presentation has no pretence of rigour and completeness 
and serves mainly the purpose of facilitating the reading.

%\vfil\eject

\subsection{\textcolor{blue}{\sffamily Generalities on the graded geometric set--up
}}\label{subsec:gradgeo}

In this paper, we adhere to a graded geometric perspective. This  
provides an especially natural language for the description the higher geometric structures
encountered along the way. The reader is referred to e.g.
ref. \ccite{Cattaneo:2010re} for a readable introduction to graded differential geometry
and its applications.
%allowing as it does to treat differential forms on any manifold $X$ as
%functions on the shifted tangent bundle $T[1]X$.
Here, we shall limit ourselves to discuss a few conceptual issues. 

Adopting the most common standpoint of the physical literature, we shall describe
a graded manifold $X$ using local coordinates. As well--known, these divide into
body and soul coordinates $x^a$ and $\xi^r$ characterized by integer degrees.
%$\deg x^a=0$ and $\deg\xi^r$, respectively,
The coordinates should be regarded as formal parameters.
The topology and geometry of $X$ is encoded in the transition functions
relating  sets of coordinates $x^a$, $\xi^r$ and $\tilde x^a$, $\tilde\xi^r$. These are 
the %certain
degree $0$ smooth functions $f^a(x)$, $f^r{}_{r_1\ldots r_p}(x)$
%with $f^r{}_{r_1\ldots r_p}(x)=0$ unless $\deg\tilde\xi^r=\deg\xi^{r_1}+\cdots+\deg\xi^{r_p}$
appearing in the coordinate change relations
\begin{align}
&\tilde x^a=f^a(x),
\vphantom{\Big]}
\label{igm1}
\\
&\tilde \xi^r=\sss_{p\geq 0}f^r{}_{r_1\ldots r_p}(x)\xi^{r_1}\cdots\xi^{r_p}.
\vphantom{\Big]}
\label{igm2}
\end{align}
The summation occurring in \ceqref{igm2} is in principle infinite.
To avoid to be embroiled in subtle problems concerning the proper treatment of such formal
expressions, we shall tacitly consider only non negatively graded manifolds unless otherwise stated.
For these, the soul coordinate degree $\deg\xi^r$ are all strictly positive
and so all the summations such as the above finite.

At several points in our analysis the distinction between ordinary and internal functions and maps,
albeit technical, will play an important role. To the reader's convenience, we recall briefly the difference
between these two types of maps.
Let $X$ and $Y$ be graded manifolds and $\varphi:Y\rightarrow X$ a smooth map. 
In terms of local coordinates $x^a$, $\xi^r$ and  
$y^i$, $\eta^h$ of $X$ and $Y$, $\varphi$ has an expression of the form 
\begin{align}
&\varphi^a(y,\eta)=\sss_{p\geq 0}\varphi^a{}_{h_1\ldots h_p}(y)\eta^{h_1}\cdots\eta^{h_p},
\vphantom{\Big]}
\label{igm3}
\\
&\varphi^r(y,\eta)=\sss_{p\geq 0}\varphi^r{}_{h_1\ldots h_p}(y)\eta^{h_1}\cdots\eta^{h_p},
\vphantom{\Big]}
\label{igm4}
\end{align}
%\vfil\eject\noindent
where $\varphi^a{}_{h_1\ldots h_p}(y)$, $\varphi^r{}_{h_1\ldots h_p}(y)$ are smooth functions. 
%such that $\varphi^a{}_{h_1\ldots h_p}(y)=0$ and $\varphi^r{}_{h_1\ldots h_p}(y)=0$
%unless respectively $\deg\varphi^a{}_{h_1\ldots h_p}(y)+\deg\eta^{h_1}+\cdots+\deg\eta^{h_p}=0$
%and $\deg\varphi^r{}_{h_1\ldots h_p}(y)+\deg\eta^{h_1}+\cdots+\deg\eta^{h_p}=\deg\xi^r$. 
If $\varphi$ is an ordinary map, the $\varphi^a{}_{h_1\ldots h_p}(y)$, $\varphi^r{}_{h_1\ldots h_p}(y)$ 
have all degree $0$. 
If instead $\varphi$ is internal, the $\varphi^a{}_{h_1\ldots h_p}(y)$, $\varphi^r{}_{h_1\ldots h_p}(y)$ 
may have non zero internal degree.
As above, the summations occurring in \ceqref{igm3}, \ceqref{igm4} have generally an infinite range.
For the non negatively graded manifolds which we restrict to, the degrees of the 
$\phi^a{}_{h_1\ldots h_p}(y)$, $\phi^r{}_{h_1\ldots h_p}(y)$ are allowed
to take only non negative values and the summations are again all \linebreak finite. 
We distinguish the ordinary and internal map sets
by employing the generic notation $\Fun$ and $\Map$ for the former and 
$\iFun$ and $\iMap$ for the latter. In sect. \cref{sec:derform},
the derived set--up is formulated mostly in the internal case, since internal functions
provide a broader function range. The ordinary case can be treated anyway essentially in the same way. 

We can associate a clone manifold $X^+=\iMap(*,X)$ with any graded manifold $X$, where $*$ is the
singleton manifold. Inspection of \ceqref{igm3}, \ceqref{igm4} for the case where $Y=*$ reveals that
$X^+$ provides a description of the `range' of any assigned local coordinate
system $x^a$, $\xi^r$ of $X$, since the points of the range are in bijective correspondence
with a subset of maps of $X^+$. Working with $X^+$ furnishes in this manner an index free 
way of encoding coordinate expressions. While this does not constitute by itself a sufficient
reason for a systematic application of cloning in general, it is when $X$ is endowed with a linear
structure that the usefulness of cloning becomes apparent as we illustrate next.

Let $E$ be a negatively graded vector space. Then, the dual $E^*$ of $E$ is a positively
graded vector space. The coordinate functions of $E$, as elements of $E^*$, have thus positive
degrees. With any basis $x_a$ of $E$, there is associated a full set of coordinate functions
of $E$, namely the basis $x^a$ of $E^*$ dual to $x_a$. A basis change $x_a\mapsto\tilde x_a$
in $E$ induces a linear transformation $x^a\mapsto\tilde x^a$.
In this way, upon treating the $x^a$ as formal parameters,
$E$ can be viewed as a positively graded manifold. So, while $E$ is negatively graded as a vector space,
it is positively graded as a manifold. Because of this sign mismatch between vector and geometric grading,
the geometric structure of $E$ cannot be directly described through the vector structure. However,
since the clone manifold $E^+$ of $E$ is naturally a positively graded vector space with a vector 
grading content matching the geometric one of $E$
\footnote{$\vphantom{\dot{\dot{\dot{a}}}}$  The difference between $E$ and $E^+$
can be illustrated somewhat more explicitly as follows. 
Let $E^0$ the ungraded vector space underlying $E$ and
$x^0{}_a$ the basis of $E^0$ corresponding to a basis $x_a$ of $E$.
Then, $E$ consists of the linear combinations $c^ax_a$ with $c^a\in\bbR$, while
$E^+$ of the formal expressions $x^a\otimes x^0{}_a$.},
the geometric properties of $E$ can be expressed
in principle as vector properties of $E^+$ in a very convenient and natural manner.  
Similar considerations apply to negatively graded vector bundles. 

%$E$ and $E^+$ can so be connected through a sequence of (de)suspension maps. 

Though the difference between a vector space or bundle $E$ and its clone $E^+$ mat\-ters if one is to have
the correct grading matching as explained above, $E$ and $E^+$ may be harmlessly confused in practice
in most instances. We decided so, albeit with some hesitation, that keeping the distinction between
$E$ and $E^+$ manifest may burden notation unnecessarily. Throughout this paper, therefore, we shall not 
discriminate notationally between $E$ and $E^+$. It will be clear from context which is which.
In our analysis, we shall mainly focus on graded geometric aspects 
and correspondingly it will be the clone space that will be tacitly considered. 
At any rate, the reader not interested in fine technical points such as this
and content with mere formal manipulations can ignore cloning altogether.

%Though one may naively think that the clone manifold $X^+$ of a graded manifold
%$X$ is essentially the same kind of object as $X$ is, it turns out in several contexts that the
%distinction between $X$ and $X^+$ matters if one is to have the correct grading matching
%and for this reason we rely extensively on cloning in our analysis.
%The reader not interested in finely technical points such as this
%and content with mere formal manipulations can simply ignore cloning altogether. 
%Else, he/she can get an intuitive idea of its scope from the following example. 

%Because of this grading mismatch, geometric properties
%of $E$ cannot be expressed through vector relations of $E$ as a rule, but they can provided $E$
%is replaced by its clone $E^+$, which unlike $E$ is a positively graded vector space
%with the correct grading values.

%It would be natural to express the geometric properties of the manifold $E$ by means of algebraic relations
%of the vector space $E$, but this is not possible because the grading mismatch that we found. 
%It becomes possible, if we replace the vector space $E$ with the clone $E^+$ of the manifold $E$,
%which is a vector space as $E$ is, but it is positively graded and with the correct grading content. 

%An ordinary manifold, viewed as a graded manifold concentrated in degree $0$,
%can be identified with its clone manifold, so that the distinction between the manifold
%and its clone can be dropped.

%\vfil\eject

\subsection{\textcolor{blue}{\sffamily Differential forms as graded functions
}}\label{subsec:difform}

Differential forms on a manifold $X$ can be described as graded functions on the shifted tangent bundle
$T[1]X$. We have indeed a graded algebra isomorphism $\Omega^*(X)\simeq\Fun(T[1]X)$. This well--known
property will be extensively exploited in the present work. In the graded geometric
framework we are adopting, it is definitely more natural to treat forms as graded functions.
This standpoint has also the advantage of extending the range of manipulations which can be
performed with forms. We shall also consider internal differential forms and deal with these
as internal functions exploiting the isomorphism ${\bf\Omega}^*(X)\simeq\iFun(T[1]X)$.
%wherever required.

%\vfil\eject

\subsection{\textcolor{blue}{\sffamily The operational approach}}\label{subsec:clkks}

The operational approach underlies many of the constructions of ordinary and derived
KKS theory as well as the related ordinary and derived TCO models for its aptness
to describe the basic geometry of principal bundles, in particular homogeneous spaces,
and provides a efficient framework for the application of standard cohomological methods
such as transgression. For this reason, we review its basic notions here.
 An exhaustive exposition of the subject and its applications to
topology and geometry can be found in ref. \ccite{Greub:1973ccc}. 

Suppose that $\fkf$ is a Lie algebra and $\msA$ an associative commutative graded algebra. 
An $\fkf$--operation on $\msA$ consists of a degree $-1$ derivation $j_x$ and a degree $0$ derivation $l_x$ 
for every $x\in\fkf$ and a degree $1$ derivation $d$ of $\sfA$
obeying the relations
\begin{align}
&[j_x,j_y]=0,
\vphantom{\Big]}
\label{hkksop1}
\\
&[l_x,j_y]=j_{[x,y]}, \qquad [l_x,l_y]=l_{[x,y]}, 
\vphantom{\Big]}
\label{hkksop2}
\\
&[j_x,d]=l_x, \qquad [l_x,d]=0,
\vphantom{\Big]}
\label{hkksop3}
\\
&[d,d]=0
\vphantom{\Big]}
\label{hkksop4}
\end{align}
for $x,y\in\fkf$. All commutators shown above are graded. 

Suppose $P$ is a manifold, $\Vect(P)$ the Lie algebra of vector fields of $P$
and $\Fun(T[1]P)$ the graded algebra of functions on the shifted tangent bundle
$T[1]P$ of $P$. Then, there exists a canonical $\Vect(P)$--operation on $\Fun(T[1]P)$,
customarily called Cartan calculus of $P$. Under the isomorphism 
$\Fun(T[1]P)\simeq\Omega^*(P)$ recalled in app. \cref{subsec:difform},
this consists of the familiar degree $-1$
contractions $j_V$ and degree $0$ Lie derivatives $l_V$ along the vector fields $V\in\Vect(P)$
and the de Rham differential $d$, all realized as graded vector fields on $T[1]P$.
The action of such derivations extends to the mapping spaces $\Map(T[1]P,X)$, where $X$
is a graded manifold, %through the mappings' coordinate expression. by reducing it to the
as an action on functions on $T[1]P$ using coordinates of $X$.

Suppose that $\msF$ is a Lie group, $B$ is a manifold and 
$P$ is a principal $\msF$--bundle over $B$. With the right action
of $\msF$ on $P$, there is associated an $\mathfrak{f}$--operation. 
%, where $\mathfrak{f}$ is the Lie algebra of $\msF$.
comprising the contractions $j_x$ and Lie derivatives
$l_x$ along the vertical vector fields $S_x\in\Vect(P)$ of the right $\msF$--action of 
$P$ associated with the Lie algebra elements 
$x\in\mathfrak{f}$ and the degree $1$ de Rham differential $d$ of $P$. 
Since $B=P/\msF$, the operational set--up allows for an elegant description of the
differential geometry of $B$. The function algebra $\Fun(T[1]B)$ of $T[1]B$
can be identified with the basic subalgebra of 
$\Fun(T[1]P)$, the joint kernel of the derivations $j_x$ and $l_x$ with $x\in\fkf$. 
Similarly, the mapping space $\Map(T[1]B,X)$ can be identified with a basic subspace of 
$\Map(T[1]P,X)$.

%Derived Lie groups are instances of positively graded manifolds. 
The operational approach, 
can be adapted straightforwardly to the case where the principal bundle $P$, its base $B$ and
its structure group $\msF$ are positively graded manifolds. 
One important aspect distinguishing an operational set--up of graded manifolds from an ordinary one 
is that by consistency the underlying graded algebra is the internal function algebra $\iFun(T[1]P)$
rather than the ordinary function algebra $\Fun(T[1]P)$, because of the
graded nature of $\fkf$. Besides, the whole formal apparatus works much in the same way.

\end{appendix}

\vfil\eject

\markright{\textcolor{blue}{\sffamily Acknowledgements}}

\noindent
\textcolor{blue}{\sffamily Acknowledgements.}
The author thanks E. Meinrenken for correspondence. 
The author  acknowledges financial support from INFN Research Agency
under the provisions of the agreement between University of Bologna and INFN. 

\vspace{1cm}

\vfil\eject

\noindent
\textcolor{blue}{\bf\sffamily References}
%\section{\textcolor{blue}{\sffamily References}}\label{ref:ref}

\end{document}